\documentclass{aa}
\usepackage{graphicx,color}

\def\mb#1{\setbox0=\hbox{$#1$}\kern-.025em\copy0\kern-\wd0
\kern-0.05em\copy0\kern-\wd0\kern-.025em\raise.0233em\box0}

\begin{document}

   \title{Dynamical stability of infinite homogeneous
   self-gravitating systems: application of the Nyquist method}

\titlerunning{Dynamical stability of infinite homogeneous self-gravitating systems}

   \author{Pierre-Henri Chavanis}

\institute{ Laboratoire de Physique Th\'eorique (UMR 5152 du CNRS), Universit\'e Paul
Sabatier, 118 route de Narbonne 31062 Toulouse, France\\
\email{chavanis@irsamc.ups-tlse.fr}}

   \date{\today}

\abstract{We complete classical investigations concerning the dynamical
stability of an infinite homogeneous gaseous medium described by the
Euler-Poisson system or an infinite homogeneous stellar system
described by the Vlasov-Poisson system (Jeans problem). To determine
the stability of an infinite homogeneous stellar system with respect to a
perturbation of wavenumber $k$, we apply the Nyquist method. We first
consider the case of single-humped distributions and show that, for
infinite homogeneous systems, the onset of instability is the
same in a stellar system and in the corresponding barotropic gas,
contrary to the case of inhomogeneous systems. We show that this
result is true for any symmetric single-humped velocity distribution,
not only for the Maxwellian.  If we specialize on isothermal and
polytropic distributions, analytical expressions for the growth rate,
damping rate and pulsation period of the perturbation can be
given. Then, we consider the Vlasov stability of symmetric
and asymmetric double-humped distributions (two-stream stellar
systems) and determine the stability diagrams depending on the degree
of asymmetry. We compare these results with the Euler stability of two
self-gravitating gaseous streams. Finally, we determine the
corresponding stability diagrams in the case of plasmas and compare
the results with self-gravitating systems.

 }

\maketitle

\section{Introduction}
\label{sec_introduction}

The dynamical stability of self-gravitating systems is an important
problem in astrophysics.  This problem was first considered by Jeans
(1902,1929) who studied the linear dynamical stability of an infinite
homogeneous collision-dominated gas described by the Euler
equations. He found that this system becomes unstable if the
wavelength of the perturbation exceeds a critical value
$\lambda_c=(\frac{\pi}{G\rho})^{1/2}c_s$ (where $c_s$ is the velocity
of sound) nowadays called the Jeans length. In that case, the
perturbation grows exponentially rapidly without oscillating. By
contrast, for small wavelengths $\lambda<\lambda_c$, the perturbation
oscillates without attenuation and behaves like a gravity-modified
sound wave. As is well-known, the Jeans approach suffers from a
mathematical inconsistency at the start. Indeed, an infinite
homogeneous gravitating system cannot be in static equilibrium since
there are no pressure gradients to balance the gravitational
force. Jeans removed this inconsistency by assuming that the Poisson
equation describes only the relationship between the perturbed
gravitational potential and the perturbed density. However, this
assumption is {\it ad hoc} and is known as the {\it Jeans swindle}
(Binney \& Tremaine 1987). A Jeans-type analysis can however be
justified in certain situations: (i) In cosmology, when we work in the
comoving frame, the expansion of the universe introduces a sort of
``neutralizing background'' in the Poisson equation (like in the
Jellium model of plasma physics). In that case, an infinite
homogeneous self-gravitating medium can be in static equilibrium
(Peebles 1980). Therefore, the Jeans instability mechanism is relevant to
understand the formation of galaxies from the homogeneous early
universe. (ii) If we consider a uniformly rotating system, the
gravitational attraction can be balanced by the centrifugal
force. Therefore, a homogeneous system can be in static equilibrium in
the rotating frame provided that the angular velocity and the density
satisfy a well-defined relation (Chandrasekhar 1961). (iii) The Jeans
procedure is approximately correct if we only consider perturbations
with small wavelengths, much smaller than the Jeans length (Binney \&
Tremaine 1987).

The dynamical stability of an infinite homogeneous encounterless
stellar system described by the Vlasov equation was considered much
later by Lynden-Bell (1962)\footnote{See also Simon (1961) and Liboff
(1963).} who used methods similar to those introduced by Landau (1946)
in plasma physics. There exists indeed many analogies between
self-gravitating systems and plasmas since the Coulombian and
Newtonian interactions both correspond to an inverse square
law. However, there exists also crucial differences. First of all,
gravity is attractive whereas electricity is repulsive. On the other
hand, as already indicated, the self-gravity of a uniform gravitating
system is usually not neutralized, contrary to a plasma where the
presence of electrons and ions provide electrical neutrality.  In
order to circumvent the Jeans swindle, Lynden-Bell considered the case
of a rotating system and noted that the angular velocity plays a role
similar to the magnetic field in plasma physics (Bernstein 1958). He
found that the system becomes unstable above a critical
wavelength. For a Maxwellian distribution, this critical wavelength
$\lambda_c=(\frac{\pi}{G\rho})^{1/2}\sigma$ is equivalent to the Jeans
length if we identify the r.m.s. velocity of the stars $\sigma$ in one
direction with the velocity of sound $c_s$ in an isothermal gas. In
that case, the perturbation grows exponentially rapidly without
oscillating. By contrast, for small wavelengths $\lambda<\lambda_c$,
the perturbation is damped exponentially. This is similar to the
Landau damping in plasma physics that is derived in the complete
absence of collisions.

Sweet (1963) considered the gravitational instability of a system of
gas and stars in relative motion and provided a detailed analysis of
the dispersion relation in different cases. He considered in
particular the situation where the gas is at rest ($U=0$) and the star
system is made of two identical interpenetrating streams with
Maxwellian distribution moving in opposite direction with equal
velocity $\pm v_a$. This corresponds to the Kapteyn-Eddington
two-stream representation in the solar neighborhood. He mentioned that
the two humps could be asymmetric but he did not study the consequence
of this asymmetry in detail. One important conclusion of his work is
that the critical Jeans wavelength is reduced by the relative motion
of the gas and stars and that it vanishes when the relative velocity
$v_a$ exceeds a limit of the order of the effective velocity of sound
$c_s$ in the gas. In particular, the star streaming in the solar
neighborhood can cause the gas to be unstable at all
wavelengths. A similar conclusion was reached by Talwar \& Kalra (1966) who
considered the contrastreaming instability of two self-gravitating
gaseous streams with velocity $\pm U$ described by fluid equations. In
the subsonic regime $U<c_s$, they found that the critical Jeans
wavelength is reduced and tends to zero as the streaming velocity $U$
approaches the velocity of sound $c_s$ (by contrast, in the supersonic
regime $U>c_s$, the critical wavelength is larger than the Jeans
length without streaming). The works of Sweet (1963) and Talwar \&
Kalra (1966) were further developed by Ikeuchi {et al.} (1974) who
presented various stability diagrams for two-stream stellar systems,
gaseous-stellar systems, two gaseous systems and plasmas. For a purely
two-stream stellar systems (without gas), they found that the critical
Jeans length becomes larger due to the stabilization effect of
relative velocity, contrary to the case where a gas component is
present. The two-stream instability was also discussed by Araki (1987)
for infinite homogeneous and uniformly rotating stellar systems.

The seminal works of Jeans (for gaseous systems) and Lynden-Bell (for
stellar systems) have been continued in several directions. For
example, the Jeans instability of a self-gravitating gas in the
presence of a magnetic field or an overall rotation has been studied
in detail by Chandrasekhar (1961). On the other hand, the
gravitational stability of a disk of stars is treated in the classical
paper of Toomre (1964). Other interesting contributions are reviewed
by Binney \& Tremaine (1987). However, this problem is largely
academic because real stellar systems (like stars and galaxies) are
not spatially homogeneous and are limited in space. Now, when spatial
inhomogeneity is properly taken into account, the picture becomes very
different. The dynamical stability of stars has been considered by
various authors such as Eddington (1918), Ledoux (1945) and
Chandrasekhar (1964) and different stability criteria for
the Euler-Poisson system have been obtained. For example, it can be
shown that polytropic stars are Euler stable if their index $n\le 3$ while they
are unstable if $3<n\le 5$ (polytropic stars with $n>5$, including
isothermal stars $n\rightarrow +\infty$, have infinite mass). On the
other hand, the Vlasov stability of stellar systems has been studied by
Antonov (1960a), Doremus et al. (1971), Kandrup \& Sygnet (1985)
and others. They have shown that all stellar systems with a
distribution function $f=f(\epsilon)$ which is a monotonically
decreasing function of the energy are linearly dynamically stable with
respect to the Vlasov-Poisson system. In particular, all polytropic
galaxies with index in the range $3/2\le n\le 5$ (such that the total
mass is finite) are stable\footnote{The difference between the
dynamical stability of gaseous stars and stellar systems, which is
related to the Antonov (1960b) first law, can also be related to a notion of
``ensemble inequivalence'' like in thermodynamics (see Chavanis
2006a).}. Very recently, it was proven rigorously by Lemou et
al. (2009) that these systems are also {\it nonlinearly} dynamically
stable with respect to the Vlasov-Poisson system. These results show
that, when the spatial inhomogeneity of the system is properly
accounted for, the Jeans instability is suppressed\footnote{A form of
Jeans instability is recovered for spatially inhomogeneous isothermal
($n=\infty$) and polytropic ($n>5$) systems confined within a box
(Padmanabhan 1990, Semelin et al. 2001, Chavanis 2002a,2002b, Taruya \& Sakagami 2003a).}.

Recently, there was a renewed interest for this classical problem due
to analogies with other physical systems. For example, the community
of statistical mechanics involved in the dynamics and thermodynamics
of systems with long-range interactions (Dauxois et al. 2002) has
studied a toy model called the Hamiltonian Mean Field (HMF) model
which presents many features similar to self-gravitating
systems\footnote{In fact, this model is directly inspired by
astrophysics. It was introduced by Pichon (1994) as a simplified model
to describe the formation of bars in disk galaxies (see a detailed
historic in Chavanis et al. (2005)).}. In particular, this model
presents an instability below a critical temperature that is similar
to the Jeans instability in astrophysics. Interestingly, a spatially
homogeneous phase exists for this model at any temperature (it is
stable for $T>T_c$ and unstable for $T<T_c$). Therefore, there is no
mathematical inconsistency (i.e. no ``Jeans swindle'') when we perform
the stability analysis of the homogeneous phase of this system
(Inagaki \& Konishi 1993, Choi \& Choi 2003, Yamaguchi et al. 2004,
Chavanis \& Delfini 2009). On the other hand, in biology, several
researchers in physics and applied mathematics have studied the
phenomenon of chemotaxis using the Keller-Segel (1970) model in order
to explain the self-organization of bacterial colonies, cells or even
social insects. Recently, it was shown that the chemotactic
instability in biology bears some deep analogies with the Jeans
instability in astrophysics (Chavanis 2006b)\footnote{Curiously, Keller \& Segel (1970) did not point out this analogy and interpreted instead the chemotactic instability in relation to B\'enard convection.}. Interestingly, in the
biological problem, there is no ``Jeans swindle'' because the Poisson
equation is replaced by a more complex reaction-diffusion equation
that allows for the existence of infinite and homogeneous
distributions.

These analogies were a source of motivation to study anew the
classical Jeans problem in astrophysics (for the Euler-Poisson and
Vlasov-Poisson systems). In fact, we discovered that the study of this
classical problem was still incomplete. For example, the case of
(spatially homogeneous) polytropic distribution functions has not been
investigated in detail and the case of a symmetric double-humped
distribution considered by Ikeuchi et al. (1974) 
needs further discussion. On the
other hand, the stability of an asymmetric double-humped distribution
has never been considered in the astrophysical literature (although
the interest of such distributions was mentioned early by Sweet
(1963)) and could form an interesting theoretical problem. The
stability criteria for homogeneous stellar systems can be established
very easily with the powerful Nyquist method (Nyquist 1932) used in
plasma physics. This method can be readily applied to homogeneous
self-gravitating systems. However, since the gravitational interaction
is attractive while the electric interaction is repulsive, a crucial
sign changes in the dispersion relation and the problem must be
reconsidered. In particular, this change of sign is the reason for the
Jeans instability in astrophysics and the Nyquist method gives a nice
graphical illustration of this instability. This method does not seem
to be well-known by astrophysicists and is not exposed in classical
monographs\footnote{As mentioned by the referee, an early application
of the Nyquist method was made by Lynden-Bell (1967) in a not very easily accessible paper. Nyquist's method has also been used by Weinberg (1991) to study the stability of spherical stellar systems. We are not aware of other references.}.  Although it essentially has an academic interest since
real stellar systems are not spatially
homogeneous\footnote{In fact, the Nyquist method can be generalized to apply to spatially {\it inhomogeneous} self-gravitating
systems (see Weinberg 1991).}, we think that it deserves a detailed pedagogical
exposition. We thus propose an exhaustive study of the linear
dynamical stability of infinite and homogeneous gaseous (Euler) and
stellar (Vlasov) systems that uses the Nyquist method and completes
previous works on that problem.

The paper is organized as follows. In Sec. \ref{sec_gas}, we consider
a self-gravitating barotropic gas described by the Euler-Poisson
system. We recall the classical Jeans instability criterion and, for
future comparison, consider explicitly the case of isothermal
and polytropic equations of state. In Sec. \ref{sec_stellar}, we
consider a collisionless stellar system described by the
Vlasov-Poisson system.  Using the Nyquist method, we derive the Jeans
instability criterion for an infinite and homogeneous distribution. We
show that, for any single humped distribution function of the form
$f=f(v^2)$ with $f'(v^2)<0$, the Jeans instability criterion for a
stellar system is equivalent to the Jeans instability criterion for
the corresponding barotropic gas with the same equation of state. Of
course, the dispersion relation and the evolution of the perturbation
is different in the two systems (gas and stellar system) but the
threshold of the instability is the same. This generalizes the result
obtained by Lynden-Bell (1962) for the Maxwellian distribution. In
Secs. \ref{sec_vb} and
\ref{sec_vb2}, we specifically consider the case of isothermal and
polytropic distribution functions and derive analytical expressions
for the growth rate and damping rate of the perturbation by adapting
methods of plasma physics to the gravitational context. In
Sec. \ref{sec_vhq}, we consider a symmetric double-humped distribution
made of two counter-streaming Maxwellians with velocities $\pm v_a$
and establish the stability diagram. In Sec. \ref{sec_vha}, we
generalize our results to the case of an asymmetric double-humped
distribution. We find that: (i) The system is stable for
$\lambda<(\lambda_c)_*$ and unstable for $\lambda>(\lambda_c)_*$ where
the critical wavelength $(\lambda_c)_*$ is a non-trivial function of
the relative velocity $v_a$ and asymmetry parameter $\Delta$. (ii) The
critical wavelength $(\lambda_c)_*$ is always larger than the critical
wavelength in the absence of streaming ($v_a=0$) so that the
instability is delayed. (iii) The phase velocity of the unstable mode
corresponds to the global maximum of the distribution function.  In
Sec. \ref{sec_plasmas}, we consider the case of plasmas. We recall the
well-known fact that single-humped distributions are always
stable. Then, we consider the case of symmetric and asymmetric
double-humped distributions. For an asymmetry parameter
$\Delta<\Delta_*=3.3105$: (i) the system is stable for
$v_a^2<y_c(\Delta)T$ where $y_c(\Delta)$ depends on the asymmetry
($y_c=1.708$ for a symmetric distribution); (ii) for
$v_a^2>y_c(\Delta)T$, the system is stable for $\lambda<(\lambda_c)_*$
and unstable for $\lambda>(\lambda_c)_*$ where $(\lambda_c)_*$ depends
on the relative velocity $v_a$ and on the asymmetry parameter
$\Delta$. For an asymmetry parameter $\Delta>\Delta_*=3.3105$: (i) the
system is stable for $v_a^2<y_*(\Delta)T$ where $y_*(\Delta)$ depends
on the asymmetry; (ii) for $y_*(\Delta)T<v_a^2<y_c(\Delta)T$, the system is stable
for $\lambda<(\lambda_c)_1$, unstable for
$(\lambda_c)_1<\lambda<(\lambda_c)_2$ and stable for
$\lambda>(\lambda_c)_2$ where $(\lambda_c)_1$ and $(\lambda_c)_2$
depend on the relative velocity $v_a$ and on the asymmetry parameter
$\Delta$; (iii) for $v_a^2>y_c(\Delta)T$, the system is stable
for $\lambda<(\lambda_c)_1$ and unstable for
$\lambda<(\lambda_c)_1$.  Throughout the paper, we also compare our results obtained
for stellar systems (Vlasov) to those obtained for gaseous systems
(Euler).

\section{Self-gravitating barotropic gas}
\label{sec_gas}

\subsection{The Euler-Poisson system}
\label{sec_ep}

We consider a self-gravitating gas described by the
Euler-Poisson system
\begin{equation}
{\partial\rho\over\partial t}+\nabla \cdot (\rho {\bf u})=0,
\label{ep1}
\end{equation}
\begin{equation}
{\partial {\bf u}\over\partial t}+({\bf u}\cdot \nabla) {\bf u}=-{1\over\rho}\nabla p-\nabla\Phi,
\label{ep2}
\end{equation}
\begin{equation}
\Delta\Phi=4\pi G\rho.
\label{ep3}
\end{equation}
These are the usual equations used to study the dynamical stability of
a gaseous medium like a molecular cloud or a star. We assume that the
gas is a perfect gas in local thermodynamic equilibrium (L.T.E).
Therefore, at each point, the distribution function of the particles
(atoms or molecules) is of the form
\begin{equation}
f({\bf r},{\bf v},t)=\biggl \lbrack {1\over 2\pi T({\bf r},t)}\biggr\rbrack^{3/2}\rho({\bf r},t)e^{-{ \lbrack {\bf v}-{\bf u}({\bf r},t)\rbrack^{2}\over 2 T({\bf r},t)}}.
\label{ep4}
\end{equation}
Note that the kinetic temperature $T({\bf r},t)=\frac{1}{3}\langle
w^2\rangle$ where ${\bf w}={\bf v}-{\bf u}({\bf r},t)$ is the relative
velocity\footnote{Throughout this paper, by an abuse of language, we
shall define the kinetic temperature by $T({\bf
r},t)\equiv \frac{1}{d}\langle ({\bf v}-{\bf u}({\bf r},t))^2\rangle$ where
$d$ is the dimension of space. Therefore, the kinetic temperature is a
measure of the local velocity dispersion of the particles in one
dimension. It is related to the real kinetic temperature by $T({\bf
r},t)=k_B T_{real}({\bf r},t)/m$. We do that in order to extend this
definition to the case of collisionless stellar systems described by
the Vlasov equation where the individual mass of the stars does not
appear.}. Introducing the pressure $p={1\over 3}\int f w^{2}d{\bf v}$, we find that the local equation of state is
\begin{equation}
p({\bf r},t)=\rho({\bf r},t)T({\bf
r},t).
\label{ep5}
\end{equation}
The system of equations (\ref{ep1})-(\ref{ep3}) is not closed since it
must be supplemented by an equation for the energy or, equivalently,
for the temperature $T({\bf r},t)$. Here, we shall restrict ourselves
to the case of a {\it barotropic gas} for which the pressure depends
only on the density
\begin{equation}
p({\bf r},t)=p[\rho({\bf r},t)].
\label{ep5b}
\end{equation}
The local velocity of sound is
\begin{equation}
c_{s}^{2}({\bf r},t)=p'(\rho({\bf r},t)).
\label{ep6}
\end{equation}
A steady state of the Euler-Poisson system satisfies ${\bf u}={\bf 0}$
and the condition of hydrostatic equilibrium
\begin{equation}
\nabla p+\rho\nabla\Phi={\bf 0}.
\label{ep7}
\end{equation}
Combining the condition of hydrostatic equilibrium (\ref{ep7}) and
the equation of state $p=p(\rho)$, we find that the density is a function $\rho=\rho(\Phi)$ of the gravitational potential given by
\begin{eqnarray}
\int^{\rho}{p'(\rho')\over\rho'}d\rho'=-\Phi.
\label{ep8}
\end{eqnarray}
We note that $p'(\rho)=-\rho/\rho'(\Phi)$. Since $p'(\rho)=c_s^2\ge 0$, we
find that $\rho'(\Phi)\le 0$. The density is a monotonically
decreasing function of the gravitational potential. The total energy
of the star is
\begin{eqnarray}
{\cal W}[\rho,{\bf
u}]=\int\rho\int_{0}^{\rho}{p(\rho')\over\rho^{'2}}d\rho'd{\bf r}
+{1\over 2}\int \rho\Phi d{\bf r}+\int \rho
{{\bf u}^{2}\over 2}d{\bf r},\nonumber\\
 \label{ep9}
\end{eqnarray}
including the internal energy, the potential energy and the kinetic
energy of the mean motion. The mass $M[\rho]$ and the energy ${\cal
W}[\rho,{\bf u}]$ are conserved by the barotropic Euler-Poisson
system: $\dot M=\dot{\cal W}=0$. Therefore, a minimum of the energy
functional at fixed mass determines a steady state of the barotropic
Euler-Poisson system that is nonlinearly dynamically stable.  Here, we
have stability in the sense of Lyapunov. This means that the size of
the perturbation is bounded by the size of the initial perturbation
for all times. We are led therefore to considering the minimization
problem
\begin{equation}
\min_{\rho,{\bf u}} \ \left\lbrace {\cal W}[\rho]\quad |\quad M[\rho]=M\right \rbrace.\label{ep10}
\end{equation}
The critical points of energy at fixed mass are determined by the
Euler-Lagrange equation $\delta{\cal W}-\mu\delta M=0$ where $\mu$ is
a Lagrange multiplier. This yields the condition of hydrostatic
equilibrium (\ref{ep7}). Then, a critical point of energy at fixed
mass is a (local) energy minimum iff
\begin{equation}
\delta^2{\cal W}=\int \frac{p'(\rho)}{2\rho}(\delta\rho)^2\, d{\bf r}+\frac{1}{2}\int \delta\rho\delta\Phi\, d{\bf r}\ge 0
\label{ep10b}
\end{equation}
for all perturbations $\delta\rho$ that conserve mass: $\delta M=0$.

\subsection{The Jeans instability criterion}
\label{sec_jeans}

We consider the linear dynamical stability of an infinite homogeneous
gaseous medium described by the hydrodynamic equations (\ref{ep1}),
(\ref{ep2}), (\ref{ep3}) and (\ref{ep5b}). This is the classical Jeans
(1902,1929) problem. Linearizing Eqs. (\ref{ep1})-(\ref{ep3}) around a
solution $\rho=\Phi={\rm const.}$ and ${\bf u}={\bf 0}$, making the
Jeans swindle and decomposing the perturbations in normal modes of the
form $e^{i({\bf k}\cdot {\bf r}-\omega t)}$, we obtain the classical
dispersion relation (Binney \& Tremaine 1987):
\begin{eqnarray}
\omega^{2}=c_{s}^{2}k^{2}-4\pi G\rho,
\label{jeans1}
\end{eqnarray}
where $c_{s}^{2}=dp/d\rho$ is the velocity of sound. Since $\omega^2$ is real, the complex
pulsation $\omega$ is either real or purely imaginary. The condition $\omega=0$
determines a critical wavenumber
\begin{eqnarray}
k_{c}^{2}=\frac{4\pi G\rho}{c_{s}^{2}},
\label{jeans2}
\end{eqnarray}
called the Jeans wavenumber for a barotropic gas.
The dispersion relation can be rewritten
\begin{eqnarray}
\frac{\omega^{2}}{4\pi G\rho}=\frac{k^{2}}{k_{c}^{2}}-1.
\label{jeans3}
\end{eqnarray}
For $k>k_{c}$, we have $\omega=\omega_r=\pm
\sqrt{\omega^2}$ so that the perturbation oscillates like $e^{-i\omega_r t}$ with a pulsation
\begin{eqnarray}
\omega_r=\pm\sqrt{c_{s}^{2}k^{2}-4\pi G\rho},
\label{jeans4}
\end{eqnarray}
without attenuation. This corresponds to gravity-modified sound waves. In that case the system is stable.
For $k<k_{c}$, we have $\omega=\omega_{i}=\pm
i\sqrt{-\omega^2}$ so that the perturbation grows  like $e^{\omega_i t}$  with a growth rate
\begin{eqnarray}
\omega_i=\sqrt{4\pi G\rho-c_{s}^{2}k^{2}},
\label{jeans5}
\end{eqnarray}
without oscillation (the second mode is attenuated exponentially
rapidly so that the growing mode dominates). In that case, the system
is unstable. This is the so-called Jeans instability.  The growth rate
is maximum for $k=0$ and its value is $\omega_i(k=0)=\sqrt{4\pi
G\rho}$.

In conclusion, a perturbation with wavenumber $k$ is stable if
\begin{eqnarray}
k>k_{c}=\frac{\sqrt{4\pi G\rho}}{c_{s}},
\label{jeans6}
\end{eqnarray}
and unstable otherwise. These results are valid for an
arbitrary barotropic equation of state $p=p(\rho)$. We now specialize
on particular cases, namely isothermal and polytropic equations of
state.

\subsection{Isothermal equation of state}
\label{sec_ip}

We first consider an isothermal equation of state
\begin{equation}
p({\bf r},t)=\rho({\bf r},t)T,
\label{ip1}
\end{equation}
where the temperature $T({\bf r},t)=T$ is a uniform. At hydrostatic
equilibrium, according to Eq. (\ref{ep8}), the relation between the
density and the gravitational potential is given by the Boltzmann law
\begin{equation}
\rho=A'e^{-{\Phi\over T}}, \label{ip2}
\end{equation}
where $A'$ is a constant. The energy functional (\ref{ep9}) reads
\begin{eqnarray}
{\cal W}[\rho,{\bf u}]=k_{B}T\int {\rho}\ln {\rho}
d{\bf r}+{1\over 2}\int \rho\Phi d{\bf r} +\int \rho {{\bf
u}^{2}\over
2}d{\bf r},\nonumber\\
\label{ip3}
\end{eqnarray}
and it can be viewed as a Boltzmann free energy $F=E-TS$ where $E$ is the macroscopic energy and $S$ the Boltzmann entropy. The velocity of sound (\ref{ep6}) is uniform:
\begin{equation}
c_{s}^{2}=T.
\label{ip4}
\end{equation}
Finally, for an infinite homogeneous isothermal gas, the Jeans instability criterion takes the classical form
\begin{eqnarray}
k^{2}<k^2_{c}=\frac{4\pi G\rho}{T}.
\label{ip5}
\end{eqnarray}

\subsection{Polytropic equation of state}
\label{sec_ip2}

The equation of state of a polytropic gas\footnote{The polytropic equation of
state corresponds to an adiabatic transformation for which the
specific entropy $s({\bf r},t)=s$ is uniform. In that case,
$\gamma=c_{p}/c_{v}$ is the ratio of specific heats at constant
pressure and constant volume. For a monoatomic gas $\gamma=5/3$.
This law describes convective regions of stellar interior. An
approximate polytropic equation of state with index $\gamma\simeq
3.25$ also holds in the radiative region of a star. Finally,
classical white dwarf stars are described by a polytropic equation
of state with index $\gamma=5/3$ ($n=3/2$). This equation of state
is valid for a completely degenerate gas of electrons described by
the Fermi-Dirac statistics at $T=0$ (Chandrasekhar 1942).} is
\begin{equation}
p({\bf r},t)=K \rho({\bf r},t)^{\gamma},
\label{ip6}
\end{equation}
where $K$ is a constant.  The polytropic index $n$ is
defined by $\gamma=1+1/n$.  For $n\rightarrow +\infty$, we recover
the isothermal case with $\gamma=1$.
At hydrostatic equilibrium, according to Eq. (\ref{ep8}), the relation between
the density and the gravitational potential can be written
\begin{eqnarray}
\rho=\biggl\lbrack \lambda-{\gamma-1\over K\gamma}\Phi\biggr
\rbrack^{1\over \gamma-1}, \label{ip7}
\end{eqnarray}
where $\lambda$ is a constant. The energy
functional (\ref{ep9}) can be put in the form
\begin{eqnarray}
{\cal W}[\rho,{\bf u}]={K\over \gamma-1}\int
(\rho^{\gamma}-\rho)d{\bf r}+{1\over 2}\int \rho\Phi d{\bf r} +\int
\rho {{\bf u}^{2}\over
2}d{\bf r}.\nonumber\\
\label{ip8}
\end{eqnarray}
In the limit $\gamma\rightarrow 1$, we recover the
results (\ref{ip2}) and (\ref{ip3}) obtained for isothermal
distributions. For a polytropic gas, the local temperature $T({\bf
r},t)$ given by Eq. (\ref{ep5}) reads
\begin{equation}
T({\bf r},t)=K\rho({\bf r},t)^{\gamma-1}.
\label{ip9}
\end{equation}
We note that the temperature is position dependent (while the
specific entropy $s$ is uniform) and related to the density by a
power law (this is the local version of the usual isentropic law
$TV^{\gamma-1}={\rm const.}$ in thermodynamics). The polytropic index $n$
is related to the gradients of temperature and density according
to
\begin{equation}
{\nabla\ln T}={1\over n}\nabla\ln\rho. \label{ip10}
\end{equation}
Combining Eqs. (\ref{ip7}) and (\ref{ip9}), we note that, for a
polytropic gas at equilibrium, the relation between the kinetic
temperature and the gravitational potential is {\it linear} so that
\begin{equation}
\nabla T=-\frac{\gamma-1}{\gamma}\nabla\Phi. \label{ip11}
\end{equation}
The coefficient of proportionality is related to the adiabatic index $\gamma$.
The velocity of sound (\ref{ep6}) can be expressed as
\begin{equation}
c_{s}({\bf r},t)^{2}=\gamma T({\bf r},t),
\label{ip12}
\end{equation}
and it is usually position dependent. However, when we consider an
infinite and homogeneous distribution (making the Jeans swindle), the
velocity of sound and the temperature are uniform.  In that case, we
can speak of ``the temperature $T$ of the polytropic gas'' and write the
Jeans instability criterion as
\begin{eqnarray}
k^{2}<k^2_{c}=\frac{4\pi G\rho}{\gamma T}.
\label{ip13}
\end{eqnarray}
We thus find that the critical Jeans wavenumber $k_{c}^{(poly)}$ of a polytropic gas is related to the  critical Jeans  wavenumber $k_{c}^{(iso)}$ of an isothermal gas with the same temperature $T$ by
\begin{eqnarray}
k_{c}^{(poly)}=\frac{1}{\sqrt{\gamma}}k_{c}^{(iso)},
\label{ip14}
\end{eqnarray}
where the proportionality factor is the polytropic index $\gamma$.
The positivity of $c_{s}^{2}=dp/d\rho$ implies that $\gamma\ge 0$,
i.e. $n\ge 0$ or $n\le -1$ (for $-1<n<0$, the gas is always unstable,
even in the absence of gravity, since $\omega^{2}<0$). For $\gamma=0$
($n=-1$), the gas is unstable to all wavelengths since
$k_{c}^{(poly)}=+\infty$. For $0<\gamma<1$ ($n<-1$),
$k_{c}^{(poly)}>k_{c}^{(iso)}$. For $\gamma=1$ ($n=\infty$),
$k_{c}^{(poly)}=k_{c}^{(iso)}$ (isothermal case).  For $\gamma>1$
($n>0$), $k_{c}^{(poly)}<k_{c}^{(iso)}$.  For $\gamma=+\infty$ ($n=0$), the
gas is stable to all wavelengths since $k_{c}^{(poly)}=0$ (solid
medium). For an isentropic gas for which $\gamma=c_{p}/c_{v}$, the
Mayer relation $c_{p}-c_{v}=k_{B}$ implies that $\gamma>1$
($n>0$). Therefore, for an isentropic gas, the gravitational
instability is always delayed (i.e., it occurs for larger wavelengths)
with respect to an isothermal gas with the same temperature.

\section{Collisionless stellar systems}
\label{sec_stellar}

\subsection{The Vlasov-Poisson system}
\label{sec_vh}

We consider a self-gravitating system described by the Vlasov-Poisson
system
\begin{equation}
{\partial f\over\partial t}+{\bf v}\cdot {\partial f\over\partial {\bf r}}+{\bf F}\cdot {\partial f\over\partial {\bf v}}=0,\label{vh0}
\end{equation}
\begin{equation}
\Delta\Phi=4\pi G\int f d{\bf v},\label{vh1}
\end{equation}
where ${\bf F}=-\nabla\Phi$ is the self-consistent gravitational field
produced by the particles. The Vlasov description assumes that the
evolution of the system is encounterless. This is a very good
approximation for most astrophysical systems like galaxies and dark
matter because the relaxation time due to close encounters is in general
larger than the age of the universe by several orders of magnitude
(Binney \& Tremaine 1987). The Vlasov equation admits an infinite number of stationary solutions
given by the Jeans theorem (Jeans 1915). For example,
distribution functions of the form $f=f(\epsilon)$ which depend only on the individual energy
$\epsilon={v^2}/{2}+\Phi({\bf r})$ of the stars are particular steady states of the
Vlasov equation. They describe spherical stellar systems.

The Vlasov
equation conserves the mass $M[f]=\int f\, d{\bf r}d{\bf v}$, the
energy $E[f]=\frac{1}{2}\int fv^2 \, d{\bf r}d{\bf
v}+\frac{1}{2}\int\rho\Phi\, d{\bf r}$ and the Casimirs $I_{h}=\int
h(f)\, d{\bf r}d{\bf v}$ for any continuous function $h$. Let us
introduce the functionals
\begin{equation}
S=-\int C(f)\, d{\bf r}d{\bf v},\label{vh2}
\end{equation}
where $C$ is a convex function ($C''>0$). These particular Casimirs
will be called ``pseudo entropies''\footnote{The functionals $H=-\int
C(\bar{f})\, d{\bf r}d{\bf v}$ defined in terms of the coarse-grained
distribution $\bar{f}$ are called generalized $H$-functions (Tremaine
et al. 1986).}.

{\it The two constraints problem:} since the Vlasov equation
conserves $M$, $E$ and $S$, the {maximization} problem
\begin{equation}
\max_{f} \ \left\lbrace S[f]\quad |\quad E[f]=E, \quad M[f]=M\right \rbrace,\label{vh3}
\end{equation}
determines a steady state of the Vlasov-Poisson system that is
nonlinearly dynamically stable. The critical points of pseudo entropy
at fixed mass and energy are determined by the variational principle
$\delta S-\beta\delta E-\alpha\delta M=0$, where $\beta$ (pseudo
inverse temperature) and $\alpha$ (pseudo chemical potential) are
Lagrange multipliers. This yields $C'(f)=-\beta\epsilon-\alpha$. Since
$C$ is convex, this relation can be reversed to give
$f=F(\beta\epsilon+\alpha)$ where $F=(C')^{-1}(-x)$. We have
$f'(\epsilon)=-\beta/C''(f)$.  Therefore, if $\beta>0$, the critical
points of pseudo entropy at fixed mass and energy determine
distributions of the form $f=f(\epsilon)$ with $f'(\epsilon)<0$: the
distribution function is a monotonically decreasing function of the
energy\footnote{More generally, the solutions of (\ref{vh3}) are of
the form $f=f(\epsilon)$ where $f$ is monotonic, decreasing at
positive temperatures and increasing at negative temperatures. For
realistic stellar systems, the DF should decrease close to the escape
energy $\epsilon=0$. Therefore, for the class of distributions
considered, $f$ must decrease everywhere implying $\beta>0$.}.  Then, a
critical point of pseudo entropy at fixed mass and energy is a (local)
 maximum iff
\begin{equation}
\delta^2{S}=-\frac{1}{2}\int C''(f)(\delta f)^2\, d{\bf r}d{\bf v}-\frac{1}{2}\beta\int \delta\rho\delta\Phi\, d{\bf r}< 0,
\label{vh3b}
\end{equation}
for all perturbations $\delta f$ that conserve mass and energy at
first order: $\delta M=\delta E=0$.

{\it The one constraint problem:} the minimization problem
\begin{equation}
\min_{f} \ \left\lbrace F[f]=E[f]-TS[f]\quad | \quad M[f]=M\right \rbrace,\label{vh4}
\end{equation}
where $T=1/\beta>0$ is prescribed, determines a steady state of the Vlasov-Poisson system that is
nonlinearly dynamically stable. The functional $F[f]$ is a pseudo free
energy. The critical points of pseudo free energy at fixed mass are
determined by the variational principle $\delta F+\alpha T\delta M=0$,
where $\alpha$ (pseudo chemical potential) is a Lagrange
multiplier. This yields $C'(f)=-\beta\epsilon-\alpha$.  Then, a
critical point of pseudo free energy at fixed mass  is a (local)
minimum iff
\begin{equation}
\delta^2{F}=\frac{1}{2}T\int C''(f)(\delta f)^2\, d{\bf r}d{\bf v}+\frac{1}{2}\int \delta\rho\delta\Phi\, d{\bf r}> 0,
\label{vh4b}
\end{equation}
for all perturbations $\delta f$ that conserve mass: $\delta M=0$.

{\it The no constraint problem:} the maximization problem
\begin{equation}
\max_{f} \ \left\lbrace G[f]=S[f]-\beta E[f]-\alpha M[f]\right \rbrace,\label{erica1}
\end{equation}
where $\beta$ and $\alpha$ are prescribed, determines a steady state of
the Vlasov-Poisson system that is nonlinearly dynamically stable. The functional $G[f]$ is a 
pseudo grand potential. The critical points of pseudo
grand potential $G$, satisfying $\delta G=0$, are given by $C'(f)=-\beta\epsilon-\alpha$.
 Then, a
critical point of pseudo grand potential is a (local)
 maximum iff
\begin{equation}
\delta^2{G}=-\frac{1}{2}\int C''(f)(\delta f)^2\, d{\bf r}d{\bf v}-\frac{1}{2}\beta\int \delta\rho\delta\Phi\, d{\bf r}< 0,
\label{erica2}
\end{equation}
for all perturbations $\delta f$.

The optimization problems (\ref{vh3}), (\ref{vh4}) and (\ref{erica1}) have the {\it
same} critical points (canceling the first order variations). Furthermore, a solution
of (\ref{erica1}) is always a solution of the more constrained dual
problem  (\ref{vh4}). Indeed, if inequality (\ref{erica2}) is true for
all perturbations, it is true a fortiori for all
perturbations that conserve mass. Similarly, a solution of (\ref{vh4}) is always a solution of the more constrained dual
problem (\ref{vh3}). Indeed, if inequality (\ref{vh4b}) is true for
all perturbations that conserve mass, it is true a fortiori for all
perturbations that conserve mass {\it and} energy. However, the
reciprocal is wrong. A solution of (\ref{vh4}) is not necessarily a solution
of (\ref{erica1}), and a solution of (\ref{vh3}) is not necessarily a solution
of (\ref{vh4}). This is similar to the notion of ensemble inequivalence in
thermodynamics (Ellis et al. 2000, Bouchet \& Barr\'e 2005, Chavanis 2006a). Indeed, the two
constraints problem (\ref{vh3}) is similar to a condition of
microcanonical stability, the one constraint problem (\ref{vh4}) is
similar to a condition of canonical stability, and the no constraint problem (\ref{erica1})
is similar to a condition of grand canonical stability. The implication
$(\ref{erica1})\Rightarrow (\ref{vh4})\Rightarrow (\ref{vh3})$ is similar to the fact that
grand canonical stability implies canonical stability which itself implies 
microcanonical stability (but not the
converse) in thermodynamics. Therefore, (\ref{erica1}) provides just a
{\it sufficient} condition of nonlinear dynamical stability that is
less refined than (\ref{vh4}), and (\ref{vh4}) provides just a
{\it sufficient} condition of nonlinear dynamical stability that is
less refined than (\ref{vh3}).

{\it The most refined problem:} the minimization problem
\begin{equation}
\min_{f} \ \left\lbrace E[f]\quad | {\rm \ all \ the \ Casimirs\ } I_h \right\rbrace,\label{erica3}
\end{equation}
determines a steady state of the Vlasov-Poisson system that is
nonlinearly dynamically stable. A distribution function is a critical point of energy for symplectic
perturbations (i.e. perturbations that conserve all the Casimirs) iff $f({\bf r},{\bf v})$ is a steady state of the Vlasov equation (Bartholomew 1971, Kandrup 1991). Furthermore, if we consider spherical stellar systems for which $f=f(\epsilon)$, it can be shown (Bartholomew 1971, Kandrup 1991) that a critical point of energy for symplectic perturbations is a (local) minimum iff
\begin{equation}
\delta^2{E}=-\frac{1}{2}\int \frac{(\delta f)^2}{f'(\epsilon)}\, d{\bf r}d{\bf v}+\frac{1}{2}\int \delta\rho\delta\Phi\, d{\bf r}> 0,
\label{erica4}
\end{equation}
for all perturbations $\delta f$ that conserve energy and all the Casimirs at first order: $\delta E=\delta I_h=0$. Such symplectic (physically accessible) perturbations are of the form $\delta f=D\delta g$ where $D={\bf v}\cdot\nabla_{\bf r}-\nabla\Phi\cdot \nabla_{\bf v}$ is the advective operator in phase space and $\delta g({\bf r},{\bf v})$ is {\it any} perturbation. Inequality (\ref{erica4}) corresponds to the Antonov (1960) criterion of dynamical stability that was obtained by investigating the linear dynamical stability of a steady state of the Vlasov equation. Since this is the most constrained criterion, this is the most refined one. We note that inequality (\ref{erica4}) is equivalent to inequalities (\ref{vh3b}), (\ref{vh4b}) and (\ref{erica2}) if we use the identity $f'(\epsilon)=-\beta/C''(f)$ derived above. However, the classes of perturbations to consider are different: in (\ref{erica2}), we must consider all perturbations, in (\ref{vh4b}) we must consider perturbations that conserve mass, in (\ref{vh3b}) we must consider perturbations that conserve mass and energy, and in (\ref{erica4}) we must consider perturbations that conserve energy and all the Casimirs. Therefore, we have the implications $(\ref{erica1}) \Rightarrow (\ref{vh4}) \Rightarrow (\ref{vh3}) \Rightarrow (\ref{erica3})$. The connection between (\ref{erica3}) and the preceding optimization problems can be understood as follows. The minimization problem (\ref{erica3}), conserving all the Casimirs, is clearly  more refined than the minimization problem
\begin{equation}
\min_{f} \ \left\lbrace E[f]\quad | \quad M[f]=M, \quad S[f]=S\right\rbrace,\label{erica5}
\end{equation}
where only the mass and {\it one} Casimir of the form (\ref{vh2}) is conserved. Now, it is easy to show that  the minimization problem (\ref{erica5})  is  equivalent to the maximization problem (\ref{vh3}). Hence, we have the chain of relations
\begin{equation}
(\ref{erica1}) \Rightarrow (\ref{vh4}) \Rightarrow (\ref{vh3}) \Leftrightarrow (\ref{erica5}) \Rightarrow (\ref{erica3}).
\label{erica6}
\end{equation}
To summarize, the minimization problem (\ref{erica3}) is the most refined stability criterion because it tells that, in order to settle the dynamical stability of a stellar system, we just need considering symplectic (i.e. dynamically accessible) perturbations. Of course, if inequality (\ref{erica4}) is satisfied by a larger class of perturbations, as implied by problems (\ref{erica1}), (\ref{vh4}), (\ref{vh3}) and (\ref{erica5}), the system will be stable a fortiori. Therefore, we have the implications (\ref{erica6}). Problems (\ref{erica1}), (\ref{vh4}), (\ref{vh3}) and (\ref{erica5}) provide sufficient (but not necessary) conditions of dynamical stability. A steady state can be stable  according to (\ref{vh4}) while it does not satisfy (\ref{erica1}), or it can be stable according to (\ref{vh3}) and (\ref{erica5}) while it does not satisfy (\ref{vh4}), or it can be stable according to (\ref{erica3}) while it does not satisfy (\ref{vh3}) and (\ref{erica5}). Therefore, the criterion (\ref{erica3}) is more refined than (\ref{erica5}) or  (\ref{vh3}), which is itself more refined than (\ref{vh4}), which is itself more refined than (\ref{erica1}). Let us give an astrophysical illustration of all that (Chavanis 2006a). Using (\ref{vh4}), we can show that stellar polytropes with $3/2\le n\le 3$ are dynamically stable because they are minima of $F$ at fixed mass $M$. By contrast, stellar polytropes with $3<n<5$ are not minima of $F$ at fixed mass. However, using (\ref{vh3}), we can show that stellar polytropes with $3/2<n<5$  are dynamically stable because they are maxima of $S$ at fixed mass and energy. However, not all stellar systems of the form $f=f(\epsilon)$ are maxima of $S$ at fixed mass and energy. But, using (\ref{erica3}), we can show that all spherical galaxies $f=f(\epsilon)$ with $f'(\epsilon)<0$ are dynamically stable.   This last statement has been shown for linear stability by Kandrup (1991).  A rigorous mathematical proof has been given recently by Lemou et al. (2009) for nonlinear stability.

{\it Remark}: similar results are obtained in 2D fluid mechanics based on the Euler-Poisson system (see Chavanis 2009). Criterion (\ref{erica3})  is equivalent to the so-called Kelvin-Arnol'd energy principle, criterion (\ref{erica1}) is equivalent to the standard Casimir-energy method (see Holm et al. 1985) introduced by Arnol'd (1966) and criterion (\ref{vh3}) is equivalent to the refined stability criterion given by Ellis et al. (2002).

\subsection{The corresponding barotropic star}
\label{sec_vhc}

To any stellar system with $f=f(\epsilon)$ and $f'(\epsilon)<0$, we
can associate a corresponding barotropic star with the same
equilibrium density distribution (Lynden-Bell \& Sanitt 1969). Indeed,
defining the density and the pressure by $\rho=\int f d{\bf v}$ and
$p={1\over 3}\int f v^{2}d{\bf v}$, we get $\rho=\rho(\Phi)$ and
$p=p(\Phi)$. Eliminating the potential $\Phi$ between these two
expressions, we find that $p=p(\rho)$. Furthermore, taking the
gradient of the pressure and using the chain of identities
\begin{eqnarray}
\nabla p={1\over 3}\int {\partial f\over\partial {\bf r}}v^{2}d{\bf v}={1\over 3}\nabla \Phi \int f'(\epsilon) v^{2}d{\bf v}\nonumber\\
={1\over 3}\nabla \Phi \int {\partial f\over\partial {\bf v}}\cdot {\bf v}\, d{\bf v}=-\nabla \Phi \int  f \, d{\bf v}=-\rho\nabla \Phi,
\label{vhc1}
\end{eqnarray}
we obtain the condition of hydrostatic equilibrium (\ref{ep7}). Finally, we define the kinetic temperature of an isotropic stellar system by the relation
\begin{eqnarray}
\frac{3}{2}T({\bf r})=\frac{1}{2}\langle v^2\rangle.
\label{vhc2a}
\end{eqnarray}
Therefore, the quantity
\begin{eqnarray}
T({\bf r})=\frac{1}{3}\langle v^2\rangle={{1\over 3}\int f v^{2}d{\bf v}\over \int f d{\bf v}}={p({\bf r})\over \rho({\bf r})}.
\label{vhc2b}
\end{eqnarray}
measures the velocity dispersion (in one direction) of an isotropic
stellar system. In general, the kinetic
temperature is position dependent.

Finally, we can show that the variational principles $(\ref{vh4})$ and
$(\ref{ep10})$ are equivalent, i.e. a stellar system is a minimum of
pseudo free energy at fixed mass iff the corresponding barotropic gas
is a minimum of energy at fixed mass (see Appendix \ref{sec_eqa}). This leads to a
nonlinear version of the Antonov (1960b) first law: ``a stellar system with
$f=f(\epsilon)$ and $f'(\epsilon)<0$ is nonlinearly dynamically stable
with respect to the Vlasov-Poisson system if the corresponding
barotropic gas is nonlinearly dynamically stable with respect to the
Euler-Poisson system''. However, the reciprocal is wrong because, as
we have already indicated, (\ref{vh4}) provides just a {\it
sufficient} condition of nonlinear dynamical stability with respect to
the Vlasov equation. A galaxy can be dynamically stable according to
criterion (\ref{vh3}) [or even more generally (\ref{erica3})] 
while it fails to satisfy criterion
(\ref{vh4}). Therefore, the nonlinear Antonov first law is similar to
a notion of ensembles inequivalence between microcanonical and
canonical ensembles in thermodynamics (Chavanis 2006a).

\subsection{The dispersion relation}
\label{sec_lsv}

We shall study the linear dynamical stability of a spatially
homogeneous stationary solution of the Vlasov equation described by a
distribution function $f=f({\bf v})$. Like for an infinite homogeneous
gas, we make the Jeans swindle. Linearizing the Vlasov equation around
this steady state, taking the Laplace-Fourier transform of this
equation and writing the perturbations in the form $e^{i({\bf k}\cdot
{\bf r}-\omega t)}$, we obtain the classical dispersion relation
(Binney \& Tremaine 1987):
\begin{eqnarray}
\epsilon({\bf k},\omega)\equiv 1+{4\pi G\over k^{2}}\int {{\bf k}\cdot {\partial f\over\partial {\bf v}}\over {\bf k}\cdot {\bf v}-\omega}d{\bf v}=0,
\label{lsv1}
\end{eqnarray}
where $\epsilon({\bf k},\omega)$ is similar to the ``dielectric
function'' of plasma physics (with the sign $+$ instead of $-$). For a given distribution $f({\bf v})$, this equation determines the complex pulsation(s)
$\omega=\omega_{r}+i\omega_{i}$ of a perturbation with wavevector
${\bf k}$. Since the time dependence of the perturbation is $\delta f\sim e^{-i\omega_r t}e^{\omega_i t}$, the system is linearly
stable if $\omega_{i}<0$ and linearly unstable if $\omega_{i}>0$. The condition of marginal stability corresponds to $\omega_i=0$.

If we take the wavevector ${\bf k}$ along the $z$-axis\footnote{If the distribution function is isotropic, there is no
restriction in making this choice.}, the dispersion relation becomes
\begin{eqnarray}
\epsilon({k},\omega)\equiv 1+{4\pi G\over k^{2}}\int_{C} {{k}{\partial f\over\partial {v_z}}\over {k} {v_z}-\omega}d{v}_x d{v}_y d{v}_z=0,
\label{lsv2}
\end{eqnarray}
where the integral must be performed along the Landau contour $C$ (Binney \& Tremaine 1987). We define
\begin{eqnarray}
g(v_z)=\int f dv_x dv_y.
\label{lsv3}
\end{eqnarray}
In the following, we shall note $v$ instead of $v_z$ and $f$ instead of $g$. With these conventions, the dispersion relation (\ref{lsv2}) can be rewritten
\begin{eqnarray}
\epsilon({k},\omega)\equiv 1+{4\pi G\over k^{2}}\int_{C} {f'(v)\over {v}-\frac{\omega}{k}}d{v}=0.
\label{lsv4}
\end{eqnarray}
This is the fundamental equation of the problem.  For future
reference, let us recall that
\begin{eqnarray}
\int_{C} {{f'(v)}\over v-\frac{\omega}{k}}dv=\int_{-\infty}^{+\infty} {{f'(v)}\over v-\frac{\omega}{k}}dv, \quad (\omega_i>0),
\label{lsv5}
\end{eqnarray}
\begin{eqnarray}
\int_{C} {{f'(v)}\over v-\frac{\omega}{k}}dv=P\int_{-\infty}^{+\infty} {{f'(v)}\over v-\frac{\omega}{k}}dv+i\pi  f'\left (\frac{\omega}{k}\right ), \quad (\omega_i=0),\nonumber\\
\label{lsv6}
\end{eqnarray}
\begin{eqnarray}
\int_{C} {{f'(v)}\over v-\frac{\omega}{k}}dv=\int_{-\infty}^{+\infty} {{f'(v)}\over v-\frac{\omega}{k}}dv+ 2\pi i f'\left (\frac{\omega}{k}\right ), \quad (\omega_i<0),\nonumber\\
\label{lsv7}
\end{eqnarray}
where $P$ denotes the principal value.

In general, the dispersion relation (\ref{lsv4}) cannot be solved explicitly to obtain $\omega(k)$ except in some very simple cases \footnote{There exists less simple cases were explicit solutions can be obtained. We should mention in this respect the extensive set of closed-form solutions for generalized Lorentzian distributions due to Summers \& Thorne (1991) and Thorne \& Summers (1991). They correspond to Tsallis (polytropic) distributions of the form (\ref{pdr1}) with negative index $n<-1$.}. For example, for cold systems described by the distribution function $f=\rho \delta({v}-{v}_0)$, we obtain after an integration by parts
\begin{eqnarray}
\omega=v_0 k\pm i\sqrt{4\pi G\rho}.
\label{lsv8}
\end{eqnarray}
In particular, when $v_0=0$, we get
\begin{eqnarray}
\omega=\pm i\sqrt{4\pi G\rho}.
\label{lsv9}
\end{eqnarray}
The system is unstable to all wavenumbers and the perturbation grows
with a growth rate $\omega_i=\sqrt{4\pi G\rho}$. We also note that the
dispersion relation (\ref{lsv9}) coincides with the dispersion
relation (\ref{jeans5}) of a self-gravitating gas with $c_s=0$.

\subsection{The condition of marginal stability}
\label{sec_cms}

For $\omega_{i}=0$, the real and imaginary parts of the dielectric function $\epsilon(k,\omega_{r})=\epsilon_r(k,\omega_{r})+i\epsilon_i(k,\omega_{r})$ are given by
\begin{eqnarray}
\epsilon_r({k},\omega_r)= 1+\frac{4\pi G}{k^2}P\int_{-\infty}^{+\infty} {f'(v)\over {v}-{\omega_r}/{k}}d{v},
\label{cms1}
\end{eqnarray}
\begin{eqnarray}
\epsilon_i({k},\omega_r)= \frac{4\pi^2 G}{k^2}f'\left ({\omega_r}/{k}\right ).
\label{cms2}
\end{eqnarray}
The condition of marginal stability corresponds to $\epsilon(k,\omega)=0$ and $\omega_i=0$, i.e. $\epsilon_r({k},\omega_r)=\epsilon_i({k},\omega_r)=0$. We obtain therefore the equations
\begin{eqnarray}
 1+\frac{4\pi G}{k^2}P\int_{-\infty}^{+\infty} {f'(v)\over {v}-{\omega_r}/{k}}d{v}=0,
\label{cms1b}
\end{eqnarray}
\begin{eqnarray}
f'\left ({\omega_r}/{k}\right )=0.
\label{cms2b}
\end{eqnarray}
The second condition (\ref{cms2b}) imposes that the phase velocity
$\omega_r/k=v_{ext}$ is equal to a velocity where $f(v)$ is extremum
($f'(v_{ext})=0$). The first condition (\ref{cms1b}) then determines
the wavenumber(s) $k_c$ corresponding to marginal stability. It can be
written
\begin{eqnarray}
1+\frac{4\pi G}{k_c^2}\int_{-\infty}^{+\infty} {f'(v)\over {v}-v_{ext}}d{v}=0,
\label{cms3}
\end{eqnarray}
where the principal value is not needed anymore. The wavenumber(s) corresponding to marginal stability are therefore given by
\begin{eqnarray}
k_{c}=\left (-4\pi G\int_{-\infty}^{+\infty} \frac{f'(v)}{v-v_{ext}}\, dv\right )^{1/2}.
\label{cms4}
\end{eqnarray}
Finally, the pulsation(s) corresponding to marginal stability are $\omega_r=v_{ext}k_c$ and we have $\delta f\sim e^{-i\omega_r t}$. Note that the
distribution $f(v)$ can be  relatively arbitrary. There can be pure oscillations ($\omega=\omega_r\neq 0$) only if $f(v)$ has some extrema at $v\neq 0$. If $f(v)$ has a
single maximum at $v=0$, then $\omega_{r}=0$ (implying $\omega=0$) and
the condition of marginal stability becomes
\begin{eqnarray}
k_{c}=\left (-4\pi G\int_{-\infty}^{+\infty} \frac{f'(v)}{v}\, dv\right )^{1/2}.
\label{cms5}
\end{eqnarray}

\subsection{The Nyquist method}
\label{sec_n}

To determine whether the distribution $f=f(v)$ is stable or unstable
for a perturbation with wavenumber $k$, one possibility is to solve
the dispersion relation (\ref{lsv4}) and determine the sign of the
imaginary part of the complex pulsation. This can be done analytically
in some simple cases (see Secs. \ref{sec_mon}, \ref{sec_vb} and
\ref{sec_vb2}). We can also apply the Nyquist method introduced in plasma
physics. This is a graphical method that does not require to solve the
dispersion relation. The details of the method are explained by
Nicholson (1992) and we just recall how it works in practice. In the
$\epsilon$-plane, taking $\omega_i=0$, we plot the Nyquist
curve\footnote{This curve is also called an hodograph.}
$(\epsilon_{r}(k,\omega_{r}),\epsilon_{i}(k,\omega_{r}))$
parameterized by $\omega_{r}$ going from $-\infty$ to $+\infty$ (for a
given wavenumber $k$). This curve is closed and always rotates in the
counterclockwise sense. If the Nyquist curve does not encircle the
origin, the system is stable (for the corresponding wavenumber
$k$). If the Nyquist curve encircles the origin one or more times, the
system is unstable. The number $N$ of tours around the origin gives
the number of zeros of $\epsilon(k,\omega)$ in the upper half
$\omega$-plane, i.e. the number of unstable modes with $\omega_{i}>0$.
The Nyquist method by itself does not give the growth rate of the
instability.

Let us consider the asymptotic behavior of
$(\epsilon_{r}(k,\omega_{r}),\epsilon_{i}(k,\omega_{r}))$ defined by
Eqs. (\ref{cms1}) and (\ref{cms2}) for $\omega_{r}\rightarrow
\pm\infty$. Since $f(v)$ is positive and tends to zero for
$v\rightarrow \pm\infty$, we conclude that
$\epsilon_{i}(k,\omega_{r})\rightarrow 0$ for $\omega_{r}\rightarrow
\pm\infty$ and that  $\epsilon_{i}(k,\omega_{r})>0$ for
$\omega_{r}\rightarrow -\infty$ while $\epsilon_{i}(k,\omega_{r})<0$ for
$\omega_{r}\rightarrow +\infty$.  On the other hand, integrating by
parts in Eq. (\ref{cms1}), we obtain
\begin{eqnarray}
\epsilon_{r}(k,\omega_{r})=1+\frac{4\pi G}{k^2}{P}\int_{-\infty}^{+\infty} {{f(v)}\over (v-\omega_{r}/k)^{2}}dv,
\label{gp4}
\end{eqnarray}
provided that $f(v)$ decreases sufficiently rapidly. Therefore, for $\omega_{r}\rightarrow \pm\infty$, we obtain at leading order
\begin{eqnarray}
\epsilon_{r}(k,\omega_{r})\simeq 1+\frac{4\pi G\rho}{\omega_{r}^{2}}, \qquad (\omega_{r}\rightarrow \pm\infty).
\label{gp5}
\end{eqnarray}
In particular, $\epsilon_{r}(k,\omega_{r})\rightarrow 1^+$ for
$\omega_{r}\rightarrow
\pm\infty$.
From these results, we conclude that the behavior of the Nyquist curve
close to the limit point $(1,0)$ is like the one represented in
Fig. \ref{maxwell}. In addition, according to Eq. (\ref{cms2}), the
Nyquist curve crosses the $x$-axis at each value of $\omega_{r}/k$
corresponding to an extremum of $f(v)$. For $\omega_r/k=v_{ext}$, where
$v_{ext}$ is a velocity at which the distribution is extremum
$(f'(v_{ext})=0)$, the imaginary part of the dielectric function
$\epsilon_i(k,k v_{ext})=0$ and the real part of the dielectric function
\begin{equation}
\epsilon_r(k,k v_{ext})= 1+\frac{4\pi G}{k^2} \int_{-\infty}^{\infty}\frac{f'(v)}{v-v_{ext}}\, dv.
\end{equation}
Subtracting the value $f'(v_{ext})=0$ in the numerator of the integrand, and  integrating by parts, we obtain
\begin{equation}
\epsilon_r(k, k v_{ext})= 1-\frac{4\pi G}{k^2} \int_{-\infty}^{\infty}\frac{f(v_{ext})-f(v)}{(v-v_{ext})^2}\, dv.
\label{gp6}
\end{equation}
If $v_{Max}$ denotes the velocity corresponding to the global maximum of
the distribution, we clearly have
\begin{eqnarray}
\epsilon_{r}(k, k v_{Max})= 1-\frac{4\pi G}{k^2} \int_{-\infty}^{\infty}\frac{f(v_{Max})-f(v)}{(v-v_{Max})^2}\, dv<1.\nonumber\\
\label{gp7}
\end{eqnarray}
Furthermore, $\epsilon_r(k,kv_{Max})<0$ for sufficiently small
$k$. Therefore, by tuning $k$ appropriately, we can always make the
Nyquist curve encircle the origin. We conclude that a spatially
homogeneous self-gravitating system is always unstable to some
wavelengths.

\subsection{Single-humped distributions}
\label{sec_sh}

Let us assume that the distribution $f(v)$ has a single maximum at
$v=v_{0}$ (so that $f'(v_0)=0$) and tends to zero for $v\rightarrow
\pm \infty$. Then, the Nyquist curve cuts the $x$-axis  ($\epsilon_{i}(k,\omega_{r})$ vanishes) at the limit point
$(1,0)$ where $\omega_{r}\rightarrow \pm \infty$ and at the point
$(\epsilon_{r}(k, k v_0),0)$ where $\omega_{r}/k=v_{0}$. Due to its behavior
close to the limit point $(1,0)$, the fact that it rotates in the
counterclockwise sense, and the property that $\epsilon_r(k, k v_0)<1$ according to Eq. (\ref{gp7}),
the Nyquist curve must necessarily behave like in
Fig. \ref{maxwell}.  Therefore, the Nyquist curve starts on the real
axis at $\epsilon_r(k, \omega_r) =1$ for $\omega_r \rightarrow -\infty$,
then going in counterclockwise sense it crosses the real axis at the
point $\epsilon_r(k, k v_0)<1$ and returns on the real axis at
$\epsilon_r(k,\omega_r) =1$ for $\omega_r \rightarrow +
\infty$. According to the Nyquist criterion exposed in Sec. \ref{sec_n}, we
conclude that a single-humped distribution is linearly
stable with respect to a perturbation with wavenumber $k$ if
\begin{eqnarray}
\epsilon_{r}(k, k v_0)=1+\frac{4\pi G}{k^2}\int_{-\infty}^{+\infty} {{f'(v)}\over v-v_0}dv>0,
\label{sh1}
\end{eqnarray}
and linearly unstable if $\epsilon_{r}(k, k v_0)<0$. The equality
corresponds to the marginal stability condition
(\ref{cms3}). Therefore, the system is stable iff
\begin{eqnarray}
k>k_{c}=\left (-4\pi G\int_{-\infty}^{+\infty} \frac{f'(v)}{v-v_{0}}\, dv\right )^{1/2},
\label{sh2}
\end{eqnarray}
where $k_{c}$ is the critical Jeans wavenumber for a stellar system. Note that an infinite homogeneous stellar system whose DF has a single hump is always unstable to sufficiently small wavenumbers.  For  the unstable wavenumbers $k<k_c$, there is only one mode of instability $\omega_i>0$ since the Nyquist curve rotates only one time
around the origin. This stability criterion is valid for any single-humped distribution. In particular, a symmetric distribution $f=f(v)$ with a single maximum at $v_0=0$ is linearly dynamically stable to a perturbation with wavenumber $k$ iff
\begin{eqnarray}
k>k_{c}=\left (-4\pi G\int_{-\infty}^{+\infty} \frac{f'(v)}{v}\, dv\right )^{1/2}.
\label{sh3}
\end{eqnarray}
We shall make the
connection between the stability of an infinite homogeneous  stellar system and the stability
of the corresponding barotropic gas in Sec. \ref{sec_fo}. In
particular, using identity (\ref{fo3}), we will show that
Eq. (\ref{sh3}) is equivalent to
Eq. (\ref{jeans6}).

\subsection{Double-humped distributions}
\label{sec_doubh}

Let us consider a double-humped distribution with a global maximum at
$v_{Max}$, a minimum at $v_{min}$ and a local maximum at $v_{max}$. We
assume that $v_{Max}<v_{min}<v_{max}$. The Nyquist curve will cut the
$x$-axis at the limit point $(1,0)$ and at three other points
$(\epsilon_{r}(k,k v_{Max}),0)$, $(\epsilon_{r}(k,k v_{min}),0)$ and
$(\epsilon_{r}(k,k v_{max}),0)$. We also know that the Nyquist curve can only
rotate in the counterclockwise sense and that $\epsilon_{r}(k,k v_{Max}) <1$ according to Eq.
(\ref{gp7}). Then, we can convince ourselves, by making drawings, of the following results. If\footnote{In the following, in order to simplify the notations, we note $\epsilon_{r}(v_{Max})$ for $\epsilon_{r}(k,kv_{Max})$ etc.}

$(+++)$: $\epsilon_{r}(v_{Max})>0$, $\epsilon_{r}(v_{min})>0$, $\epsilon_{r}(v_{max})>0$,

$(+--)$: $\epsilon_{r}(v_{Max})>0$, $\epsilon_{r}(v_{min})<0$, $\epsilon_{r}(v_{max})<0$,

$(--+)$: $\epsilon_{r}(v_{Max})<0$, $\epsilon_{r}(v_{min})<0$, $\epsilon_{r}(v_{max})>0$,

$(+-+)$ : $\epsilon_{r}(v_{Max})>0$, $\epsilon_{r}(v_{min})<0$, $\epsilon_{r}(v_{max})>0$,

\noindent the Nyquist curve does not encircle the origin so the
system is stable. If

$(---)$: $\epsilon_{r}(v_{Max})<0$, $\epsilon_{r}(v_{min})<0$, $\epsilon_{r}(v_{max})<0$,

$(-++)$: $\epsilon_{r}(v_{Max})<0$, $\epsilon_{r}(v_{min})>0$, $\epsilon_{r}(v_{max})>0$,

$(++-)$:  $\epsilon_{r}(v_{Max})>0$, $\epsilon_{r}(v_{min})>0$, $\epsilon_{r}(v_{max})<0$,

\noindent the Nyquist curve rotates one time
around the origin so that there is one mode of instability. Finally, if

$(-+-)$: $\epsilon_{r}(v_{Max})<0$, $\epsilon_{r}(v_{min})>0$, $\epsilon_{r}(v_{max})<0$,

\noindent the Nyquist curve rotates two times around the origin
so that there are two modes of instability. Cases   $(+++)$, $(---)$, $(-++)$ and  $(-+-)$ are observed in
Sec. \ref{sec_vha} for an asymmetric double-humped distribution
made of two Maxwellians. The other cases cannot be obtained from this
distribution but they may be obtained from other distributions.

If the double-humped distribution is symmetric with respect to the
origin with two maxima at $\pm v_{*}$ and a minimum at $v=0$, only
three cases can arise.  If

$(+++)$: $\epsilon_{r}(v_{*})>0$, $\epsilon_{r}(0)>0$,

$(+-+)$: $\epsilon_{r}(v_{*})>0$, $\epsilon_{r}(0)<0$,

\noindent the Nyquist curve does not encircle the origin so the
system is stable. If

$(---)$: $\epsilon_{r}(v_{*})<0$, $\epsilon_{r}(0)<0$,

\noindent the Nyquist curve rotates one time
around the origin so that there is one mode of instability. Finally, if

$(-+-)$: $\epsilon_{r}(v_{*})<0$, $\epsilon_{r}(0)>0$,

\noindent the Nyquist curve rotates two times around the origin
so that there are two modes of instability. Cases  $(+++)$, $(---)$ and  $(-+-)$ are
observed in Sec. \ref{sec_vh} for a symmetric double-humped
distribution made of two Maxwellians.

\subsection{Particular solutions of $\epsilon(k,\omega)=0$}
\label{sec_mon}

We can look for a solution of the dispersion relation
$\epsilon(k,\omega)=0$ in the form $\omega=i\omega_{i}$ corresponding to
$\omega_{r}=0$. In that case, the perturbation grows ($\omega_{i}>0$)
or decays ($\omega_{i}<0$) without oscillating. For $\omega_{i}>0$,
the equation $\epsilon(k,\omega)=0$ becomes
\begin{eqnarray}
1+{4\pi G\over k^{2}}\int_{-\infty}^{+\infty}\frac{f'(v)}{v-i\frac{\omega_{i}}{k}}\, dv=0.
\label{mon1}
\end{eqnarray}
Multiplying the numerator by $v+i\omega_{i}/k$ and separating real and imaginary parts, we obtain
\begin{eqnarray}
1+{4\pi G\over k^{2}}\int_{-\infty}^{+\infty}\frac{v f'(v)}{v^2+\frac{\omega_{i}^2}{k^2}}\, dv=0,
\label{mon2}
\end{eqnarray}
\begin{eqnarray}
\int_{-\infty}^{+\infty}\frac{f'(v)}{v^2+\frac{\omega_{i}^2}{k^2}}\, dv=0.
\label{mon3}
\end{eqnarray}
If we consider distribution functions $f(v)$ that are symmetric with
respect to $v=0$, Eq. (\ref{mon3}) is always satisfied. Then, the
growth rate $\omega_i>0$ is given by Eq. (\ref{mon2}).

For $\omega_{i}<0$, the equation $\epsilon(k,\omega)=0$ becomes
\begin{eqnarray}
1+{4\pi G\over k^{2}}\int_{-\infty}^{+\infty}\frac{f'(v)}{v-i\frac{\omega_{i}}{k}}\, dv+i\frac{8\pi^2G}{k^2} f'\left (\frac{i\omega_{i}}{k}\right )=0.
\label{mon4}
\end{eqnarray}
Multiplying the numerator by $v+i\omega_{i}/k$ and assuming that
$f(v)$ is even, we obtain
\begin{eqnarray}
1+{4\pi G\over k^{2}}\int_{-\infty}^{+\infty}\frac{v f'(v)}{v^2+\frac{\omega_{i}^2}{k^2}}\, dv+i \frac{8\pi^2G}{k^2} f'\left (\frac{i\omega_{i}}{k}\right )=0,
\label{mon5}
\end{eqnarray}
which determines the damping rate $\omega_i<0$.

Let us introduce the function $K(x)$ defined by
\begin{eqnarray}
K(x)=\frac{1}{I}\int_{-\infty}^{+\infty}\frac{v f'(v)}{v^2+x^2}\, dv \ (x\ge 0)
\label{kk1}
\end{eqnarray}
\begin{eqnarray}
K(x)=\frac{1}{I}\left\lbrace \int_{-\infty}^{+\infty}\frac{v f'(v)}{v^2+x^2}\, dv+2\pi i  f'(ix)\right\rbrace \ (x\le 0)
\label{kk2}
\end{eqnarray}
where
\begin{eqnarray}
I=\int_{-\infty}^{+\infty}\frac{f'(v)}{v}\, dv.
\label{kk3}
\end{eqnarray}
This function is normalized such that $K(0)=1$. The dispersion relations (\ref{mon2}) and  (\ref{mon5}) can then be written
\begin{eqnarray}
1-\frac{k_c^2}{k^2}K\left (\frac{\omega_i}{k}\right )=0,
\label{kk4}
\end{eqnarray}
where $k_c$ is the maginal wavenumber corresponding to $\omega=0$ given by Eq. (\ref{cms5}). The pulsation $\omega_i(k)$ is given by
\begin{eqnarray}
\frac{\omega_i}{k_c}=\frac{k}{k_c}K^{-1}\left (\frac{k^2}{k_c^2}\right ).
\label{kk5}
\end{eqnarray}
Setting $u=\omega_i/k$, it can also be written in parametric form as
\begin{eqnarray}
\frac{\omega_i}{k_c}=u\sqrt{K(u)}, \qquad \frac{k}{k_c}=\sqrt{K(u)},
\label{kk6}
\end{eqnarray}
where $u$ goes from $-\infty$ to $+\infty$.

Let us obtain some asymptotic expansions of these relations (valid for
symmetric distributions):

(i) Let us first consider the case $\omega_i>0$ and $k\rightarrow 0$ corresponding to instability. Integrating Eq. (\ref{mon2}) by parts, we obtain
\begin{eqnarray}
1-4\pi G\int_{-\infty}^{+\infty}\frac{f(v)(\omega_{i}^{2}-k^2v^2)}{(k^2v^2+\omega_{i}^2)^{2}}\, dv=0.
\label{mon6}
\end{eqnarray}
Expanding the integrand in powers of $kv/\omega_i\ll 1$, we find that
\begin{eqnarray}
\omega_{i}^2=4\pi G\rho-3Tk^2-... \qquad (k\rightarrow 0),
\label{mon7q}
\end{eqnarray}
with $T=\langle v^2\rangle$ (where we recall that $v=v_z$ in the
present case).  This
expression can be compared with the corresponding expression
(\ref{fe16}) valid for a gas. This identification yields $c_s^2=3T$
so that large wavelength perturbations in a collisionless stellar system
correspond to one dimensional isentropic perturbations with index $\gamma=3$ in a gas (see
Appendix \ref{sec_fee}).

(ii) The case $\omega_i<0$ and $k\rightarrow +\infty$ corresponding to
stability cannot be treated at a general level because the result
depends on the behavior of the distribution function for large
velocities. The case of a Maxwellian distribution is specifically
considered in Sec. \ref{sec_vb}.

(iii) Let us finally consider the behavior of the dispersion relation
(\ref{mon2}) or (\ref{mon5}) close to the point of marginal
stability $k=k_c$. For $\omega_i=0$, we obtain the critical wavenumber $k_c$
given by Eq. (\ref{cms5}). Let us consider $\omega_i\rightarrow 0^+$
and $k\rightarrow k_c^{-}$ in Eq. (\ref{mon2}). We introduce the
function
\begin{eqnarray}
F(x)=\int_{-\infty}^{+\infty}\frac{v f'(v)}{v^2+x^2}\, dv,
\label{mon8}
\end{eqnarray}
for any real $x$. For $x\rightarrow 0$, we have the Taylor expansion $F(x)=F(0)+F'(0)x+...$ with
\begin{eqnarray}
F(0)=\int_{-\infty}^{+\infty}\frac{f'(v)}{v}\, dv,
\label{mon9}
\end{eqnarray}
\begin{eqnarray}
F'(x)=-\int_{-\infty}^{+\infty}\frac{2xv f'(v)}{(v^2+x^2)^2}\, dv=-\int_{-\infty}^{+\infty}\frac{x f''(v)}{v^2+x^2}\, dv,\nonumber\\
\label{mon10}
\end{eqnarray}
where we have used an integration by parts to get the last expression. Under this form, we cannot take the limit $x\rightarrow 0$ in the integral because the integral is not convergent for $x=0$. However, if we write Eq. (\ref{mon10}) in the equivalent form
\begin{eqnarray}
F'(x)=-x\int_{-\infty}^{+\infty}\frac{f''(v)-f''(0)}{v^2+x^2}\, dv-\int_{-\infty}^{+\infty}\frac{ x f''(0)}{v^2+x^2}\, dv,\nonumber\\
\label{mon11}
\end{eqnarray}
we obtain
\begin{eqnarray}
F'(0)=-\lim_{x\rightarrow 0}x f''(0)\int_{-\infty}^{+\infty}\frac{dv}{v^2+x^2}=-\pi f''(0) {\rm sign}(x).\nonumber\\
\label{mon12}
\end{eqnarray}
Regrouping the previous results, we find that the dispersion relation
(\ref{mon2}) becomes  for $\omega_i\rightarrow 0^+$:
\begin{eqnarray}
1+\frac{4\pi G}{k^2}\left\lbrace \int_{-\infty}^{+\infty}\frac{f'(v)}{v}\, dv-\pi f''(0)\frac{\omega_i}{k}\right\rbrace=0.
\label{mon13}
\end{eqnarray}
Similarly,  the dispersion relation
(\ref{mon5}) becomes for $\omega_i\rightarrow 0^-$:
\begin{eqnarray}
1+\frac{4\pi G}{k^2}\left\lbrace \int_{-\infty}^{+\infty}\frac{f'(v)}{v}\, dv+\pi f''(0)\frac{\omega_i}{k}\right\rbrace\nonumber\\
-\frac{8\pi^2G}{k^2}f''(0)\frac{\omega_i}{k}=0,
\label{mon14}
\end{eqnarray}
which takes the same form as Eq. (\ref{mon13}). Therefore,
Eq. (\ref{mon13}) is valid for $\omega_i\rightarrow 0$ whatever its
sign. To leading order, we obtain\footnote{In the derivation, we have
assumed that $f''(0)\neq 0$. If $f''(0)=0$, we need to develop the
Taylor expansion to the next order.}
\begin{eqnarray}
\omega_i=\frac{-k_c^3}{4\pi^2 G f''(0)}\left (1-\frac{k^2}{k_c^2}\right ), \qquad (k\rightarrow k_c).
\label{mon15}
\end{eqnarray}
This formula leads to the following result. First of all, we recall that the mode of marginal stability that we consider corresponds to $\omega_r/k=v_{ext}=0$, i.e. to the extremum value of the distribution function $f(v)$ at  $v_{ext}=0$. If the distribution is maximum at $v=0$, so that $f''(0)<0$, we find that the mode $\omega=i\omega_i$ is stable for $k>k_c$ and unstable for $k<k_c$. Alternatively, if the distribution is minimum at $v=0$, so that $f''(0)>0$, we find that  the mode $\omega=i\omega_i$ is stable for $k<k_c$ and unstable for $k>k_c$. This result will be  illustrated  in connection to Fig. \ref{GRAVdiagphasesym} for the symmetric double humped distribution.

In this section, we have discussed particular solutions of the
dispersion relation of the form $\omega=i\omega_i$. Of course, there
may exist other solutions to the equation $\epsilon(k,\omega)=0$ with
$\omega_r\neq 0$\footnote{The modes $\omega=\omega_r+i\omega_i$ with $\omega_i>0$ and $\omega_r\neq 0$ are sometimes called {\it overstable}.}. However, for single-humped distributions and
unstable wavenumbers $k<k_c$, the Nyquist curve encircles the origin
only once (see Sec. \ref{sec_sh}) so that, when it exists,
the solution $\omega=i\omega_i$ with $\omega_{i}>0$ is the only solution of
$\epsilon(k,\omega)=0$ (for single-humped distributions and stable
wavenumbers $k>k_c$, there may exist other solutions of
$\epsilon(k,\omega)=0$ with $\omega_{r}\neq 0$ and
$\omega_{i}<0$ as discussed in Sec. \ref{sec_mn}). Explicit solutions of the dispersion relation
$\epsilon(k,\omega)=0$ with $\omega=i\omega_{i}$ are given in Secs. \ref{sec_vb} and \ref{sec_vb2}
for isothermal and polytropic distributions.

\subsection{Equivalence between the Jeans instability criterion for a stellar system and the Jeans instability criterion for the corresponding barotropic gas}
\label{sec_fo}

Lynden-Bell (1962) first observed that the critical Jeans length for a
stellar system described by a Maxwellian distribution function is
equal to the critical Jeans length for an isothermal gas if we replace
the velocity of sound by the velocity dispersion in one direction. In
this section, we provide the appropriate generalization of this result
for an {\it arbitrary} distribution of the form $f=f(v^2)$ with $f'<0$.

As indicated in Sec. \ref{sec_vhc}, to any stellar system with a
distribution function $f=f(\epsilon)$ and $f'(\epsilon)<0$, we can
associate a corresponding barotropic gas with an equation of state
$p=p(\rho)$. According to Eq. (\ref{ep8}), the condition of
hydrostatic equilibrium can be written
\begin{eqnarray}
p'(\rho)=-\frac{\rho}{\rho'(\Phi)}.
\label{fo1}
\end{eqnarray}
Now, using Eq. (\ref{ep6}) and the fact that $\rho=\int f(\epsilon)\,
d{\bf v}$, we get
\begin{eqnarray}
c_{s}^2({\bf r})=-\frac{\rho({\bf r})}{\int f'(\epsilon)\, d{\bf v}}=-\frac{\rho({\bf r})}{\int \frac{\frac{\partial f}{\partial v_z}}{v_z} d{\bf v}}=-\frac{\rho({\bf r})}{\int_{-\infty}^{+\infty}  \frac{\frac{\partial g}{\partial v_z}}{v_z} d{v}_{z}},\nonumber\\
\label{fo2}
\end{eqnarray}
where $g({\bf r},v_z)=\int f(\epsilon) \, dv_xdv_y$.  For a spatially homogeneous system, we obtain the identity
\begin{eqnarray}
c_{s}^2=-\frac{\rho}{\int_{-\infty}^{+\infty} \frac{f'(v)}{v}\, d{v}},
\label{fo3}
\end{eqnarray}
where we have noted $v$ for $v_z$ and $f$ for $g$ like in
Sec. \ref{sec_lsv}. This identity is  explicitly checked in Appendix \ref{sec_cali} for the isothermal and polytropic distributions. Since $f(v)$ is symmetric with respect to $v=0$ and
has a single maximum at $v=0$, the Jeans instability criterion can be written (see Sec. \ref{sec_sh}):
\begin{eqnarray}
k^2<k_{c}^2=-4\pi G\int_{-\infty}^{+\infty} \frac{f'(v)}{v}\, dv.
\label{fo4}
\end{eqnarray}
Using identity (\ref{fo3}), it can be rewritten
\begin{eqnarray}
k^2<k_{c}^2=\frac{4\pi G\rho}{c_{s}^{2}}.
\label{fo5}
\end{eqnarray}
Therefore, the criterion of dynamical stability for a spatially homogeneous stellar system
coincides with the criterion of dynamical stability for the
corresponding barotropic gas (see Sec. \ref{sec_jeans}).  We insist on the fact
that this equivalence is valid for an arbitrary distribution function
of the form $f=f(v^2)$ with $f'<0$, not only for the
Maxwellian. We conclude that: an infinite homogeneous stellar system
with $f=f(v^2)$ and $f'<0$ is dynamically stable with respect to a
perturbation with wavenumber $k$ if and only if the corresponding
barotropic gas is dynamically stable with respect to this
perturbation. This is the proper formulation of the Antonov first law
for spatially homogeneous systems: in the present case, we have
equivalence\footnote{This ``equivalence'' for
the dynamical stability of a homogeneous stellar system and the
corresponding barotropic gas differs from the case of inhomogeneous
systems where the limits of stability of stars and galaxies are generically different (see,
e.g., Chavanis 2006a).}.  Of course, although the thresholds of
stability/instability coincide, the evolution of the perturbation is
different in a stellar system and in a fluid system (see Secs. \ref{sec_vb} and \ref{sec_vb2}).  We
should also emphasize that, in general, the velocity of sound $c_s$ in
the corresponding barotropic gas is {\it not} equal to the velocity
dispersion $\langle v^2\rangle^{1/2}$ in the stellar system, except
when $f(v)$ is the Maxwellian distribution leading to Lynden-Bell's result.

\section{Isothermal stellar systems}
\label{sec_vb}

\subsection{The equation of state}
\label{sec_mes}

We consider an isothermal stellar system described by the distribution function
\begin{eqnarray}
f=A e^{-\beta \epsilon}, \label{mes1}
\end{eqnarray}
where $\beta=1/T$ is a pseudo inverse temperature.
We justify here this distribution as a particular steady state of the
Vlasov equation\footnote{The statistical equilibrium state of a
self-gravitating system (resulting from a ``collisional'' relaxation)
is also described by an isothermal distribution of the form
(\ref{mes1}).  In that case, $f=Ae^{-\beta m \epsilon}$ is the
Boltzmann distribution, $\beta=1/(k_B T)$ is the inverse thermodynamic
temperature and $S_B[f]=-k_B\int
\frac{f}{m}\ln \frac{f}{m} d{\bf r}d{\bf v}$ is the Boltzmann
entropy (Padmanabhan 1990).}. The associated pseudo entropy is
\begin{eqnarray}
S[f]=-\int f\ln (f/f_0) d{\bf r}d{\bf v},\label{mes2}
\end{eqnarray}
where $f_0$ is a constant introduced to make the term in the logarithm
dimensionless (it will play no role in the following since it appears
in an additional constant term).  The distribution (\ref{mes1}) is
obtained by extremizing the pseudo entropy (\ref{mes2}) at fixed mass
and energy, writing $\delta S-\beta\delta E-\alpha\delta M=0$.  For an
isothermal distribution, the kinetic temperature (velocity dispersion)
defined by Eq. (\ref{vhc2b}) is spatially uniform, and it coincides
with the pseudo temperature: $T({\bf r})=T$. The barotropic gas
corresponding to the isothermal stellar system defined by
Eq. (\ref{mes1}) is the isothermal gas with an equation of state
$p({\bf r})=\rho({\bf r}) T$.  The velocity of sound is also spatially
uniform and it coincides with the velocity dispersion in one
direction: $c_{s}^{2}({\bf r})=T$. The density is related to the
gravitational potential by Eq. (\ref{ip2}).  It can be obtained by
integrating Eq. (\ref{mes1}) on the velocities leading to
Eq. (\ref{ip2}) with $A'=A(2\pi/\beta)^{3/2}$. It can also be obtained
by using Eq. (\ref{ep8}) with $p({\bf r})=\rho({\bf r}) T$, or by
extremizing the functional (\ref{ip3}) at fixed mass (Chavanis 2006a).
Finally, combining Eqs (\ref{mes1}) and (\ref{ip2}), we can express
the distribution function in terms of the density profile according to
\begin{eqnarray}
f=\biggl ({\beta\over 2\pi}\biggr )^{3/2}\rho({\bf r}) \ e^{-\beta {v^{2}\over 2}}.
\label{mes4}
\end{eqnarray}

{\it Remark}: the stability of box-confined isothermal stellar systems has been studied by Antonov (1962), Lynden-Bell \& Wood (1968), Padmanabhan (1990) and Chavanis (2006a).

\subsection{The dispersion relation}
\label{sec_md}

Let us now consider an infinite homogeneous isothermal system described by the Maxwellian distribution function (\ref{mes4}) with uniform
density $\rho({\bf r})=\rho$. The reduced distribution (\ref{lsv3}) is
\begin{eqnarray}
f=\biggl ({\beta\over 2\pi}\biggr )^{1/2}\rho\ e^{-\beta {v^{2}\over 2}}.
\label{md1}
\end{eqnarray}
The Maxwellian distribution has a single maximum at $v=0$. Therefore,
the condition of marginal stability (\ref{cms2b}) implies $\omega_{r}=0$. From
Eq. (\ref{cms1b}), we find that the Maxwellian distribution is marginally stable
for $k=k_c$ where we have introduced the critical wavenumber
\begin{equation}
\label{md2}
k_{c}^2=\frac{4\pi G\rho}{T}.
\end{equation}
According to the criterion (\ref{sh2}), the Maxwell distribution is
linearly dynamically stable if $k>k_{c}$ and linearly dynamically
unstable if $k<k_{c}$. The critical Jeans wavenumber (\ref{md2}) for
an isothermal stellar system is the same as the critical Jeans
wavenumber (\ref{ip5}) for an isothermal gas. This is to be expected
on account of the {general} result of Sec. \ref{sec_fo}.

The dielectric function (\ref{lsv4}) associated to the Maxwellian distribution  is
\begin{eqnarray}
\epsilon(k,\omega)=1-\frac{4\pi G}{k^2}\left (\frac{\beta}{2\pi}\right )^{1/2}\rho\int_{C}\frac{\beta v}{v-\frac{\omega}{k}}e^{-\beta {v^{2}\over 2}}\ dv.
\label{md3}
\end{eqnarray}
Introducing the critical
Jeans wavenumber (\ref{md2}), it can be rewritten
\begin{eqnarray}
\epsilon({k},\omega)=1-{k_{c}^{2}\over k^{2}} W\biggl ({\sqrt{\beta}\omega\over k}\biggr ),
 \label{md4}
\end{eqnarray}
where
\begin{eqnarray}
W(z)={1\over\sqrt{2\pi}}\int_{C}{x\over x-z}e^{-{x^{2}\over 2}}dx,
 \label{md5}
\end{eqnarray}
is the $W$-function of plasma physics (Ichimaru 1973). We note that $W(0)=1$. For any complex number $z$, we have the analytical expression
\begin{eqnarray}
W(z)=1-z e^{-{z^{2}\over 2}}\int_{0}^{z} e^{x^{2}\over 2}\, dx+i{\sqrt{\pi\over 2}}z e^{-{z^{2}\over 2}}.
\label{md6}
\end{eqnarray}

\subsection{Growth rate and damping rate}
\label{sec_mg}

\begin{figure}[htbp]
\centerline{
\includegraphics[width=8cm,angle=0]{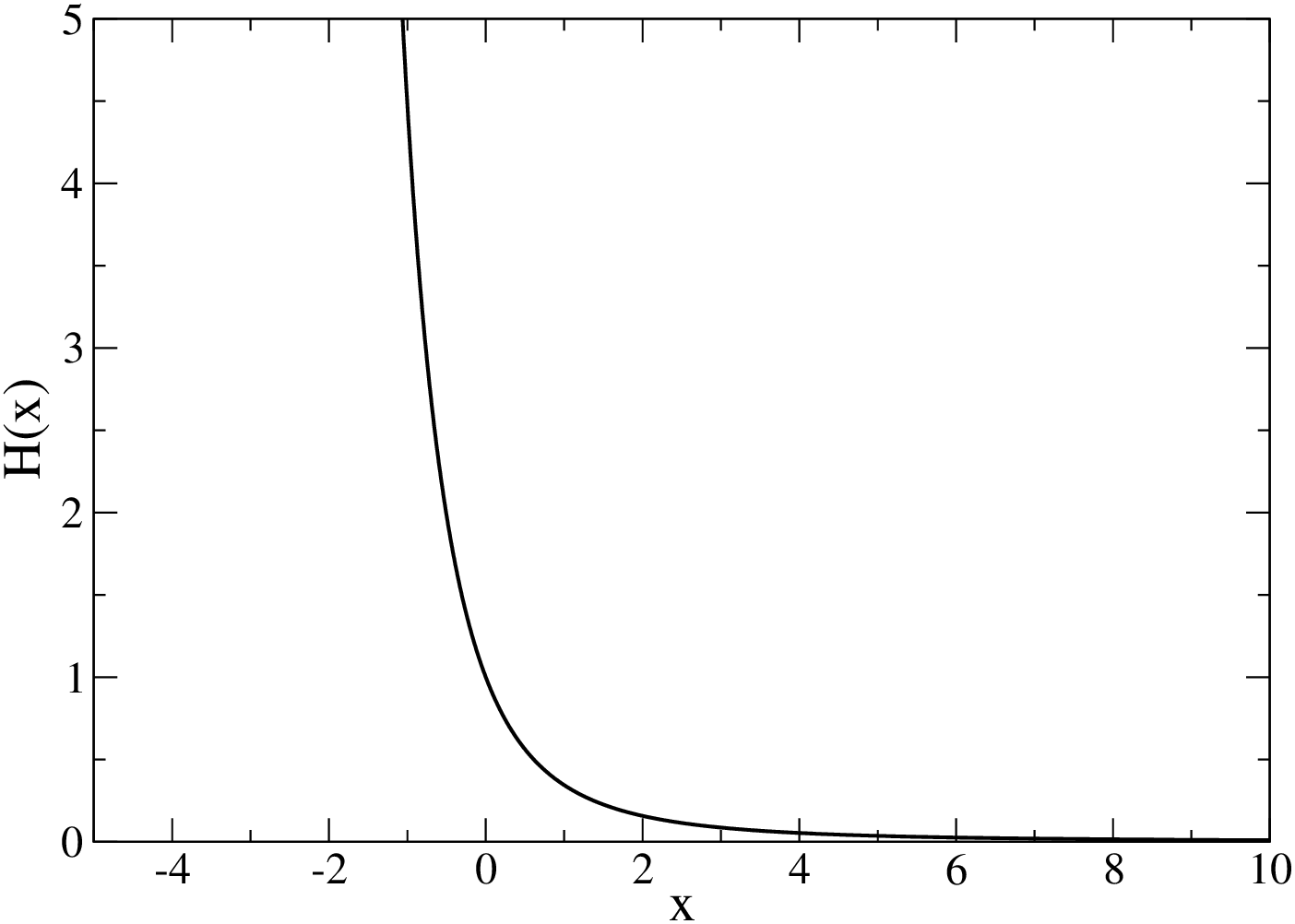}
} \caption[]{The function $H(x)$.}
\label{h}
\end{figure}

\begin{figure}[htbp]
\centerline{
\includegraphics[width=8cm,angle=0]{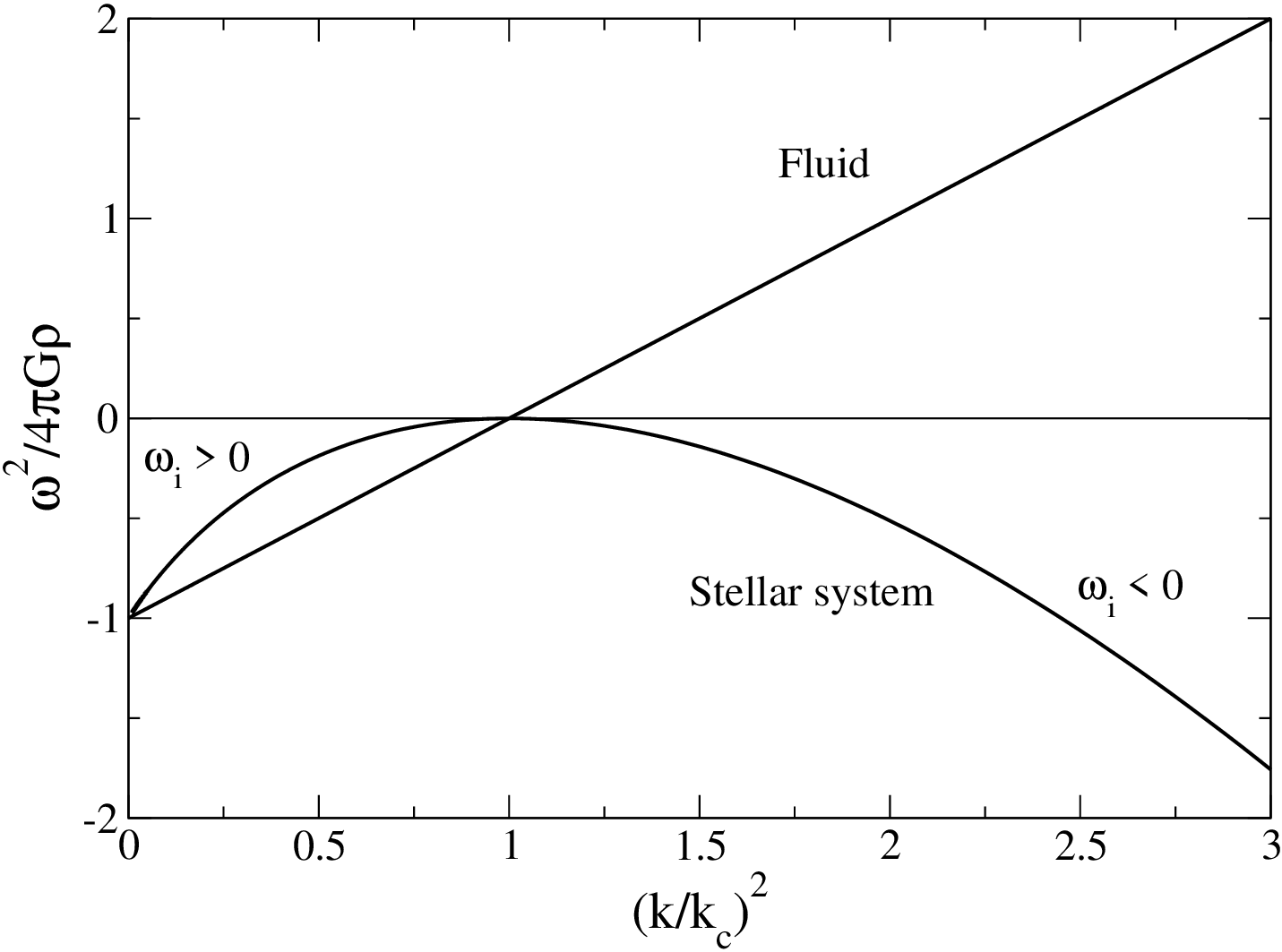}
} \caption[]{Pulsation $\omega$ as a function of the wavenumber $k$
for an isothermal stellar system. For $k<k_J$, the system is unstable
and $\omega=i\omega_i$ with $\omega_i>0$. For $k>k_J$, the system is
stable. There exists many branches of solutions
$\omega=\omega_r+i\omega_i$ with $\omega_i<0$ (see Sec. \ref{sec_mn}) but we
have only represented the branch corresponding to
$\omega_r=0$.  We have also compared these results with the case of an isothermal gas. For $k<k_J$, the system is unstable
and $\omega=\pm i\omega_i$. For $k>k_J$, the system is
stable and $\omega=\pm\omega_r$.}
\label{pulsation}
\end{figure}

We look for particular solutions of the dispersion relation $\epsilon({
k},\omega)=0$ in the form $\omega=i\omega_i$ where $\omega_i$ is
real. Using Eq. (\ref{md6}), we note that
\begin{eqnarray}
\epsilon({k},i\omega_i)=1-{k_{c}^{2}\over k^{2}}H\biggl (\frac{\sqrt{\beta}\omega_i}{k}\biggr ),
\label{mg1}
\end{eqnarray}
where we have introduced the function
\begin{equation}
\label{mg2} H(x)\equiv W(ix)= 1-\sqrt{\frac{\pi}{2}}x e^{\frac{x^{2}}{2}}{\rm erfc}\left (\frac{x}{\sqrt{2}}\right ).
\end{equation}
This function has the following asymptotic behaviors: (i) For
$x\rightarrow 0$, $H(x)=1-\sqrt{\frac{\pi}{2}}x+...$. (ii) For
$x\rightarrow +\infty$, $H(x)=\frac{1}{x^2}(1-{3\over
x^{2}}+...)$. (iii) For $x\rightarrow -\infty$, $H(x)\sim
-\sqrt{2\pi}x e^{\frac{x^{2}}{2}}$ (see Fig. \ref{h}). Using
$\epsilon(k,i\omega_i)=0$, the relation between $\omega_i$ and $k$
(for fixed $T$) can be written
\begin{eqnarray}
1-{k_{c}^{2}\over k^{2}}H\biggl ({\sqrt{\beta}\omega_i\over k}\biggr )=0,
 \label{mg3}
\end{eqnarray}
or more explicitly\footnote{This relation is established by Binney \& Tremaine (1987) for $\omega_i\ge 0$. The present analysis shows that it is also valid for $\omega_i<0$.}
\begin{equation}
\label{mg6}1-{k_{c}^{2}\over k^{2}}\biggl \lbrace 1-\sqrt{\pi\beta \over 2}{\omega_i\over k}e^{{\beta\omega_i^{2}\over 2k^{2}}}{\rm erfc}\biggl(\sqrt{\frac{\beta}{2}} {\omega_i\over k}\biggr )\biggr\rbrace=0.
\end{equation}
This equation can also be obtained directly from Eqs. (\ref{lsv5}) and
(\ref{lsv7}) (see Appendix \ref{sec_wi}).  If we set
$x=\sqrt{\beta}\omega_i/k$, we can rewrite Eq. (\ref{mg3}) in the
parametric form
\begin{eqnarray}
\frac{\omega_i}{\sqrt{4\pi G\rho}}=x\sqrt{H(x)}, \qquad \frac{k^2}{k_{c}^2}={H(x)}.
 \label{mg4}
\end{eqnarray}
By varying $x$ between $-\infty$ and $+\infty$, we obtain the full
curve giving $\omega_i$ as a function of the wavenumber $k$ (see
Fig. \ref{pulsation}). Since the time dependence of the perturbation
is $\delta f\sim e^{\omega_i t}$, the case of neutral stability
$\omega_i=0$ corresponds to $k=k_{c}$, the case of instability
$\omega_i>0$ corresponds to $k<k_{c}$ and the case of stability
$\omega_i<0$ corresponds to $k>k_{c}$. We have the asymptotic
behaviors
\begin{eqnarray}
{\omega_i\over \sqrt{4\pi G\rho}}\sim 1-\frac{3}{2}{k^{2}\over k_{c}^{2}},\qquad (k/k_{c}\rightarrow 0),
\label{mg5}
\end{eqnarray}
\begin{eqnarray}
{\omega_i\over \sqrt{4\pi G\rho}}\sim \sqrt{\frac{2}{\pi}}\biggl (1-{k^{2}\over k_{c}^{2}}\biggr ), \qquad (k/k_c\rightarrow 1),
\label{mg6b}
\end{eqnarray}
\begin{eqnarray}
{\omega_i\over \sqrt{4\pi G\rho}}\sim -\sqrt{{2k^{2}\over k_{c}^{2}}\ln\biggl ({k^{2}\over k_{c}^{2}}\biggr )},\qquad (k/k_c\rightarrow +\infty).
\label{mg7}
\end{eqnarray}

Equation (\ref{mg4}) provides a particular solution of the dispersion
relation $\epsilon(k,\omega)=0$ of the form $\omega=i\omega_i$ with
$\omega_r=0$. The dispersion relation may have other solutions with
$\omega_r\neq 0$. However, for single humped distributions, we know
that there is only one unstable mode with $\omega_i>0$ for given
$k<k_c$ (see Sec. \ref{sec_sh}). Since the solutions $\omega=i\omega_i$ given by
Eq. (\ref{mg4}) exist for any $k<k_c$, we conclude that they are the
only solutions in that range of wavenumbers. Therefore, for the
unstable wavenumbers $k<k_c$, the perturbation grows exponentially
rapidly without oscillating. In other words, there are no overstable modes for the Maxwell distribution\footnote{Binney \& Tremaine (1987) show this result by a different method.}.
This is the same behavior as in a fluid
system (see Sec. \ref{sec_gas}) except that the growth rate $\omega_i>0$ in the
stellar system [see Eq. (\ref{mg4})] and in the fluid system [see
Eq. (\ref{jeans5})] are different\footnote{They only coincide for a
cold system $T=c_s^2=0$ (see Sec. \ref{sec_lsv}) or for $k=0$ (see
Sec. \ref{sec_mon}).}. For the stable wavenumbers $k>k_c$, the perturbations in a
stellar system are damped exponentially rapidly ($\omega_i<0$). We
have exhibited particular solutions (\ref{mg4}) that are damped
without oscillating $\omega_r=0$ but these are not the only solutions
of the dispersion relation. There also exists modes that are damped
exponentially ($\omega_i<0$) and oscillate $\omega_r\neq 0$ (see
asymptotic results in Sec. \ref{sec_mn}). This form of damping for collisionless
stellar systems is similar to the Landau damping for a
plasma\footnote{In plasma physics, for the Maxwellian distribution, there is no solution to the dispersion relation of the form $\omega=i\omega_i$. The pulsation
$\omega_r$ is non zero (see Sec. \ref{sec_pvp}).}. The situation is
very different in a fluid system. In that case, the stable modes with
wavenumbers $k>k_c$ correspond to gravity-modified sound waves that
propagate without attenuation ($\omega_r\neq 0$, $\omega_i=0$).

The pulsation of the perturbations in an infinite homogeneous
isothermal stellar system is plotted in Fig. \ref{pulsation} as a
function of the wavenumber $k$. For comparison, we have also indicated
the pulsation of the perturbations in an infinite homogeneous
isothermal gas.

\subsection{Other branches for $k\rightarrow +\infty$}
\label{sec_mn}

Let us solve the dispersion relation $\epsilon(k,\omega)=0$ for an
isothermal distribution in the limit $k\rightarrow
+\infty$\footnote{We here adapt the method of plasma physics developed by
Landau (1946), Jackson (1960) and Balescu (1963).}. We look for a solution of the equation
$\epsilon(k,\omega)=0$ of the form $\omega=\omega_r+i\omega_i$ with
$\omega_i<0$ (damping) and $\omega_i\gg\omega_r$. This corresponds to
heavily damped perturbations. We shall check this approximation a
posteriori. Using Eq. (\ref{lsv7}) for $\omega_{i}<0$ ,
Eq. (\ref{lsv4}) can be written
\begin{eqnarray}
1+\frac{4\pi G}{k^2}\int_{-\infty}^{+\infty} {{f'(v)}\over v-\frac{\omega}{k}}\, dv+\frac{8\pi^2 G}{k^2}i f'\left (\frac{\omega}{k}\right)=0.
\label{gh1}
\end{eqnarray}
If $f(v)$ decreases like $e^{-\beta v^2/2}$ for real $v\rightarrow \pm\infty$, then for complex $\omega=\omega_r+i\omega_i$ with $\omega_i\gg \omega_r$, $f'(\omega/k)$ will increase like $e^{\beta\omega_i^2/2k^2}$. Therefore, to leading order, the foregoing equation reduces to
\begin{eqnarray}
1+\frac{8\pi^2 G}{k^2}i f'\left (\frac{\omega}{k}\right)=0.
\label{gh2}
\end{eqnarray}
Separating real and imaginary parts, we obtain two transcendant equations
\begin{eqnarray}
{\rm Re}\left\lbrack\frac{8\pi^2 G}{k^2}i f'\left (\frac{\omega_r+i\omega_i}{k}\right)\right\rbrack=-1,
\label{gh3}
\end{eqnarray}
\begin{eqnarray}
{\rm Im}\left\lbrack i f'\left (\frac{\omega_r+i\omega_i}{k}\right)\right\rbrack=0,
\label{gh4}
\end{eqnarray}
which crucially depend on the form of the distribution. For the Maxwellian (\ref{md1}), they can be rewritten to leading order in the limit $\omega_i/\omega_r\gg 1$ as
\begin{eqnarray}
\frac{8\pi^2 G}{k^3}\left (\frac{\beta}{2\pi}\right )^{1/2}\rho\beta\omega_i e^{\frac{\beta\omega_{i}^{2}}{2k^2}}\cos\left (\frac{\beta\omega_r\omega_i}{k^2}\right )=-1,
\label{gh5}
\end{eqnarray}
\begin{eqnarray}
\sin\left (\frac{\beta\omega_r\omega_i}{k^2}\right )=0.
\label{gh6}
\end{eqnarray}
Equation (\ref{gh6}) implies $\beta \omega_r\omega_i/k^2=m\pi$. Eq.  (\ref{gh5}) will have a solution provided that $m$ is even. Then, Eq.  (\ref{gh5}) gives
\begin{eqnarray}
\frac{8\pi^2 G}{k^3}\left (\frac{\beta}{2\pi}\right )^{1/2}\rho\beta\omega_i e^{\frac{\beta\omega_{i}^{2}}{2k^2}}=-1,
\label{gh7}
\end{eqnarray}
which determines $\omega_i$. By a graphical construction, it is easy to see that $|\omega_i|$ is an increasing function of $k$. For $k\rightarrow +\infty$, we find the asymptotic behaviors
\begin{eqnarray}
\omega_i=-\frac{2}{\sqrt{\beta}}k\sqrt{\ln k}, \qquad \omega_r=-m\frac{\pi}{2}\frac{1}{\sqrt{\beta}}\frac{k}{\sqrt{\ln k}}.
\label{gh8}
\end{eqnarray}
Since $\omega_{i}/\omega_{r}\sim \ln k\rightarrow +\infty$, our basic assumption is satisfied. Therefore, for $k>k_c$ we have several branches of solutions parameterized by the even integer $m$. For $m=0$, we recover the results of Sec. \ref{sec_mg}.

{\it Remark}: by analogy with plasma physics, we could also look for
solutions of the dispersion relation (\ref{lsv4}) of the form
$\omega=\omega_r+i\omega_i$ with $\omega_i\ll \omega_r$. This
corresponds to weakly damped perturbations. In plasma physics, these
solutions are valid for $k\rightarrow 0$ and lead to the usual Landau
damping formula. However, a self-gravitating system is unstable for
$k<k_c$. Furthermore, it is easy to show that there is no solution of
that form to Eq. (\ref{lsv4}) whatever the form of the distribution
$f(v)$ and the wavenumber $k$. This implies that for attractive
interactions (like gravity) the perturbations are unstable for $k<k_c$
and heavily damped for $k>k_c$ while for repulsive interaction (like
plasmas) they are weakly damped for $k<k_D$ and heavily damped for
$k>k_D$.

\subsection{Nyquist curve}
\label{sec_mnb}

It will be convenient in the following to work with dimensionless quantities.
We introduce the dimensionless wavenumber and the dimensionless pulsation
\begin{eqnarray}
\eta=\frac{4\pi G\rho}{T k^2}, \qquad \Omega=\frac{\omega}{\sqrt{4\pi G\rho}},
\label{mn1}
\end{eqnarray}
Noting that $\sqrt{\beta}\omega/k=\sqrt{\eta}\Omega$, the dielectric function (\ref{md4}) can be rewritten
\begin{eqnarray}
\epsilon({\eta},\Omega)=1-\eta W(\sqrt{\eta}\Omega).
 \label{mn3}
\end{eqnarray}
When $\Omega_{i}=0$, the real and imaginary parts of the dielectric function  $\epsilon(\eta,\Omega_{r})=\epsilon_{r}(\eta,\Omega_{r})+i\epsilon_{i}(\eta,\Omega_{r})$ can be written
\begin{equation}
\label{mn4}
\epsilon_{r}(\eta,\Omega_{r})=1-\eta W_{r}\left (\sqrt{\eta}\Omega_r\right ),
\end{equation}
\begin{equation}
\label{mn5}
\epsilon_{i}(\eta,\Omega_{r})=-\eta W_{i}\left (\sqrt{\eta}\Omega_r\right ),
\end{equation}
with
\begin{equation}
\label{mn6}
W_{r}(z)=1-z e^{-\frac{z^2}{2}}\int_{0}^{z}e^{\frac{x^2}{2}}\, dx,
\end{equation}
\begin{equation}
\label{mn7}
W_{i}(z)=\sqrt{\frac{\pi}{2}}z e^{-\frac{z^2}{2}},
\end{equation}
where $z$ is here a real number. The condition of marginal stability corresponds to $\epsilon_{r}(\eta,\Omega_{r})=\epsilon_{i}(\eta,\Omega_{r})=0$. The condition $\epsilon_{i}(\eta,\Omega_{r})=0$, which is equivalent to $f'(\sqrt{\eta}\Omega_{r})=0$, implies $\Omega_{r}=0$. Then, the condition $\epsilon_{r}(\eta,\Omega_{r})=0$ leads to $\eta=\eta_{c}$ with
\begin{equation}
\label{mn8}
\eta_c=1.
\end{equation}

\begin{figure}[htbp]
\centerline{
\includegraphics[width=8cm,angle=0]{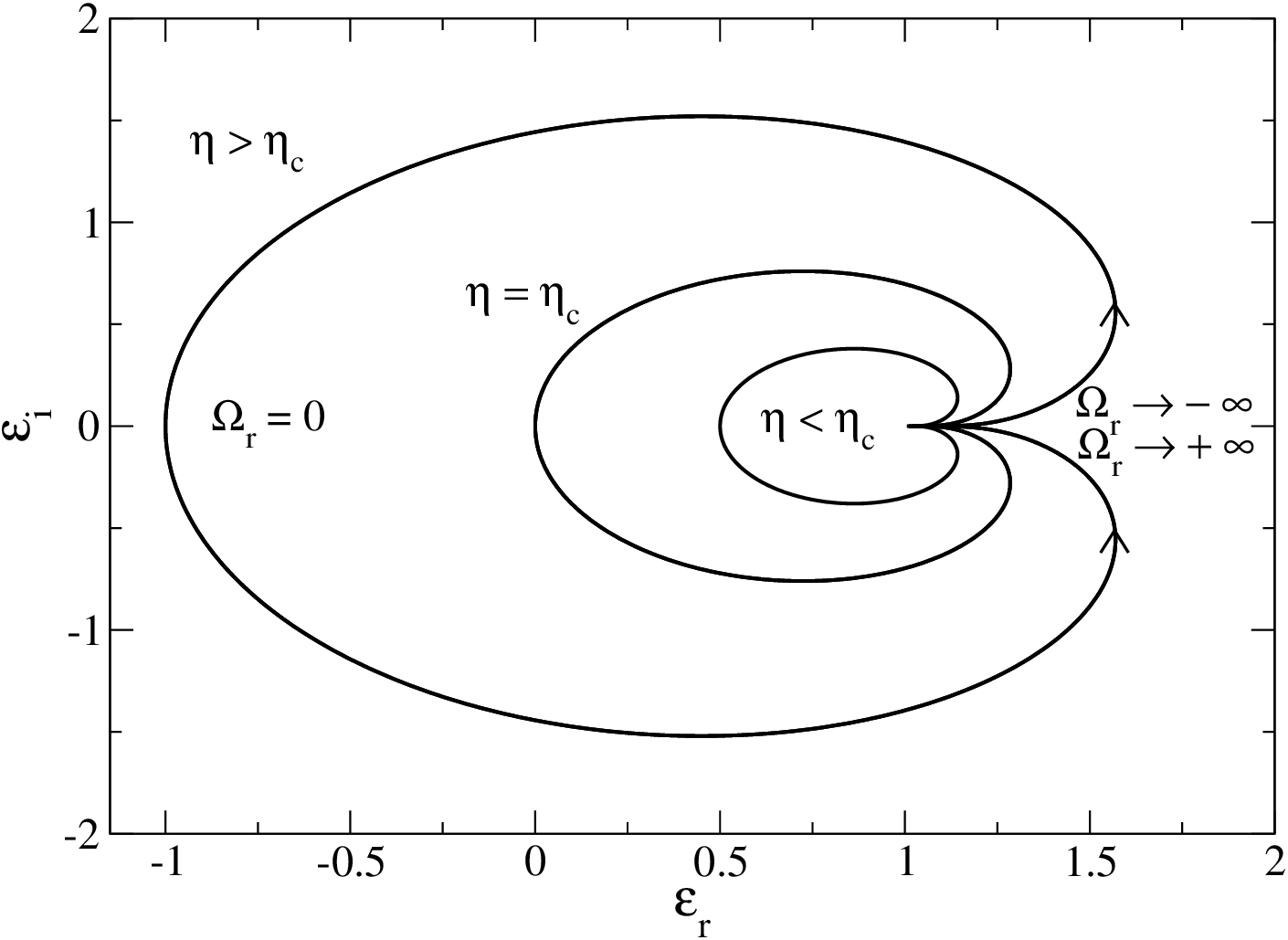}
} \caption[]{Nyquist curve for the Maxwellian distribution
(\ref{md1}). The DF is stable for perturbations with $\eta<\eta_c$
($k>k_c$), marginally stable for perturbations with $\eta=\eta_c$
($k=k_c$) and unstable for perturbations with $\eta>\eta_c$
($k<k_c$). We have taken $\eta=2,1,0.5$ from the outer to the inner
curve.}
\label{maxwell}
\end{figure}

To apply the Nyquist method, we need to plot the curve
$(\epsilon_{r}(\eta,\Omega_{r}),\epsilon_{i}(\eta,\Omega_{r}))$ in the
$\epsilon$-plane. For $\Omega_{r}\rightarrow \pm \infty$, this curve
tends to the point $(1,0)$ in the manner described in
Sec. \ref{sec_sh}. On the other hand, for $\Omega_r=0$, it crosses the
$x$-axis at $(\epsilon_r(\eta,0)=1-\eta,0)$.  The Nyquist curve is
represented in Fig. \ref{maxwell} for several values of the wavenumber
$\eta$. For $\eta<1$ (i.e. $k>k_c$), the Nyquist curve does not
encircle the origin so that the Maxwellian distribution is stable. For
$\eta>1$ (i.e. $k<k_c$) the Nyquist curve encircles the origin so that
the Maxwellian distribution is unstable. For $\eta=1$ (i.e. $k=k_c$)
the Nyquist curve passes through the origin so that the Maxwellian
distribution is marginally stable. The Nyquist method provides a nice
graphical illustration of the Jeans instability criterion for an
infinite homogeneous stellar system.

\section{Stellar polytropes}
\label{sec_vb2}

\subsection{The equation of state}

\begin{figure}[htbp]
\centerline{
\includegraphics[width=8cm,angle=0]{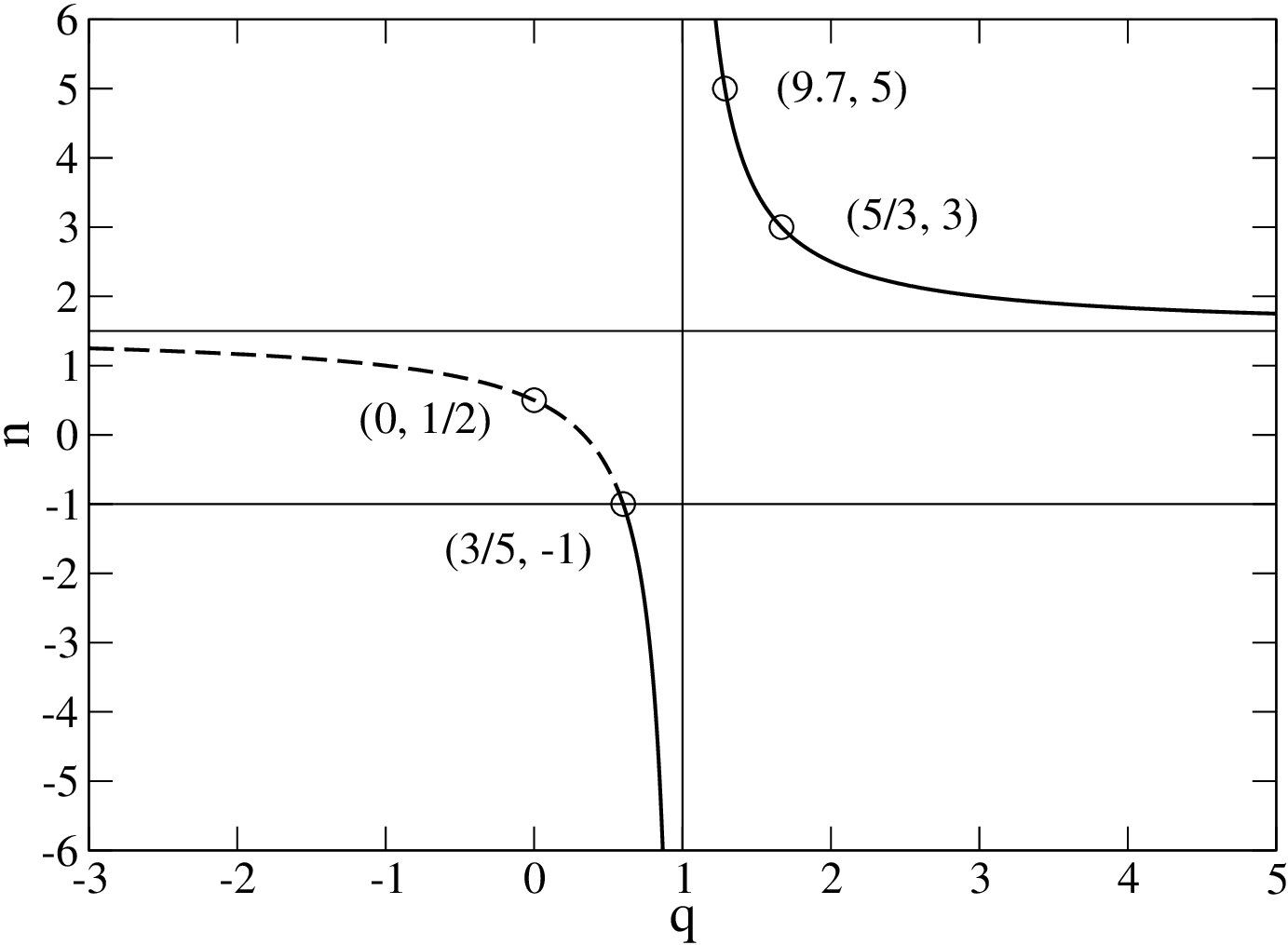}
} \caption[]{The relation between the polytropic index $n$ and the
parameter $q$. Some particular values $(q,n)$ are indicated for
reference.} \label{nq}
\end{figure}

We consider a stellar polytrope (or polytropic galaxy) described by the distribution function
\begin{eqnarray}
f=\biggl \lbrack \mu-{\beta (q-1)\over q}\epsilon\biggr \rbrack_+^{1\over q-1},
\label{vb1}
\end{eqnarray}
where $\beta=1/T$ is a pseudo inverse temperature
and $q$ is a parameter related to the
traditional polytropic index $n$ by
\begin{eqnarray}
n={3\over 2}+{1\over q-1}.
\label{vb2}
\end{eqnarray}
This relation is plotted in Fig. \ref{nq} and specific values considered
in the sequel are highlighted. We justify here the polytropic distribution function (\ref{vb1}) as a particular steady state of the Vlasov equation\footnote{Some authors (Plastino \& Plastino (1993), Lima et al. (2002), Silva \& Alcaniz (2004), Lima \& de Souza (2005), Taruya \& Sakagami (2003a), Leubner (2005), Kronberger et al. (2006), Du Julin (2006)) have interpreted the polytropic distribution  (\ref{vb1}) and the functional (\ref{vb3}) in terms of Tsallis (1988) generalized thermodynamics. Here, we use a more conventional approach (Ipser 1974, Ipser \& Horwitz 1979, Binney \& Tremaine 1987, Chavanis 2006a). We interpret the polytropic distribution  (\ref{vb1}) as a particular steady state of the Vlasov equation and the functional (\ref{vb3}) as a pseudo entropy. Its maximization at fixed mass and energy provides a condition of nonlinear dynamical stability with respect to the Vlasov equation (see Sec. \ref{sec_vh}), not a condition of ``generalized thermodynamical stability''. In particular, as argued by Chavanis \& Sire (2005) and Campa et al. (2008), the instabilities reported by Taruya \& Sakagami (2003b) correspond to Vlasov dynamical instabilities, not ``generalized thermodynamical instabilities''. In the present context, the analogies with Tsallis thermodynamics are purely coincidental. They are the mark of a {\it thermodynamical analogy} (Chavanis 2006a). Tsallis generalized thermodynamics applies in different contexts (see Chavanis 2008a).}. The associated pseudo-entropy is
\begin{eqnarray}
S=-{1\over q-1}\int (f^{q}-f_0^{q-1}f)d{\bf r}d{\bf v},
\label{vb3}
\end{eqnarray}
where $f_0$ is a constant introduced for reasons of homogeneity (it will play no role in the following since it appears in a term proportional to the mass that is conserved). The DF (\ref{vb1}) is obtained by
extremizing the pseudo entropy (\ref{vb3}) at fixed mass and energy, writing
$\delta S-\beta\delta E-\alpha\delta M=0$.  The condition that $C(f)$ must be convex imposes $q>0$. On the other hand, we shall assume that $f(\epsilon)$ is a decreasing function of the energy so that $\beta>0$. It is important to note that $1/\beta$ does {\it not} represent the kinetic temperature (or velocity dispersion) of the polytropic distribution (see Sec. \ref{sec_other}).

We need to distinguish two cases depending on the sign of $q-1$. For $q>1$ ($n>3/2$), the
distribution function can be written
\begin{eqnarray}
f=A(\epsilon_{m}-\epsilon)_+^{1\over q-1},
\label{vb4}
\end{eqnarray}
where we have set $A=\lbrack\beta(q-1)/q\rbrack^{1\over q-1}$ and
$\epsilon_{m}=q\mu/\lbrack \beta(q-1)\rbrack$. Such distributions have a compact support since they
  vanish at $\epsilon=\epsilon_m$. For $\epsilon>\epsilon_m$, we set $f=0$. Therefore, the notation $[x]_+=x$ for
$x>0$ and $[x]_+=0$ for $x<0$. At a given position, the distribution function vanishes for $v\ge v_{m}({\bf
r})=\sqrt{2(\epsilon_{m}-\Phi({\bf r}))}$. For $q\rightarrow 1$
($n\rightarrow +\infty$), we recover the isothermal distribution (\ref{mes1}) and for
$n=3/2$ the distribution function is a step function (see
Sec. \ref{sec_fermi}). This is the distribution function of a Fermi
gas at zero temperature describing classical white dwarf stars
(Chandrasekhar 1942). For $0<q<1$, the distribution function can be written
\begin{eqnarray}
f=A(\epsilon_{0}+\epsilon)^{1\over q-1},
\label{vb5}
\end{eqnarray}
where we have set $A=\lbrack\beta(1-q)/q\rbrack^{1\over q-1}$ and
$\epsilon_{0}=q\mu/\lbrack\beta(1-q)\rbrack$. Such distributions are defined for all velocities. At a given position, the distribution function behaves, for large velocities, as
$f\sim v^{2/(q-1)}\sim v^{2n-3}$. We shall only consider distribution functions for which the
density $\rho=\int f\, d{\bf v}$ and the pressure $p=\frac{1}{3}\int f v^2\, d{\bf v}$ are
finite. This implies
$3/5<q<1$ ($n<-1$)\footnote{If we allow $\beta$ to be negative, then it is possible to construct stellar polytropes with index $1/2<n<3/2$ which are mathematically well-behaved (see Binney \& Tremaine 1987). However, for those polytropes, the distribution function increases with the energy (and diverges at $\epsilon=\epsilon_m$) so they may not be physical.}.

Let us now determine the equation of state of the barotropic gas corresponding to a stellar polytrope.
For $n>3/2$, the density and
the pressure can be expressed as
\begin{eqnarray}
\rho=4\pi\sqrt{2}A(\epsilon_{m}-\Phi)^{n}{\Gamma(3/2)\Gamma(n-1/2)\over \Gamma(n+1)},
\label{vb6}
\end{eqnarray}
\begin{eqnarray}
p={1\over n+1}4\pi\sqrt{2}A(\epsilon_{m}-\Phi)^{n+1}{\Gamma(3/2)\Gamma(n-1/2)\over \Gamma(n+1)},
\label{vb7}
\end{eqnarray}
where $\Gamma(x)$ denotes the Gamma function.
For $n<-1$, the density and the pressure can be expressed as
\begin{eqnarray}
\rho=4\pi\sqrt{2}A(\epsilon_{0}+\Phi)^{n}{\Gamma(-n)\Gamma(3/2)\over \Gamma(3/2-n)},
\label{vb8}
\end{eqnarray}
\begin{eqnarray}
p=-{1\over n+1}4\pi\sqrt{2}A(\epsilon_{0}+\Phi)^{n+1}{\Gamma(-n)\Gamma(3/2)\over \Gamma(3/2-n)}.
\label{vb9}
\end{eqnarray}
Eliminating the gravitational potential between the expressions (\ref{vb6})-(\ref{vb7}) and (\ref{vb8})-(\ref{vb9}), one finds that
\begin{eqnarray}
p=K\rho^{\gamma}, \qquad \gamma=1+{1\over n},
\label{vb10}
\end{eqnarray}
where
\begin{eqnarray}
K={1\over n+1}\biggl \lbrace 4\pi\sqrt{2}A {\Gamma(3/2)\Gamma(n-1/2)\over \Gamma(n+1)}
\biggr \rbrace^{-{1\over n}} (n>3/2)\qquad
\label{vb11}
\end{eqnarray}
\begin{eqnarray}
K=-{1\over n+1}\biggl \lbrace 4\pi\sqrt{2}A{\Gamma(-n)\Gamma(3/2)\over\Gamma(3/2-n)}\biggr \rbrace^{-{1\over n}} (n<-1).
\label{vb12}
\end{eqnarray}
Therefore, a stellar polytrope has the same equation of state
(\ref{vb10}) as a polytropic star. However, they do not have the same
distribution function (compare Eq. (\ref{vb1}) to Eq.
(\ref{ep4})) except for $n\rightarrow \infty$ corresponding to the isothermal case.
The density is related to the gravitational potential by Eq. (\ref{ip7}).
It can be obtained by integrating Eq. (\ref{vb1}) on the velocity leading to Eqs. (\ref{vb6}) and (\ref{vb8}) from which, using Eqs. (\ref{vb11}) and (\ref{vb12}), we deduce Eq. (\ref{ip7}) with $\lambda=\epsilon_m/(K(n+1))$ for $n\ge 3/2$ and $\lambda=\epsilon_0/(-K(n+1))$ for $n<-1$. It can also be obtained by using Eq. (\ref{ep8}) with Eq. (\ref{vb10}) or by extremizing the functional (\ref{ip8}) at fixed mass (Chavanis 2006a). Note that Eqs. (\ref{ip7}) and (\ref{ip8})  are similar to
Eqs. (\ref{vb1}) and ({\ref{vb3})  with $\gamma$ playing the role of the
parameter $q$ and $K$ playing the role of the pseudo temperature $T=1/\beta$.
Polytropic distributions (including the isothermal one) are apparently the only
distributions for which $f(\epsilon)$ and $\rho(\Phi)$ have a similar mathematical form.

{\it Remark}: isolated stellar polytropes have a finite mass iff $3/2\le n\le 5$ and they are dynamically stable (Binney \& Tremaine 1987). The stability of box-confined polytropes is studied by  Taruya \& Sakagami (2003a) in the context of  generalized thermodynamics and by Chavanis (2006a) in the context of Vlasov dynamical stability.

\subsection{Other expressions of the distribution function}
\label{sec_other}

We can write the distribution function of stellar polytropes
(\ref{vb1}) in different forms that all have their own
interest. This will show that different notions of ``temperature'' exist for
polytropic distributions\footnote{We recall that, for collisionless stellar systems, the mass of the stars does not appear in the Vlasov equation and the different ``temperatures'' that we introduce have the dimension of a velocity squared.}.

(i) Pseudo temperature $T=1/\beta$: as indicated previously, the form (\ref{vb1}) of the polytropic distribution directly comes from the variational principle (\ref{vh3}) when we write the pseudo entropy in the form (\ref{vb3}). Therefore, $\beta=1/T$ is the Lagrange multiplier associated with the conservation of energy. It is called ``pseudo inverse temperature''. Note, however, that $T=1/\beta$ does not have the dimension of a temperature (squared velocity).

(ii) Dimensional temperature $\theta=1/b$: we can define a quantity that has the dimension of a temperature (squared velocity) by setting $b=\beta/q\mu$. If we define furthermore $f_*=\mu^{1/(q-1)}$, the polytropic distribution (\ref{vb1}) can be rewritten
\begin{eqnarray}
f=f_*\biggl \lbrack 1-{b (q-1)}\epsilon\biggr \rbrack_+^{1\over q-1}.
\label{na1}
\end{eqnarray}
Using Eqs. (\ref{vb6}) and (\ref{vb8}), the relation between the density and the gravitational potential can be written
\begin{eqnarray}
\rho=\rho_{*}\biggl \lbrack 1-{b (q-1)}\Phi\biggr \rbrack^{n},
\label{na2}
\end{eqnarray}
with
\begin{eqnarray}
\rho_*=2\pi f_{*}\left (\frac{2n-3}{b}\right )^{3/2}\frac{\Gamma(3/2)\Gamma(n-1/2)}{\Gamma(n+1)} \ (n>3/2),\nonumber\\
\label{na3}
\end{eqnarray}
\begin{eqnarray}
\rho_*=2\pi f_{*}\left (\frac{3-2n}{b}\right )^{3/2}\frac{\Gamma(3/2)\Gamma(-n)}{\Gamma(3/2-n)}  \ (n<-1).
\label{na4}
\end{eqnarray}

(iii) Polytropic temperature $K$: eliminating the gravitational potential between Eqs. (\ref{vb4}) and  (\ref{vb6}), and between  Eqs. (\ref{vb5}) and (\ref{vb8}),  we can express the distribution function in terms of the density according to
\begin{eqnarray}
f={1\over Z}\biggl \lbrack \rho({\bf r})^{1/n}-{v^{2}/2\over (n+1)K}\biggr\rbrack_+^{n-3/2},
\label{vb19}
\end{eqnarray}
\begin{eqnarray}
Z=4\pi\sqrt{2}{\Gamma(3/2)\Gamma(n-1/2)\over\Gamma(n+1)}\lbrack K(n+1)\rbrack^{3/2} \ (n>3/2)
\label{vb20}
\end{eqnarray}
\begin{eqnarray}
Z=4\pi\sqrt{2}{\Gamma(-n)\Gamma(3/2)\over\Gamma(3/2-n)}\lbrack -K(n+1)\rbrack^{3/2}\ (n<-1).
\label{vb21}
\end{eqnarray}
This is the polytropic counterpart of expression (\ref{mes4}) for the
isothermal distribution. The constant $K$ plays a role similar to the temperature $T$ in an isothermal distribution. In particular, it is uniform in a polytropic distribution as is the temperature in an isothermal system. For that reason, it is sometimes called a polytropic temperature.

(iv) Kinetic temperature $T({\bf r})$: for a polytropic distribution, the kinetic temperature (velocity dispersion) defined by Eq. (\ref{vhc2b}) is given by
\begin{eqnarray}
T({\bf r})=K\rho({\bf r})^{\gamma-1}.
\label{vb21bis}
\end{eqnarray}
For an inhomogeneous stellar polytrope, the kinetic temperature $T({\bf r})$ is position dependent and differs from the pseudo temperature $T=1/\beta$. The velocity of sound is given by
\begin{eqnarray}
c_s^2({\bf r})=K\gamma\rho({\bf r})^{\gamma-1}=\gamma T({\bf r}).
\label{vb21tris}
\end{eqnarray}
It is also position dependent and differs from the velocity dispersion (they differ by a factor $\gamma$). Using Eq. (\ref{vb21bis}), the distribution function
(\ref{vb19}) can be written
\begin{eqnarray}
f=B_{n}{\rho({\bf r})\over \lbrack 2\pi T({\bf r})\rbrack^{3/2}}\biggl\lbrack 1-{v^{2}/2\over (n+1)T({\bf r})}\biggr\rbrack_+^{n-3/2},
\label{vb23}
\end{eqnarray}
\begin{eqnarray}
B_{n}={\Gamma(n+1)\over\Gamma(n-1/2)(n+1)^{3/2}}, \quad (n>3/2)
\label{vb24}
\end{eqnarray}
\begin{eqnarray}
B_{n}={\Gamma(3/2-n)\over\Gamma(-n)\lbrack -(n+1)\rbrack^{3/2}} \quad (n<-1).
\label{vb25}
\end{eqnarray}
Note that for $n>3/2$, the maximum velocity can be expressed in terms of the kinetic temperature by
\begin{eqnarray}
v_{m}({\bf r})=\sqrt{2(n+1)T({\bf r})}.
\label{vb26}
\end{eqnarray}
Using $\Gamma(z+a)/\Gamma(z)\sim z^{a}$ for $z\rightarrow +\infty$, we recover the isothermal distribution (\ref{mes1}) for $n\rightarrow +\infty$.
On the other hand, from Eqs. (\ref{vb21bis}) and (\ref{ip7}), we
immediately get $T({\bf r})=K(\lambda-(\gamma-1)\Phi({\bf
r})/K\gamma)$ so that
\begin{eqnarray}
\nabla T=-{\gamma-1\over\gamma}\nabla\Phi.
\label{vb22}
\end{eqnarray}
This shows that, for a stellar
polytrope, the kinetic temperature (velocity dispersion) is a linear
function of the gravitational potential\footnote{For any spherical stellar system with $f=f(\epsilon)$, we have $\rho=\rho(\Phi)$ and $p=p(\Phi)$ so that the kinetic temperature (velocity dispersion) $T=p/\rho$ is a function $T=T(\Phi)$ of the gravitational potential. For a polytropic distribution function, this relation turns out to be linear.}. The coefficient of
proportionality is related
to the polytropic index by $(\gamma-1)/\gamma=1/(n+1)=2(q-1)/(5q-3)$.
This relation can also be obtained directly from Eq. (\ref{vb4}) [or Eq. (\ref{vb5})] noting that
\begin{eqnarray}
f=A(\epsilon_{m}-\Phi)^{n-3/2}\left \lbrack 1-\frac{v^{2}/2}{\epsilon_{m}-\Phi}\right\rbrack_+^{n-3/2},
\label{vb22ndl}
\end{eqnarray}
and comparing with Eq. (\ref{vb23}).

\subsection{The dispersion relation}
\label{sec_pdr}

Let us now consider an infinite homogeneous polytropic stellar system described by the polytropic distribution function (\ref{vb23}) with uniform density  $\rho({\bf r})=\rho$ and uniform kinetic temperature $T({\bf r})=T$. From now on, $T=1/\beta$ will denote the kinetic temperature (\ref{vhc2b}), not the Lagrange multiplier appearing in Eq. (\ref{vb1}). The kinetic temperature is uniform because we have assumed that the density is uniform. The reduced distribution function (\ref{lsv3}) is\footnote{If we justify the distribution function (\ref{pdr1})  from the 3D distribution function  (\ref{vb23}) integrated on $v_x$ and $v_y$, then it is valid for $n>3/2$ and $n<-1$. However, as far as mathematics is concerned, the distribution (\ref{pdr1})  is normalizable and has a finite variance in the range of indices $n\ge 1/2$ and $n<-1$.}
\begin{equation}
\label{pdr1}
f(v)=B_n \frac{\rho}{\sqrt{2\pi T}}\left [1-\frac{v^2}{2(n+1)T} \right ]_+^{n-1/2},
\end{equation}
where $\rho$ is the density, $T=1/\beta=\langle v^2\rangle$ is the velocity dispersion in one direction and $B_n$ is a normalization constant given by
\begin{eqnarray}
\label{pdr2}
B_n&=&\frac{\Gamma(n+1)}{\Gamma(n+1/2)(n+1)^{1/2}}, \qquad n>{1\over 2},\\
B_n&=&\frac{\Gamma(1/2-n)}{\Gamma(-n)[-(n+1)]^{1/2}}, \qquad n<{-1}.
\end{eqnarray}
The polytropic distribution has a single maximum at $v=0$. Therefore,
the condition of marginal stability  (\ref{cms2b}) implies $\omega_{r}=0$. From
Eq. (\ref{cms1b}), we find that the polytropic distribution is marginally stable
for $k=k_c$ where we have introduced the critical wavenumber
\begin{equation}
\label{pdr3}
k_{c}^2=\frac{4\pi G\rho}{\gamma T}.
\end{equation}
According to the criterion (\ref{sh2}), the polytropic distribution is
linearly dynamically stable if $k>k_{c}$ and linearly dynamically
unstable if $k<k_{c}$. The critical Jeans wavenumber (\ref{pdr3}) for
a stellar polytrope is the same as the critical Jeans wavenumber
(\ref{ip13}) for a polytropic gas. This is to be expected on account
of the {general} result of Sec. \ref{sec_fo}. It should be stressed
that the quantity that appears in the critical wavenumber (\ref{pdr3})
is the velocity of sound $c_s^2=\gamma T$ in the corresponding
barotropic gas, not the velocity dispersion $T$. They coincide for the
Maxwellian distribution ($\gamma=1$), but this is not general.

The dielectric function (\ref{lsv4}) associated to the polytropic distribution  is
\begin{eqnarray}
\label{pdr4}
\epsilon(k,\omega)=1- \frac{4\pi G}{k^2}\frac{\rho}{\sqrt{2 \pi T}} B_n \left (n-\frac{1}{2}\right )\frac{1}{(n+1)T} \nonumber\\
\times \int_{C} \frac{v\left [1-\frac{v^2}{2(n+1)T} \right ]_+^{n-3/2}}{v-\frac{\omega}{k}}\, dv.
\end{eqnarray}
Introducing the critical wavenumber (\ref{pdr3}), we obtain
\begin{eqnarray}
\epsilon({k},\omega)=1-{k_{c}^{2}\over k^{2}}W_{n}\biggl ({\omega\over k\sqrt{T}}\biggr ),
 \label{pdr5}
\end{eqnarray}
where
\begin{eqnarray}
\label{pdr6}
W_{n}(z)= \frac{1}{\sqrt{2\pi}}\frac{B_n}{n}\left (n-\frac{1}{2}\right )\int_{C}\frac{x[1-\frac{x^2}{2(n+1)}]_+^{n-3/2}}{x-z}dx,\nonumber\\
\end{eqnarray}
is a generalization of the $W$-function of plasma physics. We note that $W_{n}(0)=1$.  For $n\rightarrow +\infty$, we recover the
$W$-function (\ref{md5}). Equation (\ref{pdr6}) is therefore a
generalization of this function to the case of polytropic
distributions.

\subsection{Growth rate and damping rate}
\label{sec_pgd}

We look for particular solutions of the dispersion relation $\epsilon({
k},\omega)=0$ in the form $\omega=i\omega_i$ where $\omega_i$ is
real. First, we note that
\begin{eqnarray}
\epsilon({k},i\omega_i)=1-{k_{c}^{2}\over k^{2}}H_{n}\biggl (\frac{\omega_i}{k\sqrt{T}}\biggr ),
\label{pgd1}
\end{eqnarray}
where we have introduced the function $H_{n}(x)\equiv W_{n}(ix)$. For $x>0$, we have
\begin{eqnarray}
\label{pgd2}
H_{n}(x)= \frac{1}{\sqrt{2\pi}}\frac{B_n}{n}\left (n-\frac{1}{2}\right )\int_{-\infty}^{+\infty}\frac{t^2[1-\frac{t^2}{2(n+1)}]_+^{n-3/2}}{t^2+x^2}dt,\nonumber\\
\end{eqnarray}
and for $x<0$, we have
\begin{eqnarray}
\label{pgd3}
H_{n}(x)= \frac{1}{\sqrt{2\pi}}\frac{B_n}{n}\left (n-\frac{1}{2}\right )\biggl\lbrace \int_{-\infty}^{+\infty}\frac{t^2[1-\frac{t^2}{2(n+1)}]_+^{n-3/2}}{t^2+x^2}dt\nonumber\\
-2\pi x \left\lbrack1+\frac{x^2}{2(n+1)}\right\rbrack^{n-3/2}\biggr\rbrace.\qquad\qquad\qquad
\end{eqnarray}
Using $\epsilon(k,i\omega_i)=0$, the relation between $\omega_i$ and $k$ (for
fixed $T$) can be written
\begin{eqnarray}
1-{k_{c}^{2}\over k^{2}}H_{n}\biggl ({\omega_i\over k\sqrt{T}}\biggr )=0.
 \label{pgd4}
\end{eqnarray}
For $n\rightarrow +\infty$, we recover Eq. (\ref{mg3}). If we set $x=\sqrt{\beta}\omega_i/k$, we
can rewrite Eq. (\ref{pgd4}) in the parametric form
\begin{eqnarray}
\frac{\omega_i}{\sqrt{4\pi G\rho}}=x\sqrt{\frac{H_{n}(x)}{\gamma}}, \qquad \frac{k^2}{k_{c}^2}={H_{n}(x)}.
\label{pgd6}
\end{eqnarray}
By varying $x$ between $-\infty$ and $+\infty$, we obtain the full curve giving $\omega_i$ as a function of the wavenumber $k$. Since the time dependence of the  perturbation is $\delta f\sim e^{\omega_i
t}$, the case of neutral stability $\omega_i=0$ corresponds to $k=k_{c}$, the
case of instability $\omega_i>0$ corresponds to $k<k_{c}$ and the case of stability $\omega_i<0$ corresponds to $k>k_{c}$.  The discussion of the different regimes is similar to the one given in Sec. \ref{sec_mg}. For $k\rightarrow k_c$, using the general approximate formula (\ref{mon15}), the pulsation $\omega_i$ is given by
\begin{eqnarray}
\frac{\omega_i}{\sqrt{4\pi G\rho}}=\sqrt{\frac{2}{\pi}}\left (\frac{n}{1+n}\right )^{1/2}\frac{n}{B_n}\frac{1}{n-\frac{1}{2}}\left (1-\frac{k^2}{k_c^2}\right ).
\label{kkn}
\end{eqnarray}

\begin{figure}[htbp]
\centerline{
\includegraphics[width=8cm,angle=0]{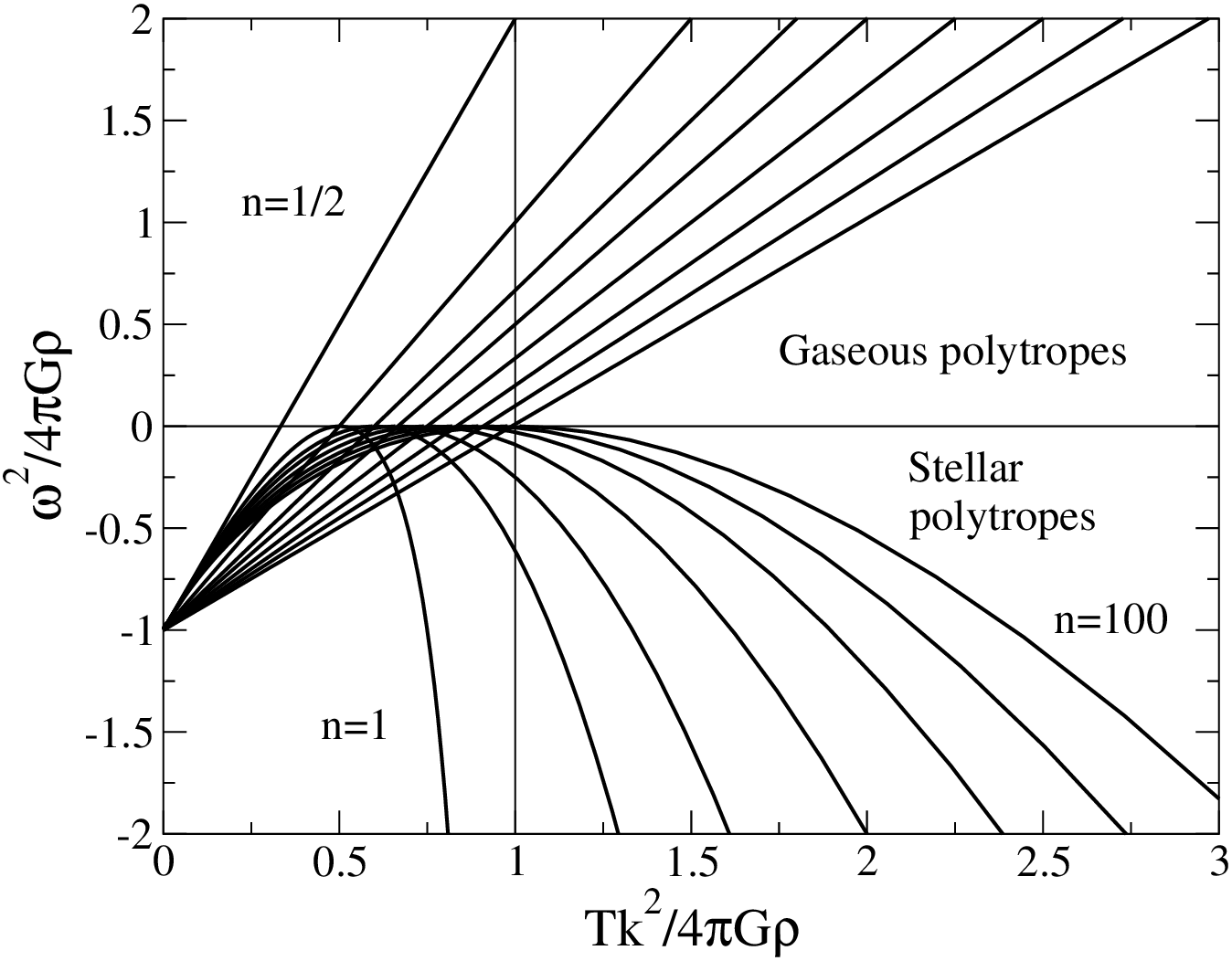}
} \caption[]{Pulsation $\omega$ as a function of the wavenumber $k$
for a stellar polytrope with $n\ge 1/2$ (we have represented $n=0.5,1,1.5,2,3,5,10,100$).
For $k<k_c$, the system is unstable
and $\omega=i\omega_i$ with $\omega_i>0$. For $k>k_c$, the system is
stable. There exists many branches of solutions
$\omega=\omega_r+i\omega_i$ with $\omega_i<0$ but we
have only represented the branch corresponding to
$\omega_r=0$.  We have also compared these results to the case of a polytropic gas. For $k<k_c$, the system is unstable
and $\omega=\pm i\omega_i$. For $k>k_c$, the system is
stable and $\omega=\pm\omega_r$. The instability occurs for $k<k_{c}^{(poly)}=k_{c}^{(iso)}/\sqrt{\gamma}$ where $\gamma=1+1/n$. For $n\ge 1/2$, $k_{c}^{(poly)}\le k_{c}^{(iso)}$. }
\label{tsallispos}
\end{figure}

\begin{figure}[htbp]
\centerline{
\includegraphics[width=8cm,angle=0]{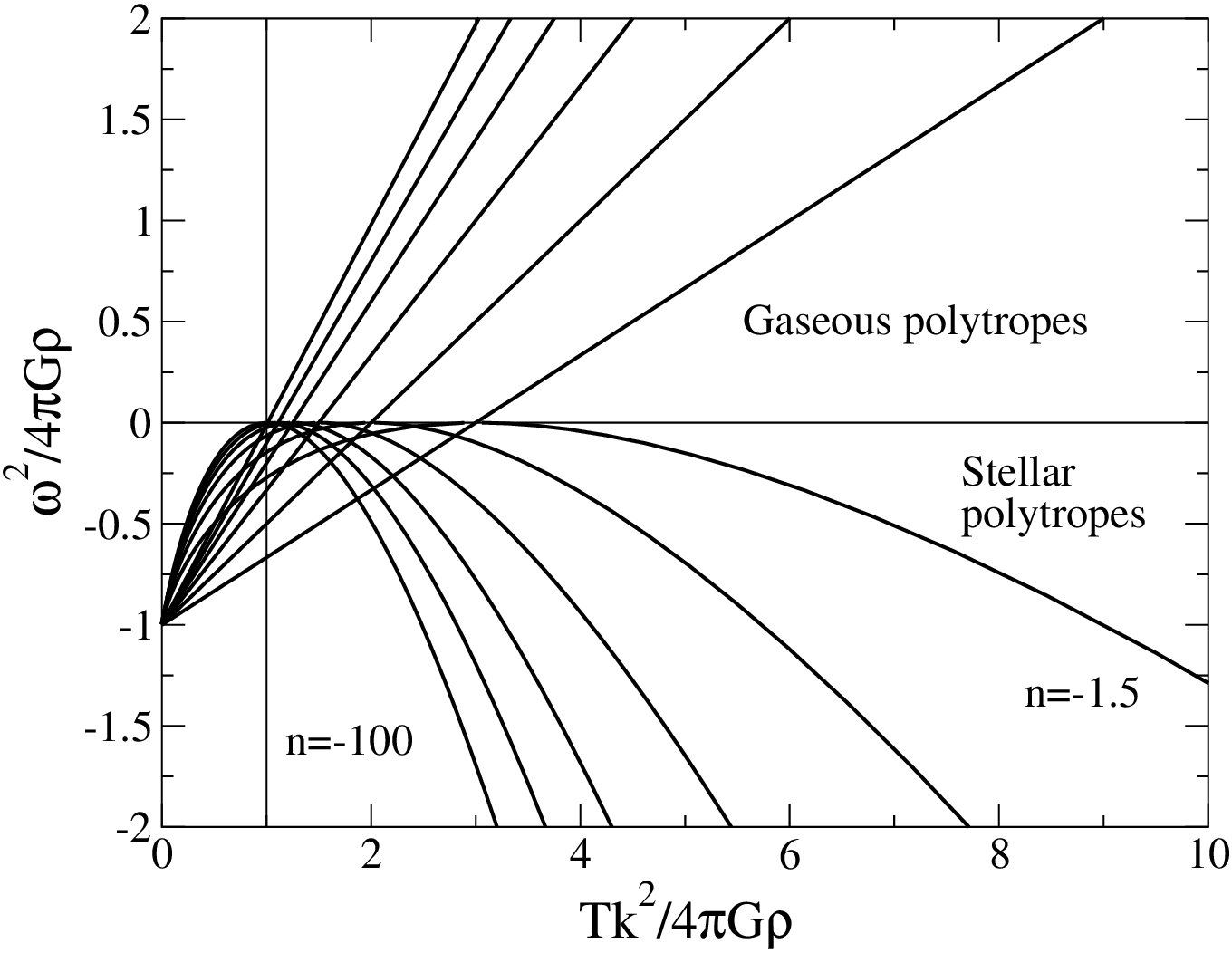}
} \caption[]{Pulsation $\omega$ as a function of the wavenumber $k$
for a stellar polytrope and a polytropic gas with $n\le -1$ (we have represented $n=-1.5,-2,-3,-5,-10,-100$). The instability occurs for $k<k_{c}^{(poly)}=k_{c}^{(iso)}/\sqrt{\gamma}$ where $\gamma=1+1/n$. For $n\le -1$, $k_{c}^{(poly)}\ge k_{c}^{(iso)}$.}
\label{tsallisneg}
\end{figure}

The pulsation of the perturbation in an infinite homogeneous stellar polytrope is plotted in Figs. \ref{tsallispos} and \ref{tsallisneg} as a function of the wavenumber $k$. For comparison, we have also indicated the pulsation of the perturbation in an infinite homogeneous polytropic  gas. For $n=\infty$, we recover the isothermal case shown in Fig. \ref{pulsation}.

\subsection{Particular cases}
\label{sec_fermi}

The case $n=3/2$ deserves a particular attention. In that case, the distribution function $f({\bf r},{\bf v})$ is a step function so that $f=\eta_{0}$ for $v\le v_{m}({\bf r})\equiv \sqrt{2(\epsilon_{m}-\Phi({\bf r}))}$ and $f=0$ otherwise. This corresponds to the distribution function of the self-gravitating Fermi gas at $T=0$ which describes classical white dwarf stars (Chandrasekhar 1942). The density is $\rho={4\pi\over 3}\eta_{0}v_{m}^{3}$ and the pressure $p={4\pi\over 15}\eta_{0}v_{m}^{5}$. Eliminating $v_{m}$ from these two relations, we get a polytropic equation of state $p=K\rho^{5/3}$ with $n=3/2$, $\gamma=5/3$ and $K={1\over 5}({3\over 4\pi\eta_{0}})^{2/3}$. The kinetic temperature is $T=\frac{1}{5}v_m^2$ and the velocity of sound is $c_{s}^{2}={1\over 3}v_{m}^{2}$. For an infinite and homogeneous medium, the reduced distribution function (\ref{pdr1}) corresponding to $n=3/2$  is a parabola
\begin{eqnarray}
f(v)=\frac{3\rho}{4\sqrt{5T}}\left (1-\frac{v^2}{5T}\right ).
 \label{par}
\end{eqnarray}
The critical Jeans wavenumber (\ref{sh3})  is $k_{c}^{2}=12\pi G\rho/5T=12\pi G\rho/v_{m}^{2}$ which is fully consistent with the expression (\ref{pdr3}) with $\gamma=1+1/n=5/3$. For $n=3/2$, we can obtain an explicit expression of the function (\ref{pgd2})-(\ref{pgd3}). For $x>0$, we have
\begin{eqnarray}
H_{3/2}(x)=1-\frac{x}{\sqrt{5}}\arctan\left (\frac{\sqrt{5}}{x}\right ),
 \label{fermi2}
\end{eqnarray}
and for $x<0$, we have
\begin{eqnarray}
H_{3/2}(x)=1-\frac{x}{\sqrt{5}}\left\lbrack \arctan\left (\frac{\sqrt{5}}{x}\right )+\pi\right\rbrack.
 \label{fermi3}
\end{eqnarray}

The index $n=1/2$ is also special and corresponds to the water-bag model. In that case, the reduced distribution  $f(v)$ is a step function so that $f=\eta_{0}$ for $|v|\le v_{m}$ and $f=0$ otherwise.  The amplitude $\eta_0$ is determined by the density according to the relation $\rho=2\eta_{0}v_{m}$. The kinetic temperature is $T=\langle v^2\rangle=\frac{1}{3}v_m^2$.
The derivative of the distribution function is $f'(v)=\eta_0\lbrack\delta(v+v_m)-\delta(v-v_m)\rbrack$.
The critical Jeans wavenumber (\ref{sh3}) is $k_{c}^{2}=4\pi G\rho/v_{0}^{2}=4\pi G\rho/3T$ which is fully consistent with the expression (\ref{pdr3}) with $\gamma=1+1/n=3$. For $n=1/2$, we can obtain an explicit expression of the dielectric function (\ref{pdr4}). We get
\begin{eqnarray}
\epsilon(k,\omega)=1-\frac{k_c^2}{k^2}W_{1/2}\left (\frac{\omega}{k\sqrt{T}}\right ),
 \label{diela}
\end{eqnarray}
with
\begin{eqnarray}
W_{1/2}=\frac{1}{1-\frac{1}{3}z^2}.
 \label{dielb}
\end{eqnarray}
The condition $\epsilon(k,\omega)=0$ determines the pulsation. For $k>k_c$, the system is stable and the perturbation presents  pure oscillations with pulsation
\begin{eqnarray}
\omega=\pm\sqrt{3T}(k^2-k_c^2)^{1/2}.
 \label{dielc}
\end{eqnarray}
For $k<k_c$, the system is unstable and the perturbation has a growth rate (and a decay rate) given by
\begin{eqnarray}
\omega=\pm i\sqrt{3T}(k_c^2-k^2)^{1/2}.
 \label{dielcb}
\end{eqnarray}
We note that, for the water-bag distribution, the general asymptotic behavior (\ref{mon7q}) becomes exact for all $k\le k_c$. We also note that for the specific index $n=1/2$ ($\gamma=3$), the dispersion relation in a stellar system takes the same form as in a gas (see Sec. \ref{sec_gas}).

\subsection{The Nyquist curve}
\label{sec_polyn}

Introducing the dimensionless wavenumber and dimensionless pulsation (\ref{mn1}), the
dielectric function (\ref{pdr5}) can be rewritten
\begin{eqnarray}
\epsilon({\eta},\Omega)=1-\frac{1}{\gamma}\eta W_{n}(\sqrt{\eta}\Omega).
 \label{pn1}
\end{eqnarray}
When $\Omega_{i}=0$, the real and imaginary parts of the dielectric function $\epsilon(\eta,\Omega_{r})=\epsilon_{r}(\eta,\Omega_{r})+i\epsilon_{i}(\eta,\Omega_{r})$ are given by
\begin{equation}
\label{pn2}
\epsilon_{r}(\eta,\Omega_{r})=1-\frac{1}{\gamma}\eta W_{r}^{(n)}\left (\sqrt{\eta}\Omega_r\right ),
\end{equation}
\begin{equation}
\label{pn3}
\epsilon_{i}(\eta,\Omega_{r})=-\frac{1}{\gamma}\eta W_{i}^{(n)}\left (\sqrt{\eta}\Omega_r\right ),
\end{equation}
with
\begin{eqnarray}
\label{pn4}
W_{r}^{(n)}(z)=\frac{1}{\sqrt{2\pi}}\frac{B_n}{n}\left (n-\frac{1}{2}\right )\nonumber\\
\times P\int_{-\infty}^{+\infty}\frac{x[1-\frac{x^2}{2(n+1)}]_{+}^{n-3/2}}{x-z}dx,
\end{eqnarray}
\begin{eqnarray}
\label{pn5}
W_{i}^{(n)}(z)=\sqrt{\frac{\pi}{2}}\frac{B_n}{n}\left (n-\frac{1}{2}\right )z\left [1-\frac{z^2}{2(n+1)}\right ]_+^{n-3/2},\nonumber\\
\end{eqnarray}
where $z$ is here a real number. The condition of marginal stability
corresponds to
$\epsilon_{r}(\eta,\Omega_{r})=\epsilon_{i}(\eta,\Omega_{r})=0$. The condition
$\epsilon_{i}(\eta,\Omega_{r})=0$, which is equivalent to
$f'(\sqrt{\eta}\Omega_{r})=0$, implies $\Omega_r=0$. Then, the relation $\epsilon_{r}(\eta,\Omega_{r})=0$ leads to $\eta=\eta_{c}$ with
\begin{equation}
\label{pn6}
\eta_c={\gamma}.
\end{equation}
To apply the Nyquist method, we need to plot the curve
$(\epsilon_{r}(\eta,\Omega_{r}),\epsilon_{i}(\eta,\Omega_{r}))$  in the $\epsilon$-plane.
We have to distinguish different cases according to the value of the index $n$. A general discussion
has been given by Chavanis \& Delfini (2009) in the context of the HMF model. This discussion can be immediately transposed to the present context.

\section{The symmetric double-humped distribution}
\label{sec_vhq}

\subsection{Determination of the extrema}
\label{sec_ext}

We consider a reduced distribution function  (\ref{lsv3}) of the form
\begin{equation}
\label{e1}
f(v)=\sqrt{\frac{\beta}{2\pi}} \frac{\rho}{2} \left [  e^{-\frac{\beta}{2}(v-v_a)^2} + e^{-\frac{\beta}{2}(v+v_a)^2}  \right ].
\end{equation}
This is a symmetric double-humped distribution corresponding to the
superposition of two Maxwellian distributions with temperature
$T=1/\beta$ centered in $v_a$ and $-v_a$ respectively (see
Fig. \ref{dbumpdist}). This distribution models two streams of
particles in opposite direction. The average velocity is $\langle
v\rangle=0$ and the kinetic temperature $T_{kin}\equiv \langle
v^2\rangle=T+v_{a}^{2}$. The velocitie(s) $v_0$ at which the
distribution function $f(v)$ is extremum satisfy $f'(v_0)=0$. They are
determined by the equation
\begin{equation}
\label{e2}
e^{-2\beta v_{a} v_0}=\frac{v_a-v_0}{v_a+v_0}.
\end{equation}
We note that $v_0\in \lbrack -v_a, +v_a\rbrack$. Introducing the dimensionless velocity and the dimensionless separation
\begin{equation}
\label{e3}
V=\sqrt{\beta}v,\qquad V_a=\sqrt{\beta}v_a,
\end{equation}
Eq. (\ref{e2}) can be rewritten
\begin{equation}
\label{e4}
1=\frac{1}{2V_a V_0}\ln\left (\frac{V_a+V_0}{V_a-V_0}\right ).
\end{equation}
It is convenient to introduce the variables
\begin{equation}
\label{e5}
x=\frac{V_0}{V_a}, \qquad  y=V_a^2.
\end{equation}
For a fixed temperature $T$, $x$ plays the role of the velocity $v_0$ at which the distribution is extremum and $y$ plays the role of separation $v_a$. Then, we have to study the function
\begin{equation}
\label{e6}
y(x)=\frac{1}{2x}\ln\left (\frac{1+x}{1-x}\right ),
\end{equation}
for $x \in ] -1, +1[$. This function is plotted in Fig. \ref{ysym}. It has the
following properties:
\begin{equation}
\label{e7}
y(-x)=y(x),
\end{equation}
\begin{equation}
\label{e8}
y(x)\sim -\frac{1}{2}\ln (1-x), \qquad (x\rightarrow 1^{-}),
\end{equation}
\begin{equation}
\label{e9}
y(0)=1.
\end{equation}
The extrema of the distribution function (\ref{e1}) can be deduced
from the study of this function. First, considering Eq. (\ref{e4}), we
note that $f(v)$ always has an extremum at $v_0=0$, for any value of
$\beta$ and $v_a$. This is a ``degenerate'' solution of Eq. (\ref{e6})
corresponding to the vertical line $x=0$ in Fig. \ref{ysym}. On the
other hand, if $y>1$, i.e $\beta v_a^2>1$, there exists two other extrema
$v_{0}=\pm v_{*}$ where $v_{*}=v_a x_*$ with $x_*=y^{-1}(\beta v_a^2)$.

In conclusion, for a given temperature $T$:

$\bullet$ if $y>1$ ($v_a^2>T$), the distribution function $f(v)$ has two maxima at $v_{0}=\pm v_{*}$  and one minimum at $v_{0}=0$.

$\bullet$ if $y\le 1$ ($v_a^2\le T$), the distribution function $f(v)$ has only one
maximum at $v_{0}=0$ (the limit case $\beta v_a^2=1$ corresponds to $f''(0)=0$).

\begin{figure}
\begin{center}
\includegraphics[clip,scale=0.3]{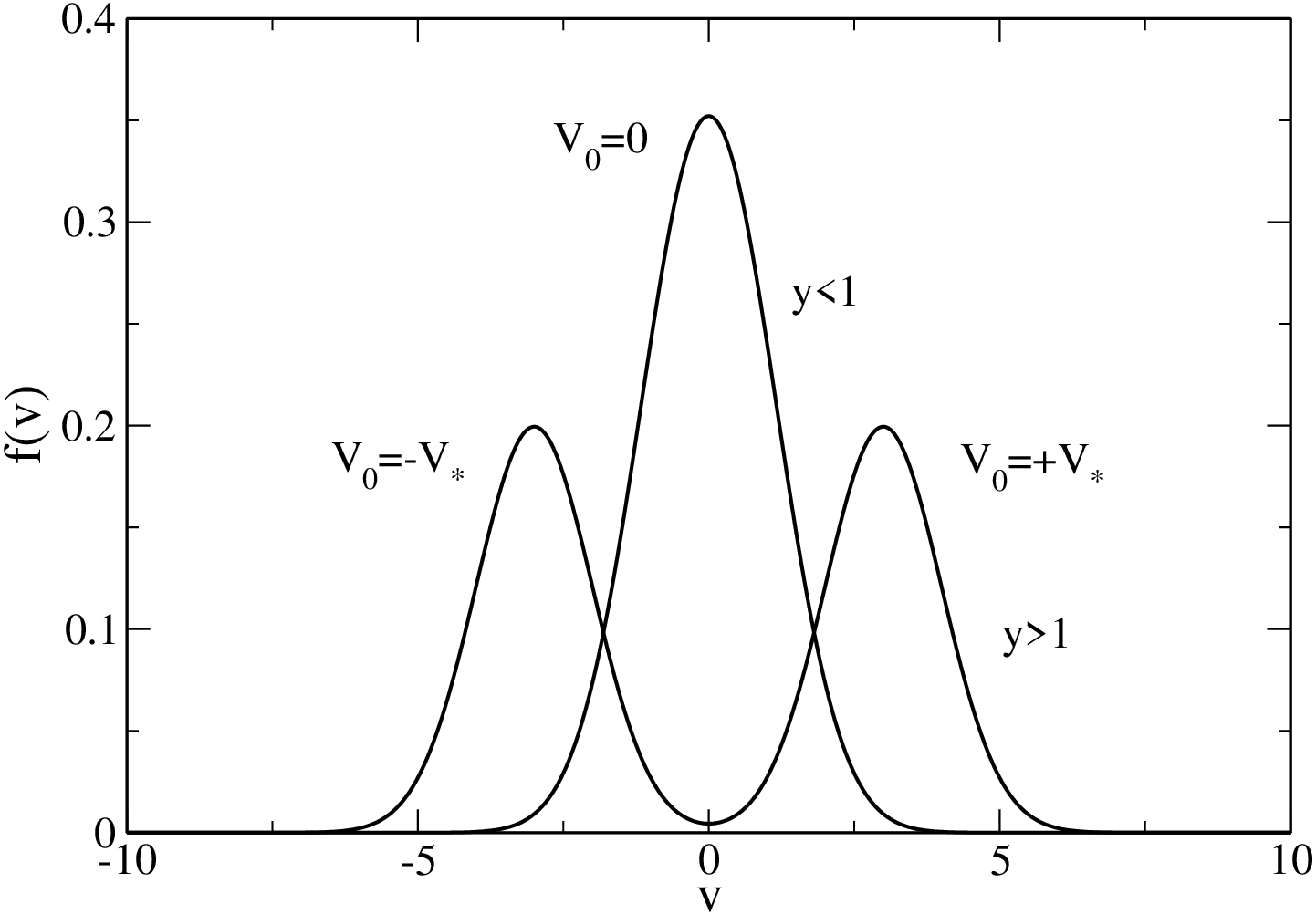}
\caption{Symmetric double-humped distribution made of two Maxwellians with separation $y$ (for a given temperature $T$). If $y>1$, the distribution has two maxima at $\pm V_{*}$ and one minimum at $V_{0}=0$ while for  $y<1$, it has only one maximum at $V_{0}=0$.}
\label{dbumpdist}
\end{center}
\end{figure}

\begin{figure}
\begin{center}
\includegraphics[clip,scale=0.3]{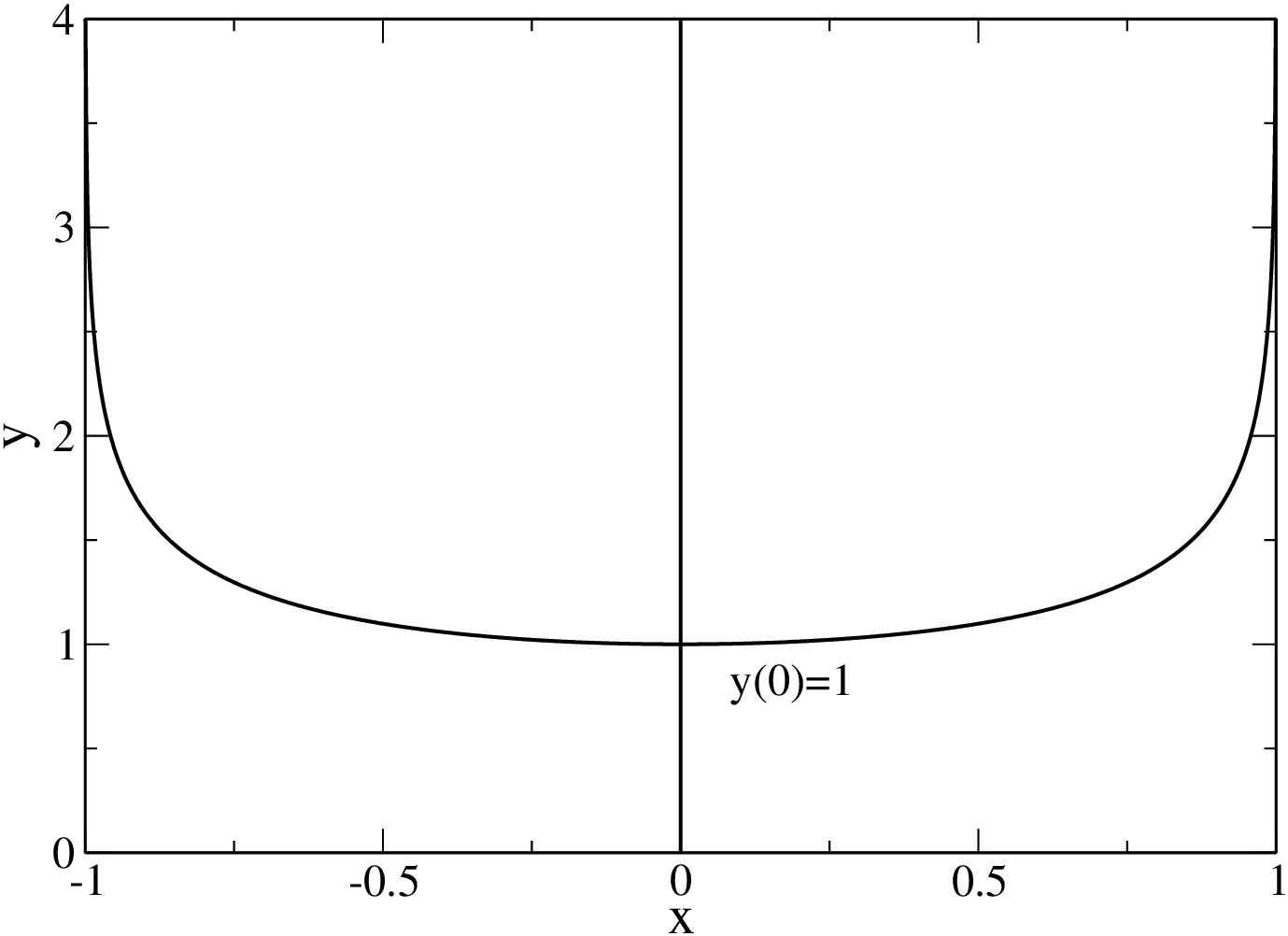}
\caption{The function $y(x)$ for the symmetric double-humped distribution.}
\label{ysym}
\end{center}
\end{figure}

\subsection{The condition of marginal stability}
\label{sec_sms}

Introducing the dimensionless variables (\ref{mn1}) and (\ref{e5}), the dielectric function associated to the symmetric double-humped distribution (\ref{e1}) is
\begin{eqnarray}
\label{sms1}
\epsilon(\eta,\Omega)= 1-\frac{\eta}{2} \left [ W(\sqrt{\eta}\Omega-\sqrt{y}))+W(\sqrt{\eta}\Omega+\sqrt{y}))  \right ],\nonumber\\
\end{eqnarray}
where $W(z)$ is defined in Eq. (\ref{md5}).  For a fixed temperature, $\eta$ plays the role of the wavenumber $k$, $\Omega$ plays the role of the pulsation $\omega$ and $y$ plays the role of the separation $v_a$. When $\Omega_{i}=0$, the real and imaginary parts of the dielectric function  $\epsilon(\eta,\Omega_{r})=\epsilon_{r}(\eta,\Omega_{r})+i\epsilon_{i}(\eta,\Omega_{r})$ can be written
\begin{eqnarray}
\label{sms2}
\epsilon_{r}(\eta,\Omega_{r})=1-\frac{\eta}{2} \left [ W_r(\sqrt{\eta}\Omega_{r} - \sqrt{y}))+W_r(\sqrt{\eta}\Omega_{r} + \sqrt{y}))   \right ],\nonumber\\
\end{eqnarray}
\begin{eqnarray}
\epsilon_{i}(\eta,\Omega_{r})=-\frac{\eta}{2} \left [ W_i(\sqrt{\eta}\Omega_{r} - \sqrt{y}))+W_i(\sqrt{\eta}\Omega_{r} + \sqrt{y}))   \right ],\nonumber\\
\end{eqnarray}
where $W_r(z)$ and $W_i(z)$ are defined in
Eqs. (\ref{mn6})-(\ref{mn7}) where $z$ is here a real number. The
condition of marginal stability corresponds to
$\epsilon_{r}(\eta,\Omega_{r})=\epsilon_{i}(\eta,\Omega_{r})=0$. The condition
$\epsilon_{i}(\eta,\Omega_{r})=0$ is equivalent to
\begin{equation}
\label{sms3}
f'(\sqrt{\eta}\Omega_{r})=0.
\end{equation}
The condition $\epsilon_{r}(\eta,\Omega_{r})=0$ leads to
\begin{equation}
\label{sms4}
1-\frac{\eta}{2} \left [ W_r(\sqrt{\eta}\Omega_{r} - \sqrt{y}))+W_r(\sqrt{\eta}\Omega_{r} + \sqrt{y}))   \right ]=0.\nonumber\\
\end{equation}
Therefore, according to Eq. (\ref{sms3}), the phase velocity
$\sqrt{\eta}\Omega_{r}$ is equal to a velocity $V_{0}$ at which the distribution
(\ref{e1}) is extremum. The second equation (\ref{sms4}) determines
the value(s) $\eta_{c}(y)$ of the wavenumber at which the
distribution is marginally stable.

\subsubsection{The case $\omega_{r}=0$}

Let us first consider the value $\Omega_{r}=0$ that is solution of
Eq. (\ref{sms3}) for any $v_a$ and $\beta$. In that case, Eq. (\ref{sms4}) becomes
\begin{equation}
\label{sms7}
\eta_{c}^{(0)}(y)=\frac{1}{W_{r}(\sqrt{y})},
\end{equation}
where we have used $W_r(-x)=W_r(x)$. For given separation $y$,
this equation determines the wavenumber $\eta_{c}^{(0)}(y)$
corresponding to a mode of marginal stability with
$\Omega_{r}=0$. The function defined by Eq. (\ref{sms7}) is plotted in
Fig. \ref{w3}. It  diverges
at $y=y_{max}=z_{c}^{2}$ where
$z_{c}=1.307$ is the zero of $W_{r}(z)$ (see Appendix A of Chavanis \& Delfini 2009). Then, using $W_r'(z_c)=-1/z_c$, we
find from Eq. (\ref{sms7}) that
\begin{equation}
\label{sms8}
\eta_{c}^{(0)}(y)\sim \frac{2y_{max}}{y_{max}-y}, \qquad (y\rightarrow y_{max}).
\end{equation}
For $y>y_{max}=1.708$, $W_{r}(\sqrt{y})$ is negative so
the branch $\eta_c^{(0)}(y)$ exists only for $y\in[0,y_{max}]$.  For
$y\rightarrow 0$, we have
\begin{equation}
\label{sms9}
\eta_{c}^{(0)}(0)=1.
\end{equation}
This result is to be expected since, for $v_a=0$, the distribution
(\ref{e1}) reduces to the Maxwellian. We thus recover the critical
wavenumber (\ref{mn8}).

In conclusion:

$\bullet$ if $y<y_{max}$, there exists a critical wavenumber
$\eta_{c}^{(0)}(y)$  determined by
Eq. (\ref{sms7}) corresponding to a marginal mode
$(\Omega_{r}=0,\Omega_{i}=0)$.

$\bullet$ if $y>y_{max}$, there is no marginal mode
$(\Omega_{r}=0,\Omega_{i}=0)$.

The curve $\eta_{c}^{(0)}(y)$ corresponding to the
marginal mode with zero pulsation $\Omega_r=0$ is plotted in Fig. \ref{GRAVdiagphasesym}.

\begin{figure}
\begin{center}
\includegraphics[clip,scale=0.3]{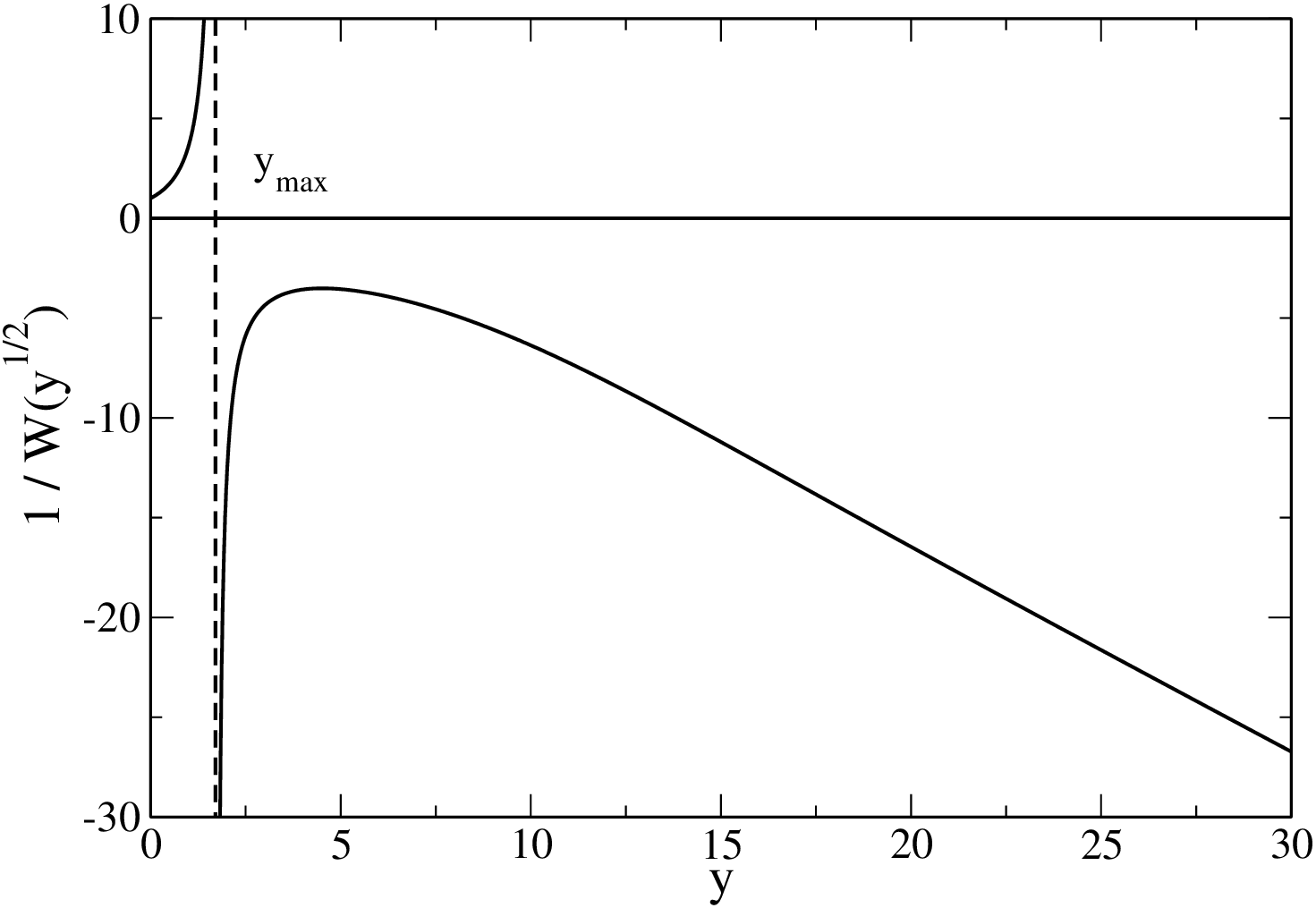}
\caption{$\eta=1/W_r(\sqrt{y})$  as a function of $y$.}
\label{w3}
\end{center}
\end{figure}

\subsubsection{The case $\omega_{r}\neq 0$}

We now consider the cases where $\sqrt{\eta}\Omega_{r}=\pm V_{*}$  is
solution of Eq. (\ref{sms3}) for $y>1$. To determine the
wavenumber(s) at which the distribution (\ref{e1}) is marginally stable, we
have to solve
\begin{equation}
\label{sms10}
1-\frac{\eta}{2} \left [ W_r(V_*- \sqrt{y}))+W_r(V_* + \sqrt{y}))   \right ]=0,\nonumber\\
\end{equation}
where $V_{*}$ is given by
\begin{equation}
\label{sms11}
1=\frac{1}{2V_* \sqrt{y}}\ln\left (\frac{\sqrt{y}+V_{*}}{\sqrt{y}-V_{*}}\right ).
\end{equation}
Eliminating $V_*$ between these two expressions, we obtain the critical wavenumber(s) $\eta_{c}^{(\pm)}(y)$ as a function of $y$. However, it is easier to proceed differently. Setting $x=V_*/V_a=V_*/\sqrt{y}$, we obtain the equations
\begin{equation}
\label{sms12}
y=\frac{1}{2x}\ln\left (\frac{1+x}{1-x}\right ),
\end{equation}
\begin{equation}
\label{sms13}
\eta=\frac{2}{\left [ W_{r}(\sqrt{y}(x - 1))+W_{r}(\sqrt{y}(x+ 1))   \right ]}.
\end{equation}
For given $x$, we can obtain $y$ from Eq. (\ref{sms12}) [see also
Fig. \ref{ysym}] and $\eta$ from Eq. (\ref{sms13}). Varying $x$ in the interval $]-1,1[$ yields the
full curve $\eta_{c}^{(\pm)}(y)$. By symmetry, we can restrict
ourselves to the interval $x\in [0,1[$.

For $x=0$, we have $y=1$ and
\begin{equation}
\label{sms18}
\eta_{c}^{(\pm)}(1)=\eta_*=\frac{1}{W_r(1)}=3.633.
\end{equation}
The branch $\eta_{c}^{(\pm)}(y)$ starts at the
point $(1,\eta_{*})$, corresponding to $\sqrt{\eta}\Omega_{r}=\pm V_{*}=0$
(i.e. $x=0$). This point is at the intersection between the branch $\eta_{c}^{(0)}(y)$ along which
$\Omega_{r}=0$ and the line $y=1$
separating the regions where the distribution has one or two maxima.

For $x\rightarrow 1$, we have $y\sim -\frac{1}{2}\ln(1-x)\rightarrow +\infty$ and
\begin{equation}
\label{sms19}
\eta_{c}^{(\pm)}(y)=2+\frac{1}{2y}+... \qquad (y\rightarrow +\infty).
\end{equation}
This result can be understood
simply. For $y\rightarrow +\infty$, the two humps are far away from
each other so that the critical wavenumber
$\eta_{c}^{(\pm)}$ coincides with the critical wavenumber of a
single Maxwellian since they do not ``see'' each other. Noting that
the mass of a single hump is $M/2$, the corresponding critical
wavenumber is $k_{c}=4\pi G(\rho/2)/T$ leading to
$\eta_{c}^{(\pm)}=2$.

In conclusion:

$\bullet$ if $y>1$, there exists a single critical wavenumber
$\eta_{c}^{(\pm)}(y)$ determined by Eqs. (\ref{sms12})-(\ref{sms13})
corresponding to a marginal mode $(\Omega_r=\pm V_*/\sqrt{\eta_c},\Omega_i=0)$. Note
that the modes $\Omega_r=+V_*/\sqrt{\eta_c}$ and $\Omega_r=-V_{*}/\sqrt{\eta_c}$ are
degenerate. This degeneracy can be raised by a small asymmetry
(symmetry breaking) in the distribution (see Sec. \ref{sec_vha}).

The curve $\eta_{c}^{(\pm)}(y)$  corresponding to the
marginal mode with pulsation $\Omega_r=\pm V_{*}/\sqrt{\eta_c}$ is plotted in Fig. \ref{GRAVdiagphasesym}.

\subsection{The stability diagram}
\label{sec_sdsy}

The critical wavenumbers $\eta_c(y)$ corresponding to marginal
stability determined previously are represented as a function of the
separation $y$ in Fig. \ref{GRAVdiagphasesym}. We have also plotted the line
$y=1$. On the left of this line, the distribution has a single maximum at $V_0=0$
and on the right of this line, the distribution has two maxima at $V_0=\pm V_*$ and a
minimum at $V_0=0$. In order to investigate the stability of the
solutions in the different regions, we have used the Nyquist method.

\begin{figure}[htbp]
\centerline{
\includegraphics[width=8cm,angle=0]{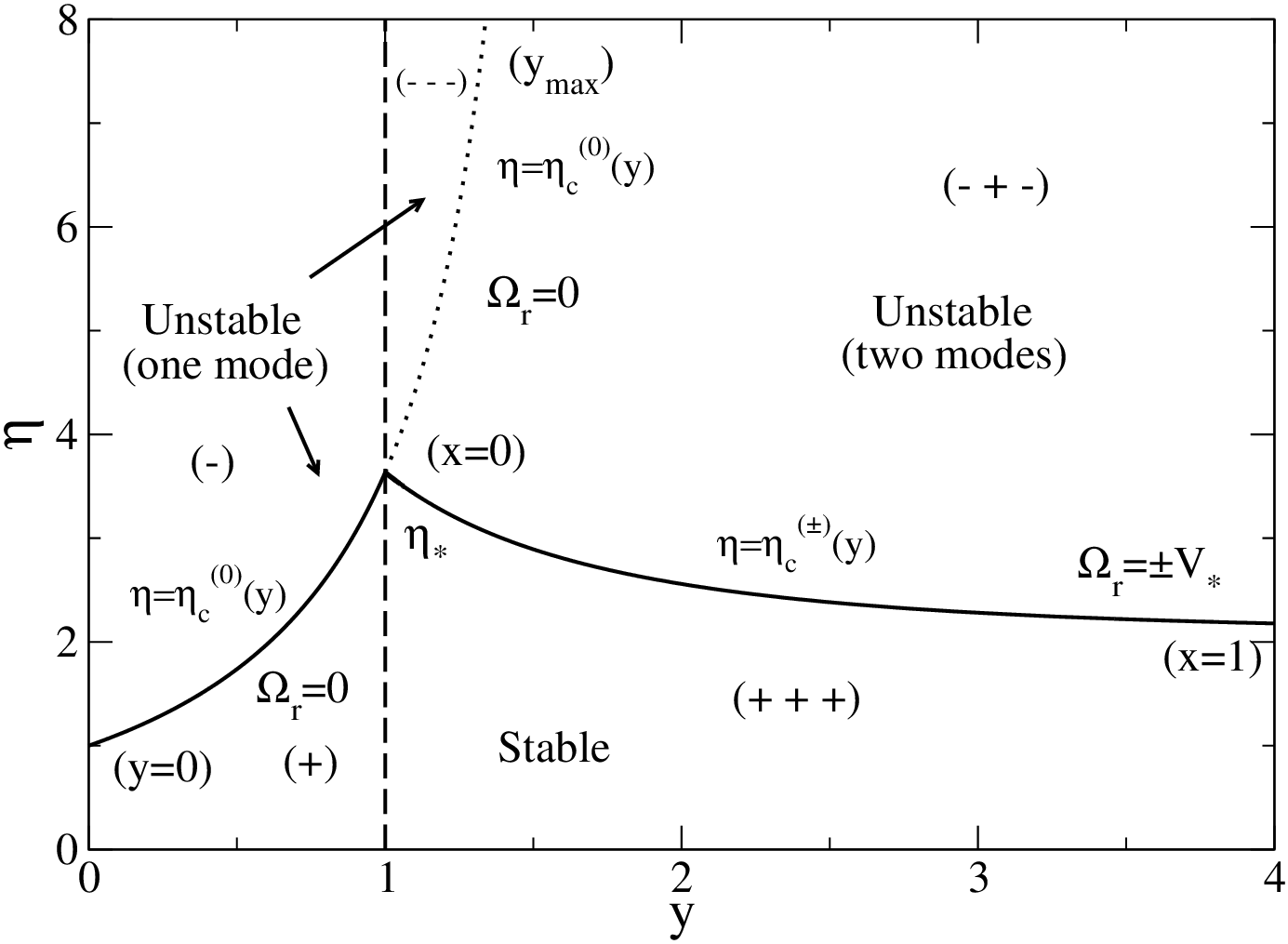}
} \caption[]{Stability diagram of the symmetric double-humped
distribution (\ref{e1}). The solid line $(\eta_c)_*$ corresponds to
the critical line: below this line the DF is stable and above this
line the DF is unstable. On the left panel (delimited by the dotted
line), there is one mode of instability and on the right panel there
are two modes of instability.  The dashed line corresponds to $y=1$:
on the left of this line the DF has one maximum and on the right of
this line the DF has two maxima and one minimum. The symbols in
parenthesis like $(x=0)$ give the values of $x$ or $y$ that
parametrize the marginal curves. The notations like $(- + -)$ give the
positions of the $\epsilon_{r}(v_{ext})$'s in the Nyquist curves as
defined in Sec. \ref{sec_doubh}.  }
\label{GRAVdiagphasesym}
\end{figure}

For $y<1$, there exists one wavenumber $\eta_{c}^{(0)}(y)$ at which
the DF is marginally stable. For $\eta=\eta_{c}^{(0)}(y)$, the DF has
one maximum at $V_0=0$. The marginal perturbation does not propagate
($\Omega_r=0$). By considering the Nyquist curves in this region (see
Figs. \ref{doublesym_y0.5_n0.5}-\ref{doublesym_y0.5_n6}), we find that
the DF is stable for $\eta<\eta_{c}^{(0)}(y)$ and unstable for
$\eta>\eta_{c}^{(0)}(y)$.

\begin{figure}[htbp]
\centerline{
\includegraphics[width=8cm,angle=0]{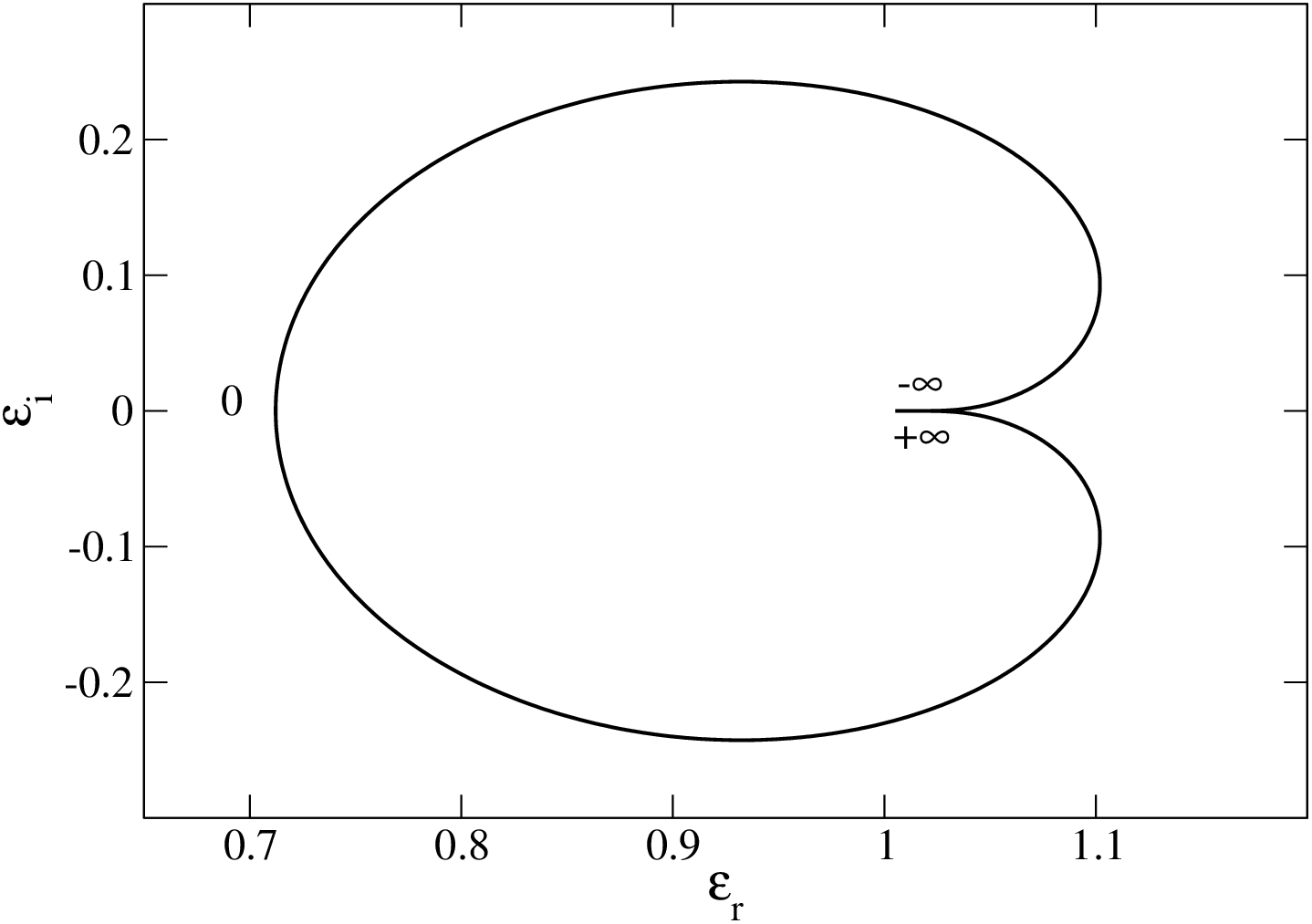}
} \caption[]{Nyquist curve for $y<1$ and $\eta<\eta_c^{(0)}$
(specifically $y=0.5$ and $\eta=0.5$). The DF has only one maximum at
$V_0=0$. It is stable (with respect to this perturbation) because the
Nyquist curve does not encircle the origin. Case (+).}
\label{doublesym_y0.5_n0.5}
\end{figure}

\begin{figure}[htbp]
\centerline{
\includegraphics[width=8cm,angle=0]{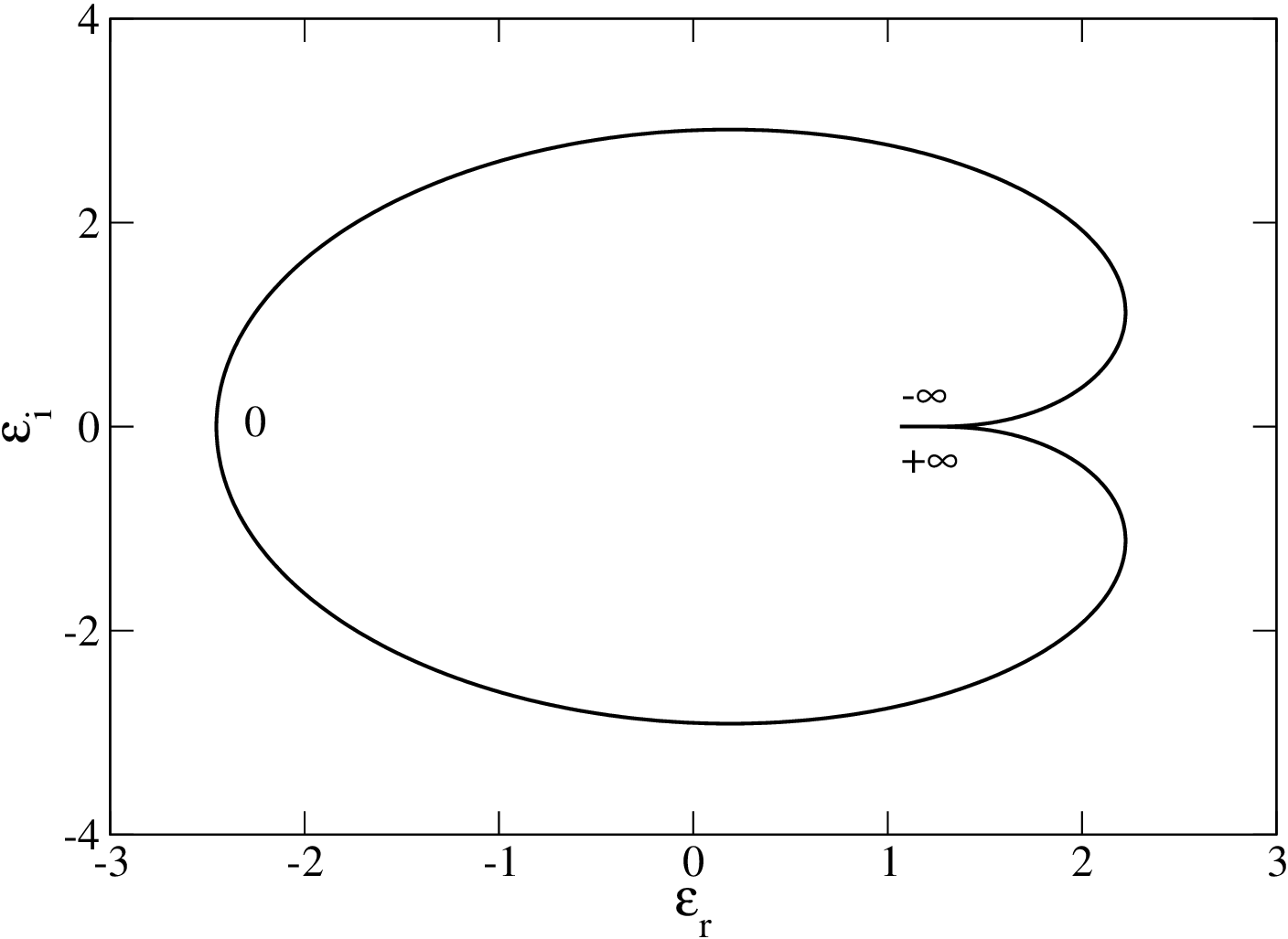}
} \caption[]{Nyquist curve for $y<1$ and $\eta>\eta_c^{(0)}$ (specifically $y=0.5$ and $\eta=6$).  The DF has only one maximum at
$V_0=0$. It is unstable (with respect to this perturbation) because the
Nyquist curve encircles the origin. Case (-).}
\label{doublesym_y0.5_n6}
\end{figure}

For $y>y_{max}$, there exists one wavenumber $\eta_{c}^{(\pm)}(y)$ at
which the DF is marginally stable. For $\eta=\eta_{c}^{(\pm)}(y)$, the
DF has two maxima at $V_0=\pm V_*$ and one minimum at $V_0=0$. The
marginal perturbation evolves with a pulsation $\Omega_r=\pm
V_*/\sqrt{\eta_c}$. By considering the Nyquist curves in this region
(see Figs. \ref{doublesym_y2n1}-\ref{doublesym_y2n6}), we find that
the DF is stable for $\eta<\eta_{c}^{(\pm)}(y)$ and unstable for
$\eta>\eta_{c}^{(\pm)}(y)$.

\begin{figure}[htbp]
\centerline{
\includegraphics[width=8cm,angle=0]{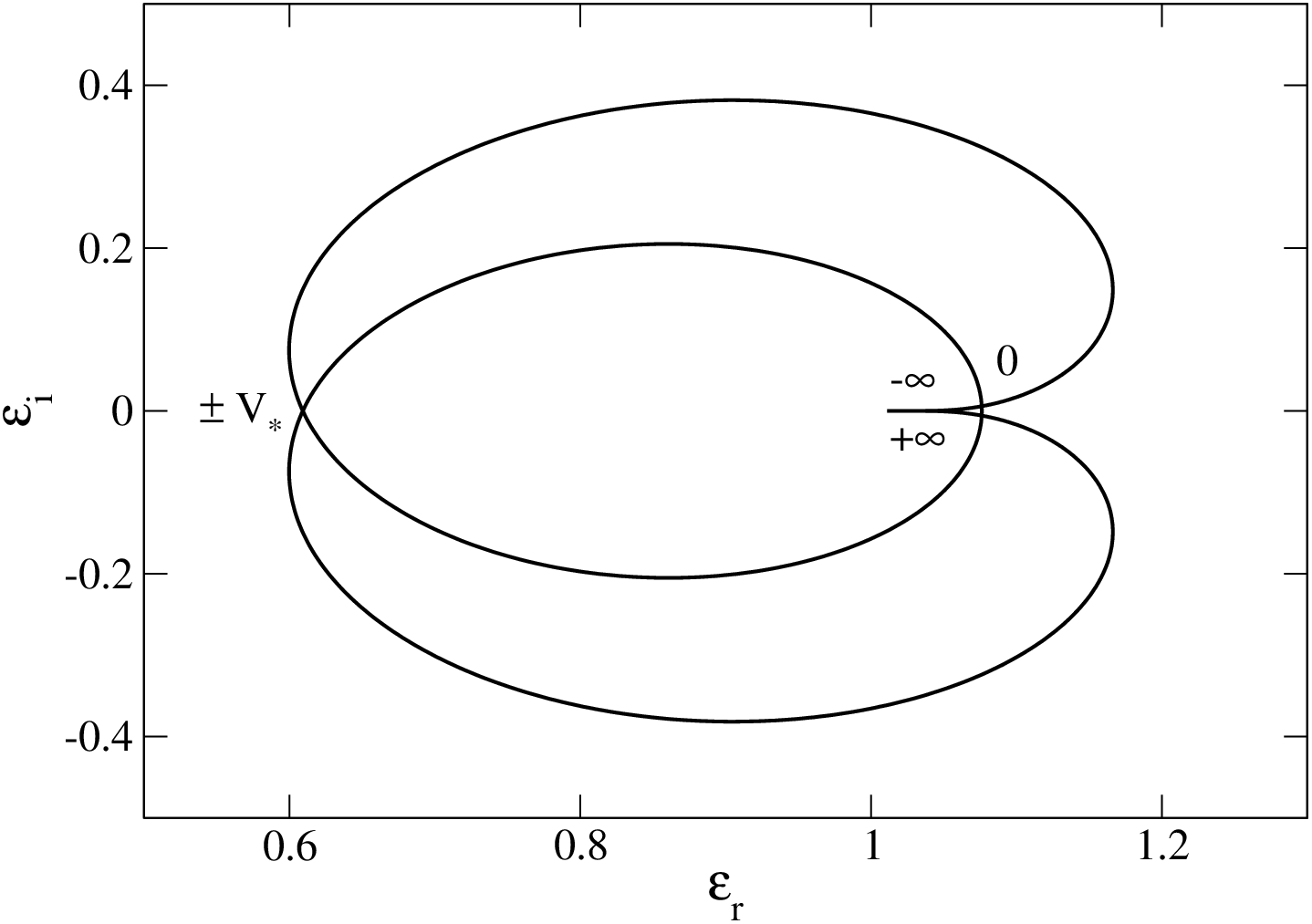}
} \caption[]{Nyquist curve for  $y>y_{max}$ and $\eta<\eta_c^{(\pm)}$ (specifically $y=2$ and $\eta=1$). The DF has two maxima at $V_0=\pm V_*$ and one minimum at $V_0=0$. The DF is stable (with respect to this perturbation) because the
Nyquist curve does not encircle the origin. Case (+ + +).}
\label{doublesym_y2n1}
\end{figure}

\begin{figure}[htbp]
\centerline{
\includegraphics[width=8cm,angle=0]{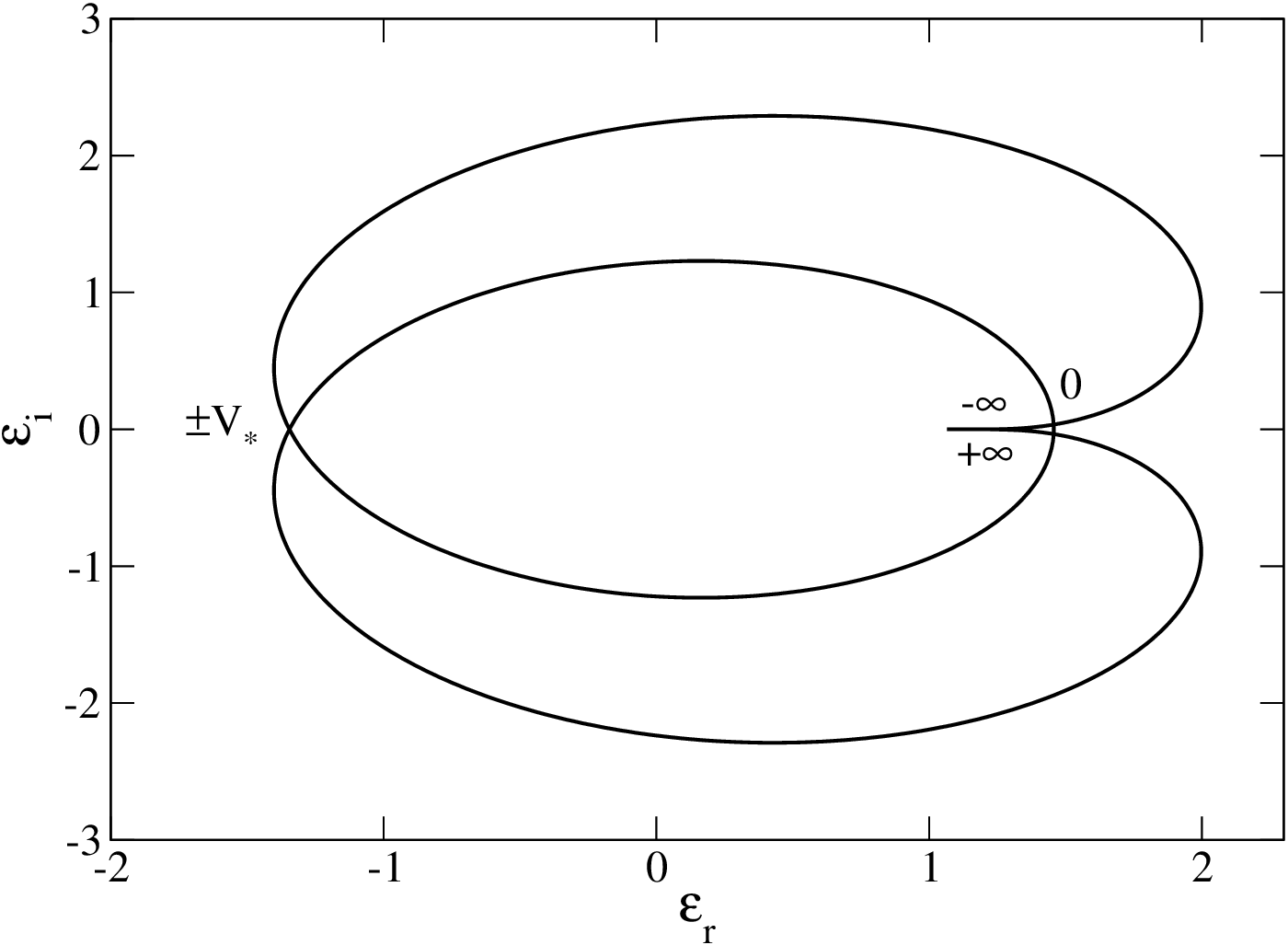}
} \caption[]{Nyquist curve for $y>y_{max}$ and $\eta>\eta_c^{(\pm)}$
(specifically $y=2$ and $\eta=6$). The DF has two maxima at $V_0=\pm
V_*$ and one minimum at $V_0=0$. The DF is unstable (with respect to
this perturbation) because the Nyquist curve encircles the
origin. Since it rotates twice around the origin, this implies that there are $N=2$ unstable modes $(\omega_r,\omega_i)$ with $\omega_i>0$. Case (- + -).}
\label{doublesym_y2n6}
\end{figure}

For $1<y<y_{max}$, there exists two wavenumbers $\eta_{c}^{(0)}(y)$
and $\eta_{c}^{(\pm)}(y)$ at which the DF is
marginally stable.  The DF has
two maxima at $V_0=\pm V_*$ and one minimum at $V_0=0$. For $\eta=\eta_{c}^{(0)}(y)$, the
marginal perturbation does not propagate ($\Omega_r=0$). For
$\eta=\eta_{c}^{(\pm)}(y)$, the marginal perturbation has a pulsation
$\Omega_r=\pm V_*/\sqrt{\eta_c}$.  By considering the Nyquist curves in this
region (see Fig. \ref{doublesym_y1.2_n6}), we find that the DF is stable
for $\eta<\eta_{c}^{(\pm)}(y)$ and unstable for $\eta>\eta_{c}^{(\pm)}(y)$.

\begin{figure}[htbp]
\centerline{
\includegraphics[width=8cm,angle=0]{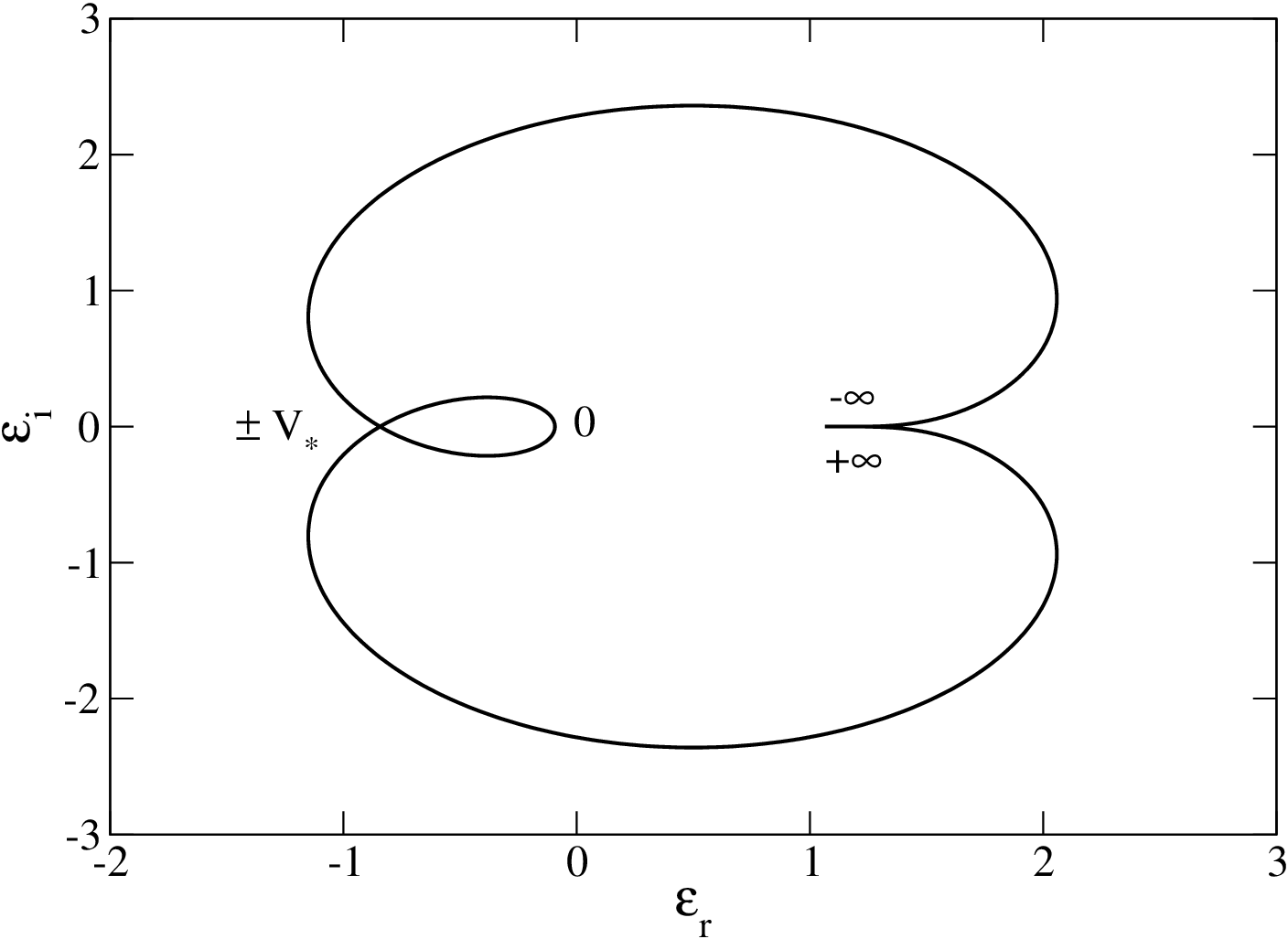}
} \caption[]{Nyquist curve for $1<y<y_{max}$ and $\eta>\eta_c^{(0)}$
(specifically $y=1.2$ and $\eta=6$). The DF has two maxima at $V_0=\pm
V_*$ and one minimum at $V_0=0$. The DF is unstable (with respect to
this perturbation) because the Nyquist curve encircles the
origin. Since it rotates only once around the origin, this implies that there is $N=1$ unstable mode $(\omega_r,\omega_i)$ with $\omega_i>0$. Case (- - -).}
\label{doublesym_y1.2_n6}
\end{figure}

A few comments are in order:

1. In Fig. \ref{GRAVdiagphasesym}, we explicitly see that the system is always unstable with respect to perturbations with sufficiently small $k$ (the case of cold systems  $T=0$ is treated in Appendix \ref{sec_cold}). This corroborates the general result given  at the end of Sec. \ref{sec_n}. More precisely, we see that the system is stable for perturbations with  $k>(k_c)_*$ (corresponding to the solid line) and unstable for perturbations with $k<(k_c)_*$. Furthermore, we note that this critical wavenumber $(k_c)_*$ corresponds to a marginal  perturbation where the phase velocity $\omega/k$ coincides with the {\it maximum} of the velocity distribution: for $v_a^2<T$, this is $v_0=0$ and for $v_a^2>T$, this is $v_0=\pm v_*$.

2. For $1<y<y_{max}$ and $\eta_c^{(\pm)}<\eta<\eta_c^{(0)}$, there are two unstable modes since the Nyquist curve encircles the origin twice. One of these two modes is $\omega=i\omega_i$ and the other is $\omega=\omega_r+i\omega_i$ with $\omega_r\neq 0$ (overstable). When we increase $\eta$ and cross the marginal line  $\eta=\eta_c^{(0)}$, the mode  $\omega=i\omega_i$ becomes stable according to the general result (\ref{mon15}). Indeed, for $y>1$, the DF is {\it minimum} at $v=0$. This is why  there is only one unstable mode for $\eta>\eta_c^{(0)}$: the mode $\omega=\omega_r+i\omega_i$ with $\omega_r\neq 0$ (overstable). On the other hand, for $y<1$, the mode $\omega=i\omega_i$ is stable for $\eta<\eta_c^{(0)}$. When we increase $\eta$ and  cross the marginal line  $\eta=\eta_c^{(0)}$, it becomes unstable according to the general result (\ref{mon15}). Indeed, for $y>1$, the DF is {\it maximum} at $v=0$.

3. In Fig. \ref{GRAVdiagphasesym}, we see that the critical Jeans length $(\lambda_c)_*$ associated to a double-humped DF  ($v_a\neq 0$) is always larger than the critical Jeans length associated to the single-humped Maxwellian ($v_a=0$). This means that the presence of streaming in a purely stellar system has a stabilizing role. This is different if there exists a gas component in the system since the critical Jeans length is reduced by the relative motion of the gas and stars (Sweet 1963).

4. For a double-humped stellar system with  $v_a<\sqrt{T}$, the critical Jeans length increases with $v_a$ due to the stabilization effect of the relative velocity.  Alternatively, in a contra-streaming self-gravitating gas with  $v_a<c_s$, the critical Jeans length decreases with $v_a$ and tends to zero as $v_a\rightarrow c_s$ (Talwar \& Kalra 1966, Ikeuchi et al. 1974). On the other hand, for a double-humped stellar system with $v_a>\sqrt{T}$ or for a contrastreaming self-gravitating gas with  $v_a>c_s$, the critical Jeans length decreases with $v_a$ (but remains larger than the classical Jeans wavelength corresponding to $v_a=0$) and  overstable modes appear.

\section{The asymmetric double-humped distribution}
\label{sec_vha}

\subsection{Determination of the extrema}
\label{sec_ae}

We now assume that the reduced distribution (\ref{lsv3}) is an asymmetric
double-humped distribution of the form
\begin{equation}
\label{ae1}
f(v)=\sqrt{\frac{\beta}{2\pi}} \frac{\rho}{1+\Delta} \left [  e^{-\frac{\beta}{2}(v-v_a)^2} + \Delta e^{-\frac{\beta}{2}(v+v_a)^2}  \right ],
\end{equation}
where $T=1/\beta$ is the temperature of the Maxwellians and $\Delta$
is the asymmetry parameter (we assume here that $\Delta>1$). This distribution  is plotted in Fig. \ref{asymdistr}. The symmetric case is recovered for
$\Delta=1$. The average velocity is $\langle
v\rangle=-[(\Delta-1)/(\Delta+1)]v_a$ and the kinetic temperature
$T_{kin}\equiv \langle (v-\langle v\rangle)^2\rangle=T+[4\Delta/(\Delta+1)^2]v_{a}^{2}$.  The velocities $v_0$ at
which the distribution function $f(v)$ is extremum satisfy
$f'(v_0)=0$. They are determined by the equation
\begin{equation}
\label{ae2}
e^{-2\beta v_{a} v_0}=\frac{1}{\Delta}\frac{v_a-v_0}{v_a+v_0}.
\end{equation}
We note that $v_0\in \rbrack -v_a,
+v_a\lbrack$. Introducing the
dimensionless velocity  and the dimensionless separation
(\ref{e3}), Eq. (\ref{ae2}) can be rewritten
\begin{equation}
\label{ae3}
1=\frac{1}{2V_a V_0}\ln\left (\frac{V_a+V_0}{V_a-V_0}\right )+\frac{\ln(\Delta)}{2 V_a V_0}.
\end{equation}
It is convenient to introduce the variables
\begin{equation}
\label{ae4}
x=\frac{V_0}{V_a}, \qquad  y=V_a^2.
\end{equation}
Then, we have to study the function
\begin{equation}
\label{ae5}
y(x)=\frac{1}{2x}\ln\left (\frac{1+x}{1-x}\right )+\frac{\ln(\Delta)}{2x},
\end{equation}
for $x \in ] -1, +1[$. This function is plotted in
Fig. \ref{yasym}. It has the following properties
\begin{equation}
\label{ae6}
y(x)\sim -\frac{1}{2}\ln (1-x), \qquad (x\rightarrow 1^{-}),
\end{equation}
\begin{equation}
\label{ae7}
y(x)\sim -\frac{1}{2}\ln (1+x), \qquad (x\rightarrow -1^{+}),
\end{equation}
\begin{equation}
\label{ae8}
y(x)\sim \frac{\ln(\Delta)}{2x}, \qquad (x\rightarrow 0).
\end{equation}
Considering the negative velocities $x<0$, we note that $y\ge 0$ iff $x\le x_0$ with
\begin{equation}
\label{ae9}
x_{0}=\frac{1-\Delta}{1+\Delta}.
\end{equation}
Considering the positive velocities $x>0$, we note that the curve $y(x)$ is
minimum at $x=x_*$ where $x_{*}$ is solution of
\begin{equation}
\label{ae10}
\frac{2x_{*}}{1-x_{*}^{2}}-\ln\left (\frac{1+x_{*}}{1-x_{*}}\right )=\ln(\Delta).
\end{equation}
This function is represented in Fig. \ref{xynew}. We note that
Eq. (\ref{ae10}) has a unique solution $x_*$ for each value of
$\Delta>1$. Therefore, the function $y(x)$ has a single minimum at
$x=x_{*}$. The value of this minimum is
\begin{equation}
\label{ae11}
y_{*}=\frac{1}{1-x_{*}^{2}}>1.
\end{equation}
The extrema of the distribution $f(V)$ can be determined from the
study of the function (\ref{ae5}). If $y<y_{*}$, the
distribution $f(V)$ has a single maximum at $V_{0}=V_{-}<0$.  If
$y>y_{*}$, the distribution $f(V)$ has two
maxima at $V_{0}=V_{-}<0$ and $V_{0}=V_{+}>0$ and one minimum at
$V_{0}=V_{p}>0$. These different values are given by $V_{0}=V_a y^{-1}(V_a^2)$.

\begin{figure}
\begin{center}
\includegraphics[clip,scale=0.3]{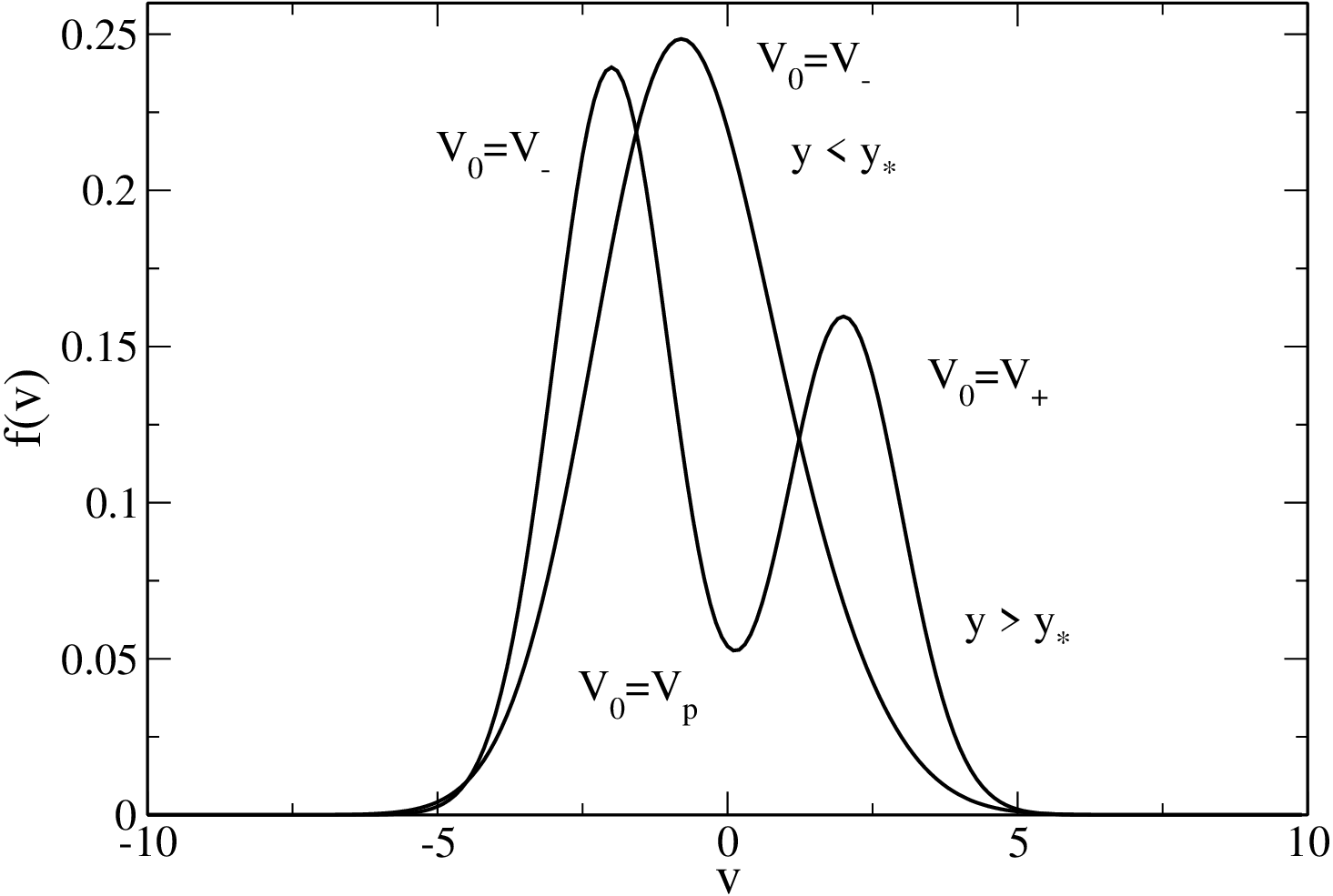}
\caption{Asymmetric double-humped distribution made of two Mawellians with separation $y$ and asymmetry $\Delta>1$. If $y>y_{*}$, the DF has one global maximum at $V_{0}=V_{-}<0$, a minimum at $V_{0}=V_{p}>0$ and a local maximum at $V_{0}=V_{+}>0$. If $y<y_{*}$, the DF has only one maximum at $V_{0}=V_{-}<0$.}
\label{asymdistr}
\end{center}
\end{figure}

\begin{figure}
\begin{center}
\includegraphics[clip,scale=0.3]{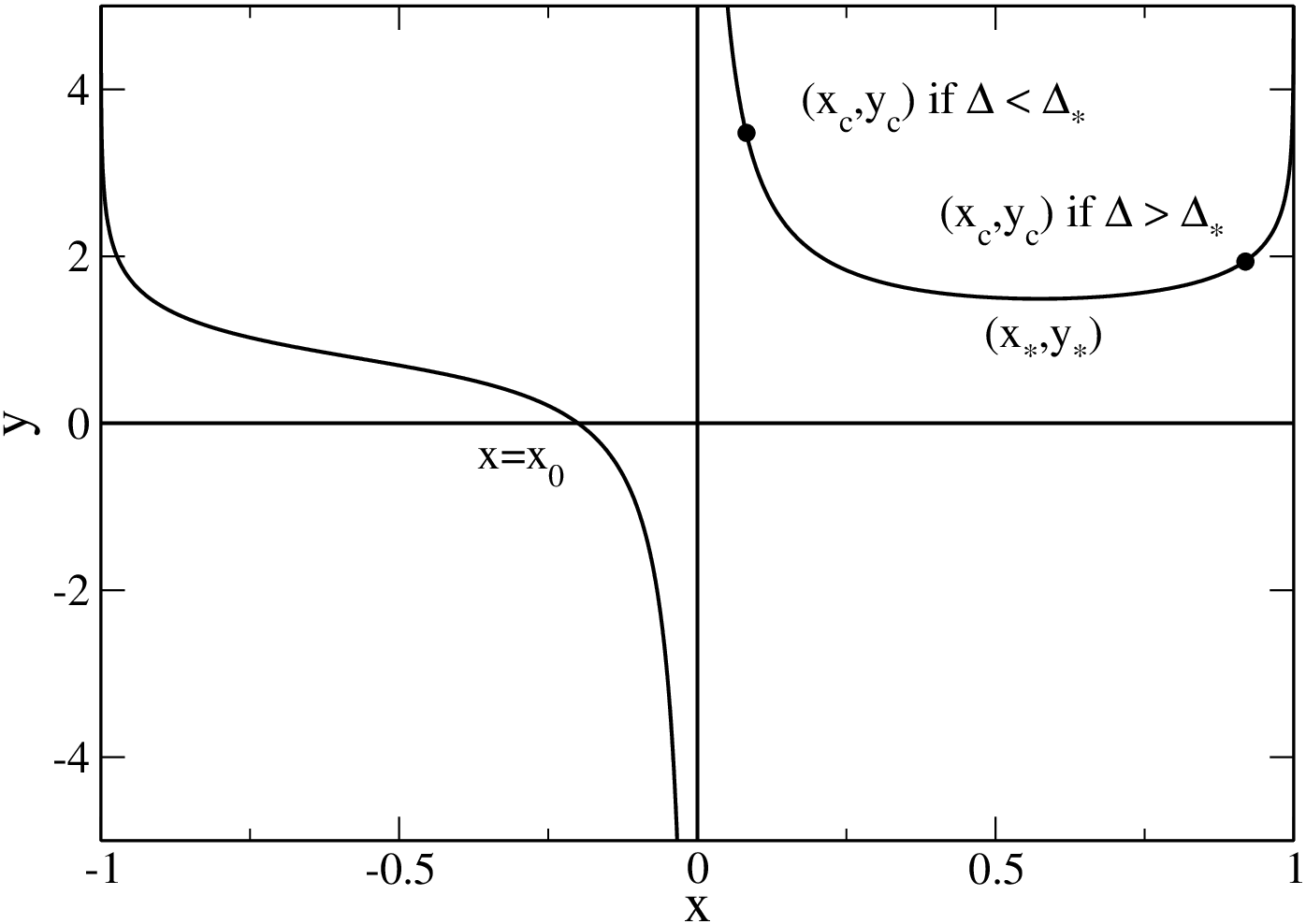}
\caption{The function $y(x)$ for the asymmetric double-humped distribution.}
\label{yasym}
\end{center}
\end{figure}

\begin{figure}
\begin{center}
\includegraphics[clip,scale=0.3]{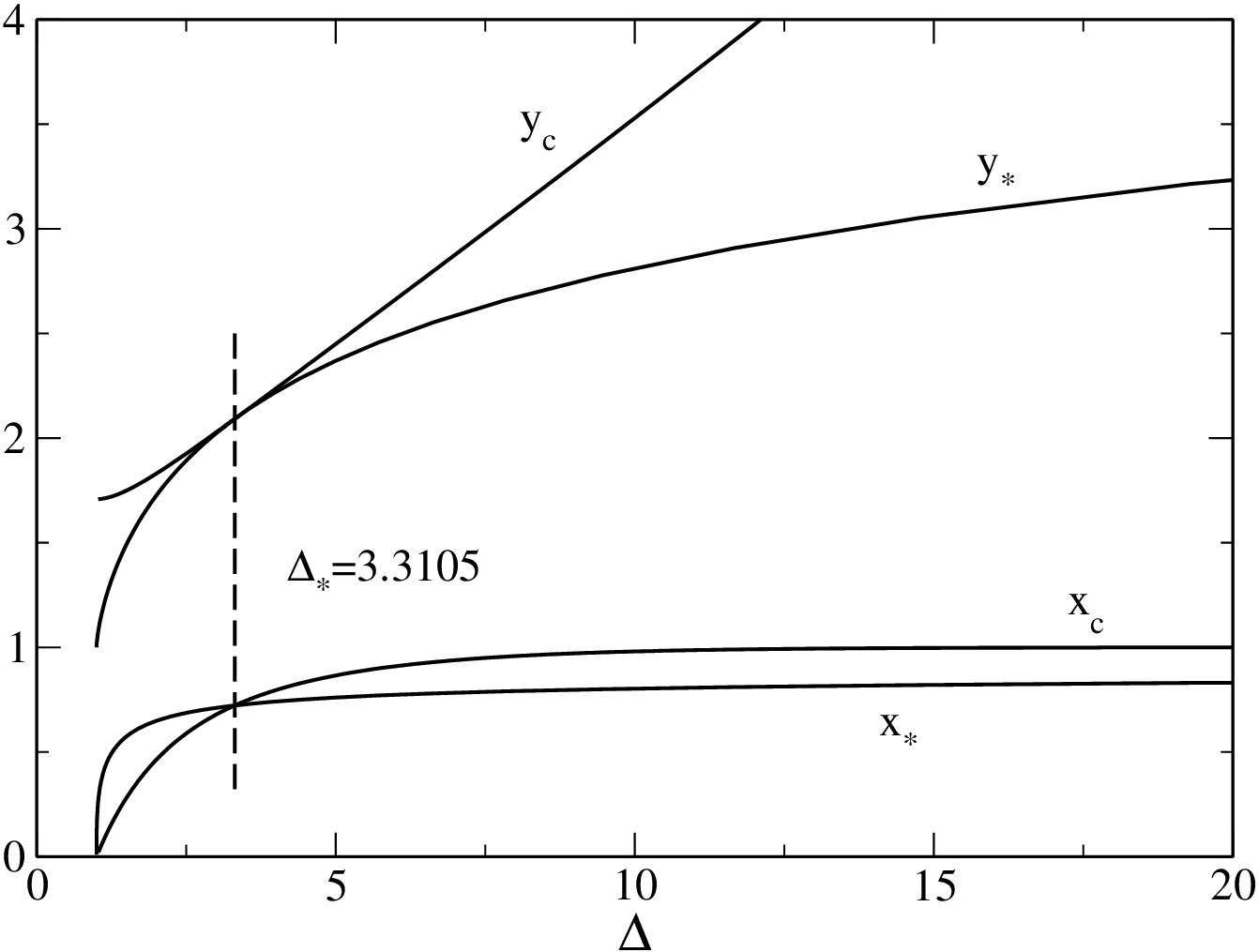}
\caption{Evolution of $x_*$, $y_*$, $x_c$ and $y_c$ as a function of $\Delta$. The curves intersect each other at $\Delta_{*}=3.3105$.}
\label{xynew}
\end{center}
\end{figure}

In conclusion, for a given asymmetry $\Delta>1$ and temperature $T$:

$\bullet$ if  $y>y_{*}$, the distribution function $f(V)$ has a global maximum at $V_{0}=V_{-}<0$, a local maximum  at $V_{0}=V_{+}>0$  and one minimum at  $V_{0}=V_{p}>0$.

$\bullet$ if  $y<y_{*}$, the distribution function $f(V)$ has only one
maximum at $V_{0}=V_{-}<0$.

\subsection{The condition of marginal stability}
\label{sec_ams}

The dielectric function associated to the asymmetric double-humped
distribution is
\begin{eqnarray}
\label{ams1}
\epsilon(\Omega)= 1-\frac{\eta}{1+\Delta} \left [ W(\sqrt{\eta}\Omega - \sqrt{y}))+\Delta W(\sqrt{\eta}\Omega + \sqrt{y}))   \right ],\nonumber\\
\end{eqnarray}
where $W(z)$ is defined in Eq.  (\ref{md5}). When $\Omega_{i}=0$, the real and imaginary parts of the dielectric function  $\epsilon(\eta,\Omega_{r})=\epsilon_{r}(\eta,\Omega_{r})+i\epsilon_{i}(\eta,\Omega_{r})$ can be written
\begin{eqnarray}
\label{ams2}
\epsilon_{r}(\eta,\Omega_{r})=1-\frac{\eta}{1+\Delta}  [ W_r(\sqrt{\eta}\Omega_{r} - \sqrt{y}))\nonumber\\
+\Delta W_r(\sqrt{\eta}\Omega_{r} + \sqrt{y}))    ],
\end{eqnarray}
\begin{eqnarray}
\label{ams3}
\epsilon_{i}(\eta,\Omega_{r})=-\frac{\eta}{1+\Delta}  [ W_i(\sqrt{\eta}\Omega_{r} - \sqrt{y}))\nonumber\\
+\Delta W_i(\sqrt{\eta}\Omega_{r} + \sqrt{y}))   ],
\end{eqnarray}
where $W_r(z)$ and $W_i(z)$ are defined in Eqs. (\ref{mn6})-(\ref{mn7})
where $z$ is here a real number. The condition of marginal stability corresponds to $\epsilon_{r}(\eta,\Omega_{r})=\epsilon_{i}(\eta,\Omega_{r})=0$. The condition $\epsilon_{i}(\eta,\Omega_{r})=0$ is equivalent  to
\begin{equation}
\label{ams4}
f'(\sqrt{\eta}\Omega_{r})=0.
\end{equation}
The condition $\epsilon_{r}(\eta,\Omega_{r})=0$ leads to
\begin{eqnarray}
\label{ams5}
1-\frac{\eta}{1+\Delta} \left [ W_{r}(\sqrt{\eta}\Omega_{r} - \sqrt{y}))+\Delta W_{r}(\sqrt{\eta}\Omega_{r} + \sqrt{y}))   \right ]=0.\nonumber\\
\end{eqnarray}
According to Eq. (\ref{ams4}), the phase velocity $\sqrt{\eta}\Omega_{r}$ is
equal to a velocity $V_{0}$ at which the distribution (\ref{ae1}) is
extremum. The second equation (\ref{ams5}) determines the value(s)
$\eta_{c}(y)$ of the wavenumber at which the distribution is marginally
stable. Therefore, we have to solve
\begin{eqnarray}
\label{ams6}
1-\frac{\eta}{1+\Delta} \left [ W_{r}(V_{0} - \sqrt{y}))+\Delta W_{r}(V_{0}+ \sqrt{y}))   \right ]=0,\nonumber\\
\end{eqnarray}
where $V_{0}$ is given by
\begin{equation}
\label{ams7}
1=\frac{1}{2\sqrt{y}V_{0}}\ln\left (\frac{\sqrt{y}+V_{0}}{\sqrt{y}-V_{0}}\right )+\frac{\ln(\Delta)}{2 \sqrt{y} V_{0}}.
\end{equation}
Eliminating $V_0$ between these two expressions yields the
critical wavenumber(s) $\eta_{c}(y)$ as a function of
$y$. However, it is easier to proceed differently. Setting $x=V_0/\sqrt{y}$, we obtain the equations
\begin{equation}
\label{ams8}
y=\frac{1}{2x}\ln\left (\frac{1+x}{1-x}\right
)+\frac{\ln(\Delta)}{2x},
\end{equation}
\begin{equation}
\label{ams9}
\eta=\frac{1+\Delta}{\left [ W_{r}(\sqrt{y}(x - 1))+\Delta W_{r}(\sqrt{y}(x+ 1))   \right ]}.
\end{equation}
For given $x$, we can obtain $y$ from Eq. (\ref{ams8}) [see also
Fig. \ref{yasym}] and $\eta$ from Eq. (\ref{ams9}). Varying $x$ in the interval $]-1,1[$ yields the
full curve $\eta_c(y)$. We have three types of solutions. For $x\in
]-1,x_0]$, we obtain a branch $\eta_c^{(-)}(y)$ where the pulsation of
the marginal mode is negative: $\Omega_{r}=V_{-}/\sqrt{\eta_c}<0$ corresponding to
the global maximum of $f(v)$. For $x\in ]0,x_*]$, we obtain a branch
$\eta_c^{(p)}(y)$ where the pulsation of the marginal mode is
positive: $\Omega_{r}=V_{p}/\sqrt{\eta_c}>0$ corresponding to the minimum of
$f(v)$. For $x\in [x_*,1[$, we obtain a branch $\eta_c^{(+)}(y)$ where
the pulsation of the marginal mode is positive: $\Omega_{r}=V_{+}/\sqrt{\eta_c}>0$
corresponding to the local maximum of $f(v)$. This leads to the curves
reported in Figs. \ref{GRAVdiagphased1.02} and \ref{GRAVdiagphased10}
for two values of the asymmetry factor $\Delta$.

For $x\rightarrow -1$, $y\rightarrow +\infty$ and
\begin{equation}
\label{ams18}
\eta_{c}^{(-)}(y)=\frac{1+\Delta}{\Delta}+\frac{1+\Delta}{4\Delta^2 y}+... \qquad (y\rightarrow +\infty).
\end{equation}
For $x\rightarrow x_0$, $y\rightarrow 0$ and
\begin{equation}
\label{ams19}
\eta_{c}^{(-)}(y)\rightarrow 1 \qquad (y\rightarrow 0).
\end{equation}
This returns the critical wavenumber (\ref{mn8}) associated with the
Maxwellian distribution (\ref{md1}) corresponding to $y=0$.
For $x\rightarrow 1$, $y\rightarrow +\infty$ and
\begin{equation}
\label{ams19b}
\eta_{c}^{(+)}(y)\sim 1+\Delta+\frac{\Delta (1+\Delta)}{4y}+...\qquad (y\rightarrow +\infty).
\end{equation}
Let $x_c$ and $y_c$ be determined by the equations
\begin{equation}
\label{ams21}
y_{c}=\frac{1}{2x_{c}}\ln\left (\frac{1+x_{c}}{1-x_{c}}\right
)+\frac{\ln(\Delta)}{2x_{c}},
\end{equation}
\begin{equation}
\label{ams22}
W_{r}(\sqrt{y_{c}}(x_{c} - 1))+\Delta W_{r}(\sqrt{y_{c}}(x_{c}+ 1))=0.
\end{equation}
For $x\rightarrow x_c$, $y\rightarrow y_c$ and
\begin{equation}
\label{ams19c}
\eta_{c}^{(s)}(y)\propto \frac{1}{y-y_c}\rightarrow  +\infty \qquad (y\rightarrow y_c),
\end{equation}
where $s=p$ if $\Delta<\Delta_{*}$ and $s=+$ if
$\Delta>\Delta_{*}$ as will become clear below.
Note that there is no physical solution to
Eqs. (\ref{ams8})-(\ref{ams9}) when $0<x<x_{c}$ ($\eta$ would be
negative) so that the branch $\eta_{c}^{(s)}(y)$ starts at
$(y_c,+\infty)$ corresponding to $x=x_{c}$.  The evolution of $x_c$ and
$y_c$ with $\Delta$ is studied in Fig. \ref{xynew} and is compared
to the evolution of $x_*$ and $y_*$. The curves intersect each other
at $\Delta_*=3.3105$.  For $\Delta<\Delta_*$, $x_c<x_*$ so that the
stability diagram displays three marginal branches
$\eta_{c}^{(-)}(y)$, $\eta_{c}^{(p)}(y)$ and $\eta_{c}^{(+)}(y)$ (see
Fig. \ref{GRAVdiagphased1.02}). The branches $\eta_{c}^{(p)}(y)$ and
$\eta_{c}^{(+)}(y)$ connect each other at $(y_{*},\eta_{*})$
corresponding to $x=x_{*}$. At that point they touch the line
$y=y_{*}$ separating distributions with one or two
maxima. For
$\Delta=\Delta_*$, $x_c=x_*$ so that the branch $\eta_{c}^{(p)}(y)$ is rejected to infinity and
only the branches $\eta_{c}^{(-)}(y)$ and
$\eta_{c}^{(+)}(y)$ remain. For $\Delta>\Delta_*$, $x_c>x_*$ so that the phase
diagram displays only two marginal branches $\eta_{c}^{(-)}(y)$ and
$\eta_{c}^{(+)}(y)$ (see Fig. \ref{GRAVdiagphased10}).

\subsection{The stability diagram}

The critical wavenumbers $\eta_c(y)$ corresponding to marginal
stability determined previously are represented as a function of the
separation $y$ in Figs. \ref{GRAVdiagphased1.02} and \ref{GRAVdiagphased10}
for two values of the asymmetry factor $\Delta$. We
recall that $\eta_{c}^{(-)}(y)$ corresponds to the wavenumber
associated with a marginal mode with pulsation $\Omega_{r}=V_{-}/\sqrt{\eta_c}<0$ (global maximum of $f$),
$\eta_{c}^{(p)}(y)$ corresponds to the wavenumber associated with a
marginal mode with pulsation $\Omega_{r}=V_{p}/\sqrt{\eta_c}>0$ (minimum of $f$) and
$\eta_{c}^{(+)}(y)$ corresponds to the wavenumber associated with a
marginal mode with pulsation $\Omega_{r}=V_{+}/\sqrt{\eta_c}>0$ (local maximum of $f$). We have also
plotted the line $y=y_*$. On the left of this line, the DF has a
single maximum at $V_{-}<0$ and on the right of this line, the DF has two
maxima at $V_{-}<0$ and $V_{+}>0$ and a minimum at $V_{p}>0$. In order
to investigate the stability of the solutions in the different regions
of the parameter space, we have used the Nyquist criterion.  The description
of the stability diagram is similar to the one given in
Sec. \ref{sec_sdsy} and the different possible cases can be understood
directly from the reading of Figs. \ref{GRAVdiagphased1.02} and \ref{GRAVdiagphased10}. The best way is to fix $y$ and progressively
increase the value of $\eta$. For $y<y_*$, the distribution
has only one maximum at $V_{0}=V_{-}$ so the Nyquist curve has one
intersection with the $x$-axis (in addition to the limit point
$(1,0)$). For $\eta\rightarrow 0$, we find that $\epsilon_{r}(V_{-})>0$
so the system is stable. As we increase $\eta$ and pass above the solid
line, we find that $\epsilon_{r}(V_{-})<0$
so the system is unstable. For $y>y_*$, the distribution
has two maxima at $V_{0}=V_{-}$ and
$V_{0}=V_{+}$ and one minimum at $V_{0}=V_{p}$. At each intersection
with a marginal line, one of the values $\epsilon_{r}(V_{-})$,
$\epsilon_{r}(V_{p})$ or $\epsilon_{r}(V_{+})$ changes sign. We have indicated by symbols like $(-++)$ the respective signs of  $\epsilon_{r}(V_{-})$,
$\epsilon_{r}(V_{p})$ and $\epsilon_{r}(V_{+})$. We
can then easily draw by hands the corresponding Nyquist curve.
Therefore, it is not necessary to show all the possibilities and we
have only indicated a few representative cases in
Figs. \ref{data_delta1.02_y1.1_n1}-\ref{data_delta1.02_y1.1_n6} for illustration.

\begin{figure}[htbp]
\centerline{
\includegraphics[width=8cm,angle=0]{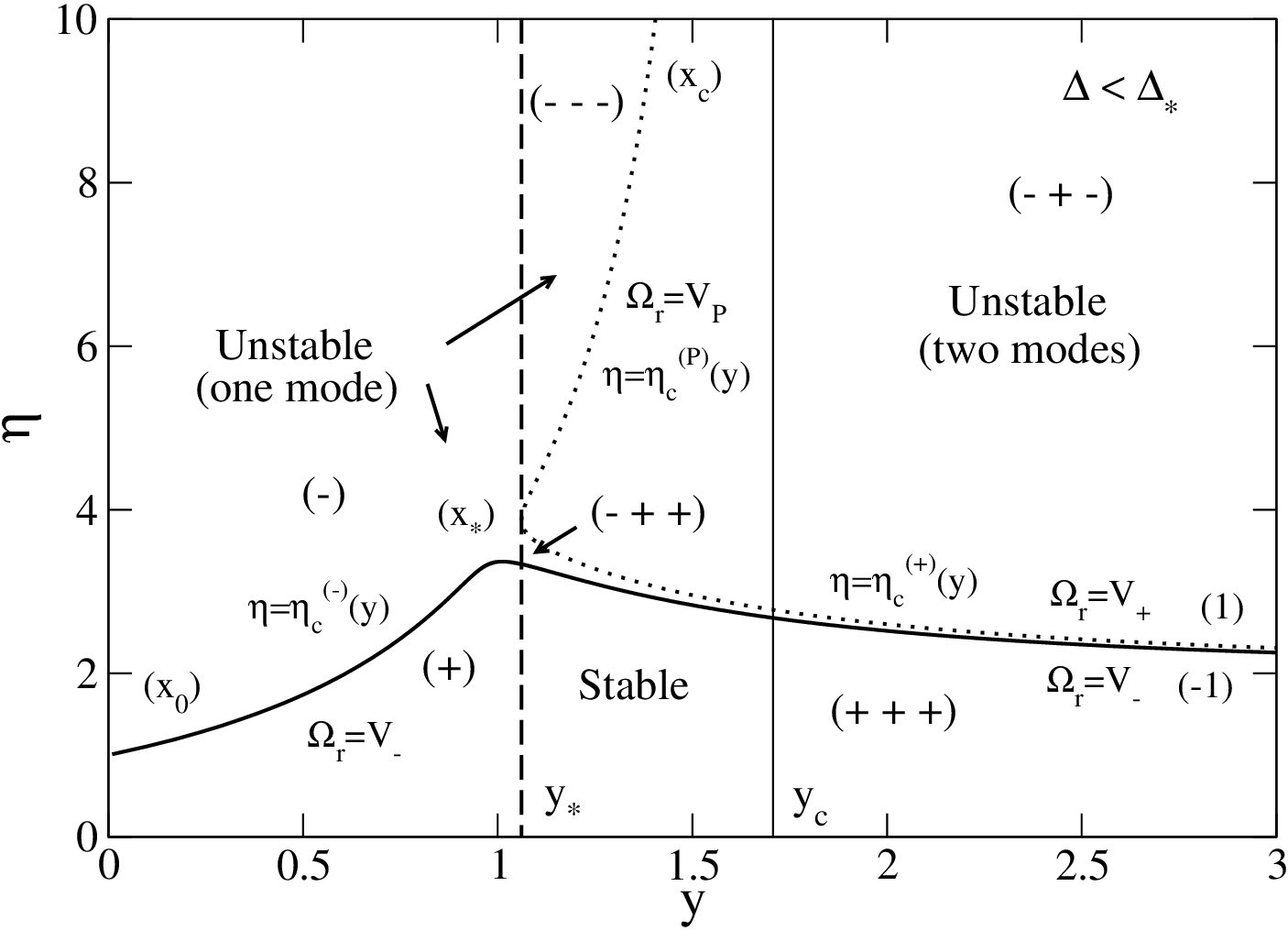}
} \caption[]{Stability diagram of the asymmetric double-humped distribution for $\Delta<\Delta_*$ (specifically $\Delta=1.02$). There exists three marginal branches $\eta_c^{(-)}(y)$, $\eta_c^{(p)}(y)$ and $\eta_c^{(+)}(y)$. The symbols have been defined in the text.}
\label{GRAVdiagphased1.02}
\end{figure}

\begin{figure}[htbp]
\centerline{
\includegraphics[width=8cm,angle=0]{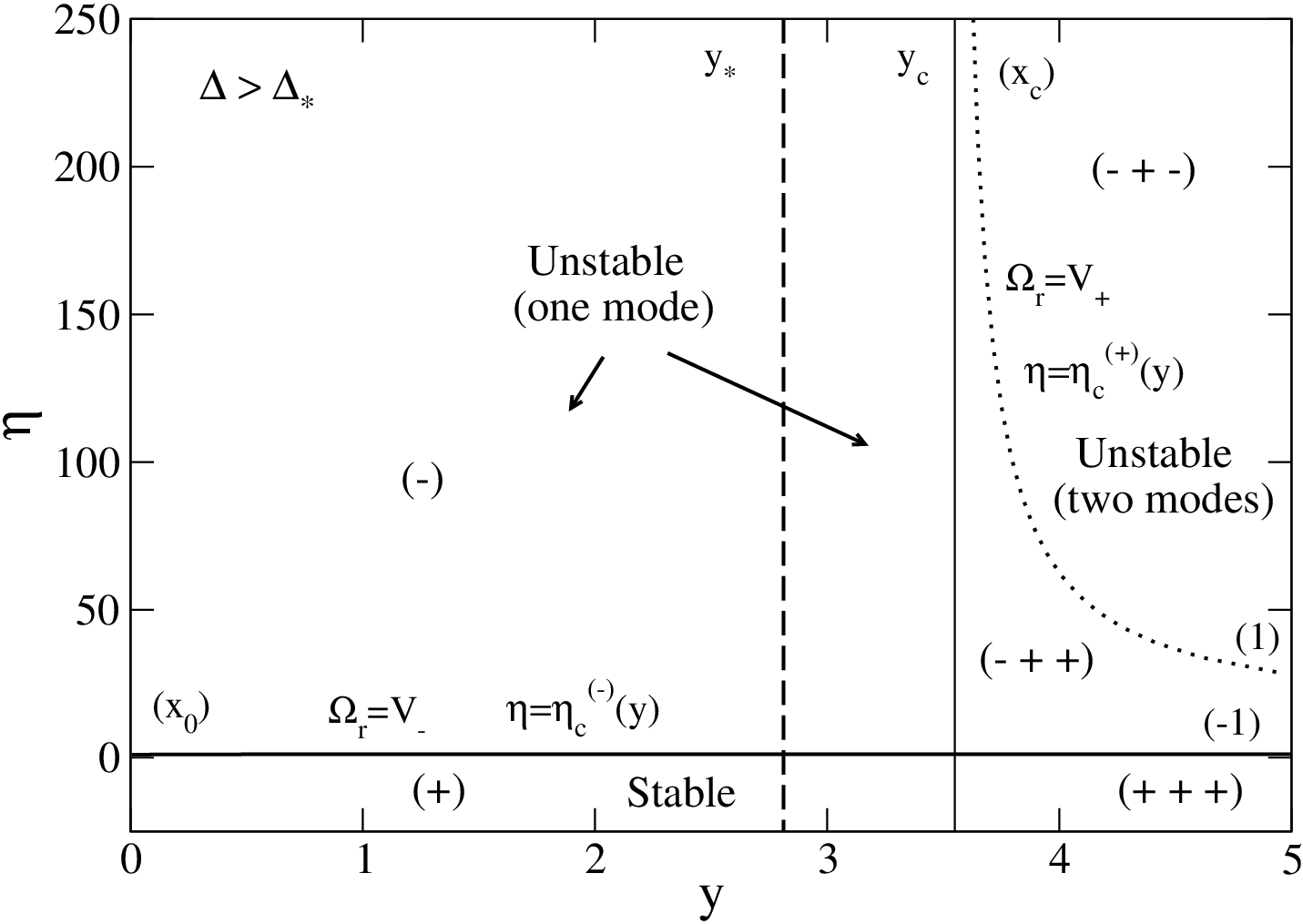}
} \caption[]{Stability diagram of the asymmetric double-humped distribution for $\Delta>\Delta_*$ (specifically $\Delta=10$). There exists two marginal branches $\eta_c^{(-)}(y)$ and $\eta_c^{(+)}(y)$. }
\label{GRAVdiagphased10}
\end{figure}

\begin{figure}[htbp]
\centerline{
\includegraphics[width=8cm,angle=0]{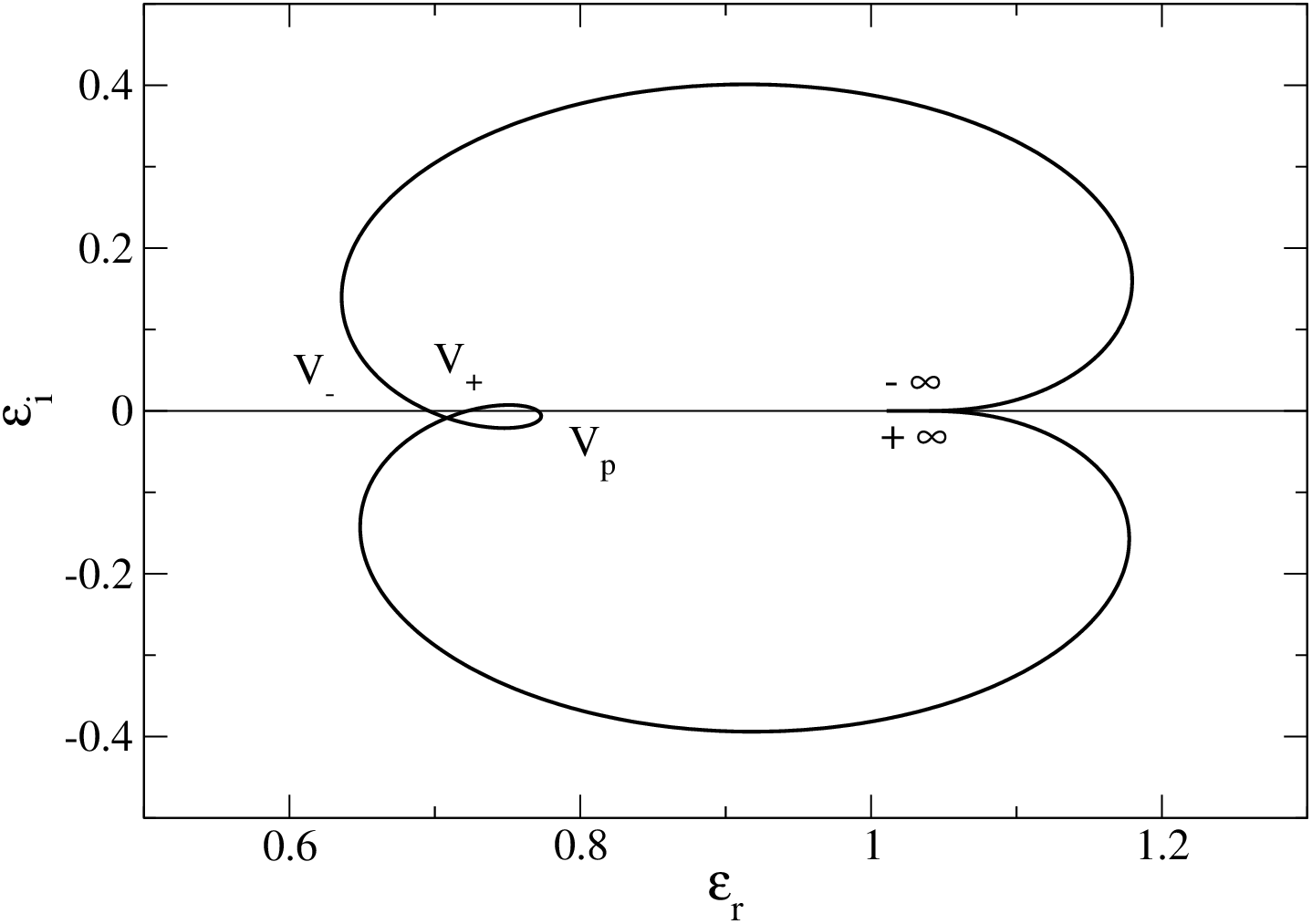}
} \caption[]{$\Delta<\Delta_*$: Nyquist curve for $y_*<y<y_c$ and $\eta<\eta_c^{(-)}$
(specifically $\Delta=1.02$, $y=1.1$ and $\eta=1$). The DF has a global maximum at $V_{-}$, a minimum at $V_p$ and a local maximum at $V_{+}$. The DF is stable (with respect to
this perturbation) because the Nyquist curve does not encircle the
origin.  Case (+ + +).}
\label{data_delta1.02_y1.1_n1}
\end{figure}

\begin{figure}[htbp]
\centerline{
\includegraphics[width=8cm,angle=0]{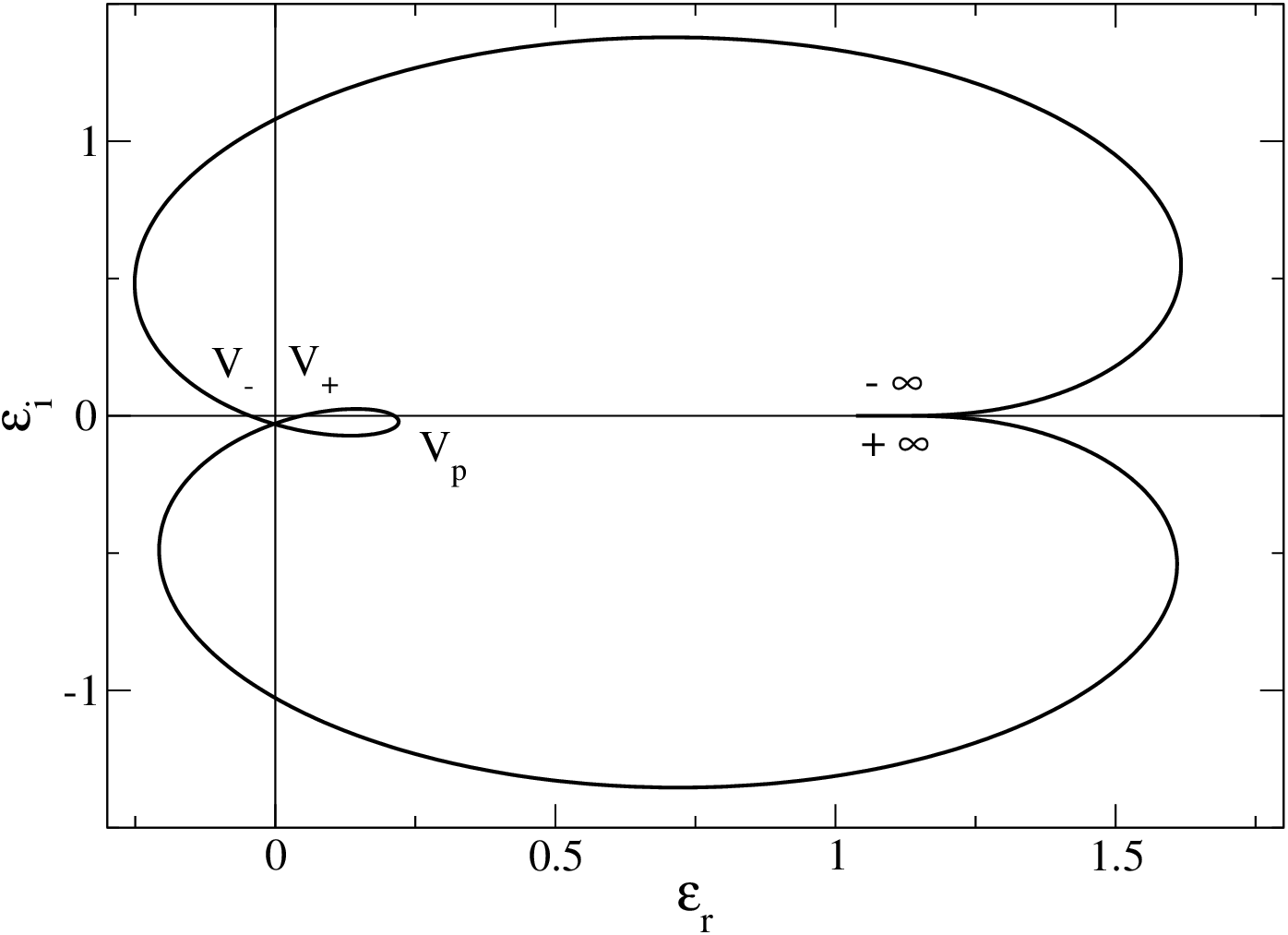}
} \caption[]{$\Delta<\Delta_*$: Nyquist curve for $y_*<y<y_c$ and $\eta_c^{(-)}<\eta<\eta_c^{(+)}$
(specifically $\Delta=1.02$, $y=1.1$ and $\eta=3.436$). The DF has a global maximum at $V_{-}$, a minimum at $V_p$ and a local maximum at $V_{+}$. The DF is unstable (with respect to
this perturbation) because the Nyquist curve encircles the
origin once. There is $N=1$ unstable mode.  Case (- + +).}
\label{data_delta1.02_y1.1_n3.436}
\end{figure}

\begin{figure}[htbp]
\centerline{
\includegraphics[width=8cm,angle=0]{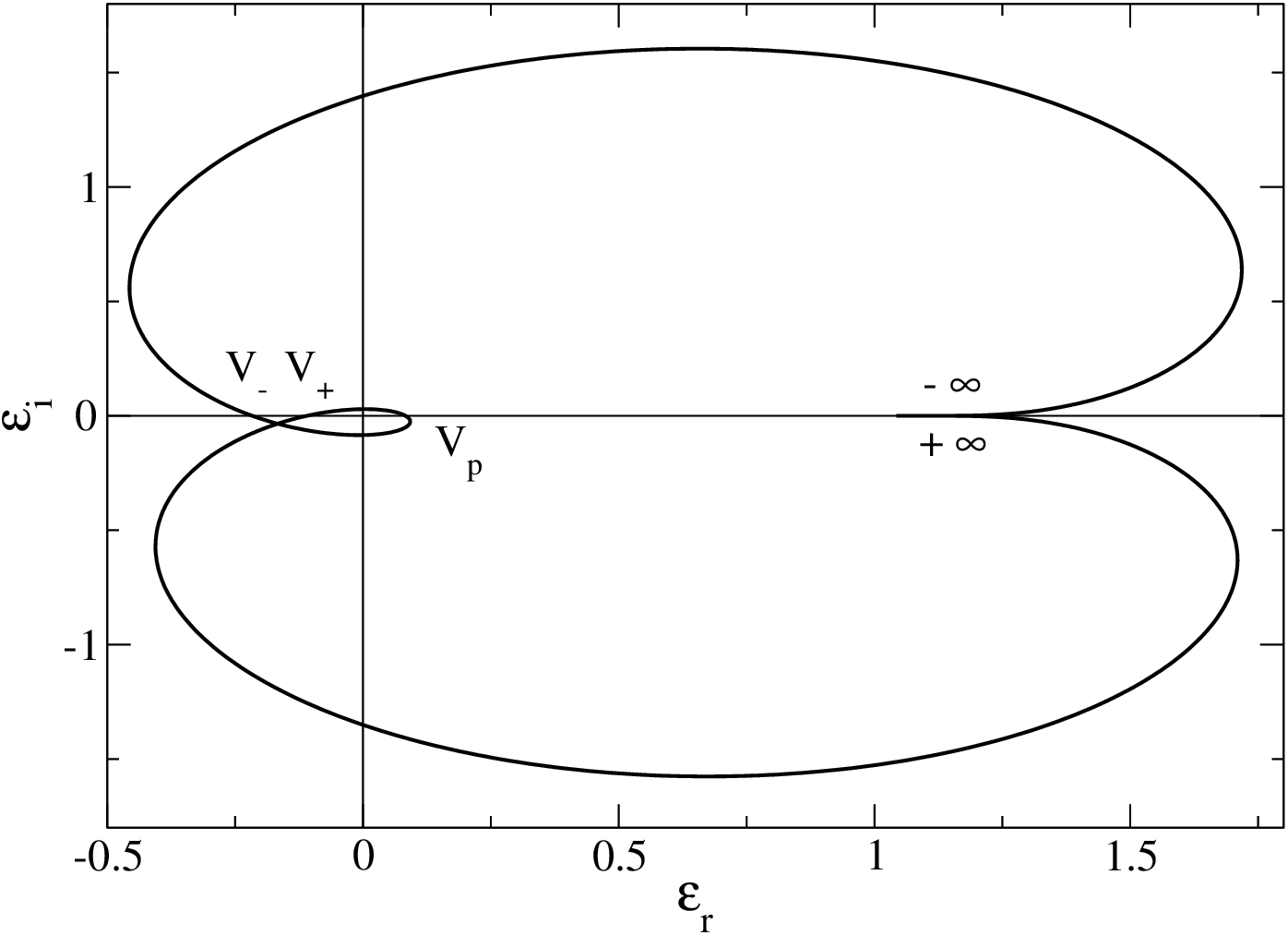}
} \caption[]{$\Delta<\Delta_*$: Nyquist curve for $y_*<y<y_c$ and $\eta_c^{(+)}<\eta<\eta_c^{(p)}$
(specifically $\Delta=1.02$, $y=1.1$ and $\eta=4$). The DF has a global maximum at $V_{-}$, a minimum at $V_p$ and a local maximum at $V_{+}$. The DF is unstable (with respect to
this perturbation) because the Nyquist curve encircles the
origin twice. There are $N=2$ unstable modes.  Case (- + -).}
\label{data_delta1.02_y1.1_n4}
\end{figure}

\begin{figure}[htbp]
\centerline{
\includegraphics[width=8cm,angle=0]{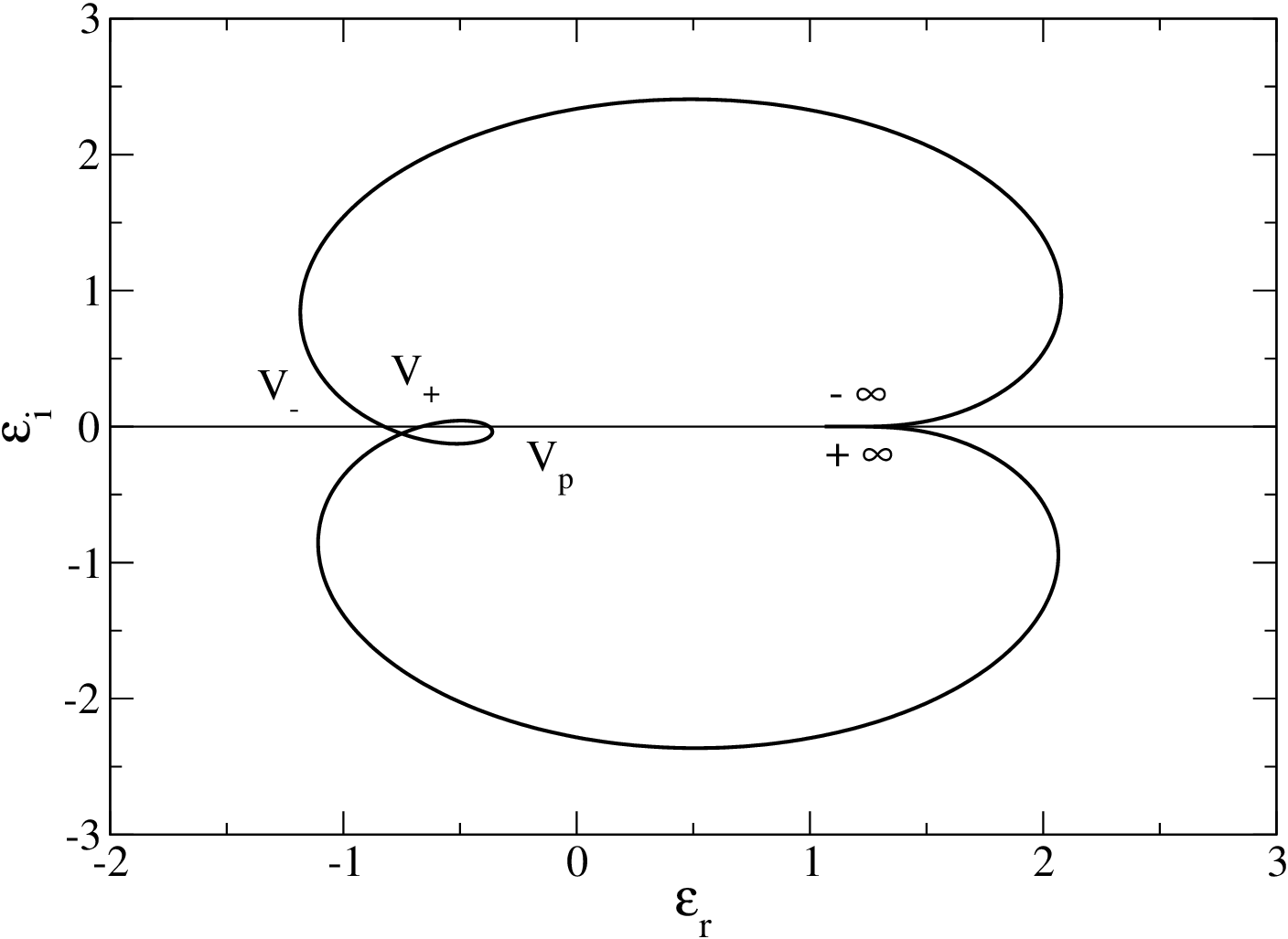}
} \caption[]{$\Delta<\Delta_*$: Nyquist curve for $y_*<y<y_c$ and $\eta>\eta_c^{(p)}$
(specifically $\Delta=1.02$, $y=1.1$ and $\eta=6$). The DF has a global maximum at $V_{-}$, a minimum at $V_p$ and a local maximum at $V_{+}$. The DF is unstable (with respect to
this perturbation) because the Nyquist curve encircles the
origin once. There is $N=1$ unstable mode.  Case (- - -).}
\label{data_delta1.02_y1.1_n6}
\end{figure}

{\it Remark}: In Figs. \ref{GRAVdiagphased1.02} and \ref{GRAVdiagphased10}, we see that the system is stable for perturbations with  $k>(k_c)_*$ (corresponding to the solid line) and unstable for perturbations with $k<(k_c)_*$. This critical wavenumber $(k_c)_*$ corresponds to a marginal  perturbation where the phase velocity $\omega/k$ coincides with the {\it global maximum} $v_0=v_{-}$ of the velocity distribution.

\section{The case of plasmas}
\label{sec_plasmas}

We shall now compare the previous results obtained for self-gravitating systems to the case of plasmas.

\subsection{A brief historic}
\label{sec_paz}

The name plasma was introduced by Tonks \& Langmuir (1929) to describe
an ionised gas made of electrons and ions. They found that cold
plasmas oscillate with a natural pulsation $\omega_p=(4\pi\rho
e^2/m^2)^{1/2}$.  In order to take into account thermal effects,
plasmas were initially described in terms of hydrodynamic equations. A
dispersion relation was derived by Thomson \& Thomson (1933) for
isothermal perturbations and by Gross (1951) for three dimensional
adiabatic perturbations. However, fluid equations do not provide a
good description of plasmas and rely on arbitrary assumptions
concerning the perturbations (see van Kampen 1957). Indeed, plasmas
are essentially collisionless so that a hydrodynamic description is
not clearly justified. Vlasov (1938,1945) was the first to attempt to
derive the dispersion relation directly from the collisionless
Boltzmann equation (nowdays called the Vlasov equation). He obtained
the expression of the pulsation $\omega(k)$ in the high wavelength
limit\footnote{The same result was derived in a different manner by
Bohm \& Gross (1949) and Jackson (1960).}. However, Landau (1946)
criticized his mathematical treatment showing that there are serious
divergences in the integrals considered by Vlasov. Landau performed a
rigorous mathematical study using an appropriate contour of
integration in the complex plane and showed that the plasma undergoes
damped oscillations. The pulsation is in agreement with the expression
given by Vlasov but, in addition, the plasma undergoes collisionless
damping (nowdays called Landau damping).  Following Landau's seminal
work, several authors studied the stability of a plasma. They showed
that single-humped distributions are always stable (Berz 1956,
Appendix by W. Newcomb in Bernstein 1958, Auer 1958, Penrose 1960,
Noerdlinger 1960, Jackson 1960). Then, they considered double humped
distributions modeling two contrastreaming beams (Haeff 1949, Nergaard
1948, Pierce
\& Hebenstreit 1949) and determined particular conditions under which
such distributions are unstable (Twiss 1952, Buneman 1958, Auer 1958,
Kahn 1959, Buneman 1959, Penrose 1960, Noerdlinger 1960, Jackson
1960). They found qualitatively that the plasma becomes unstable when
the stream velocity becomes sensibly larger than the thermal velocity.

After briefly recalling some classical results, we will provide a detailed study of the stability/instability of a double-humped distribution made of two Maxwellians.  We shall complete the study of a symmetric double-humped distribution (by explicitly computing the range of unstable wavelengths) and consider for the first time the case of asymmetric double-humped distributions. We will show that the nature of the problem changes above a critical asymmetry $\Delta_*=3.3105$.

\subsection{The Euler-Poisson system}
\label{sec_pep}

We consider a plasma made of electrons with mass $m$ and charge $-e$ and ions with mass $m_i$ and charge $+e$. We assume the ions to be infinitely massive so that they do not contribute to the motion and only provide a neutralizing background. The fluid equations describing the motion of the electrons can be written
\begin{equation}
{\partial\rho\over\partial t}+\nabla \cdot (\rho {\bf u})=0,
\label{pep1}
\end{equation}
\begin{equation}
{\partial {\bf u}\over\partial t}+({\bf u}\cdot \nabla) {\bf u}=-{1\over\rho}\nabla p-\nabla\Phi,
\label{pep2}
\end{equation}
\begin{equation}
\Delta\Phi=-\frac{4\pi e^2}{m^2}(\rho-\rho_i).
\label{pep3}
\end{equation}
These equations differ from the Euler-Poisson system (\ref{ep1})-(\ref{ep3}) describing a self-gravitating barotropic gas only in the sign of the interaction and in the presence of a neutralizing background. This neutralizing background avoids the Jeans swindle since an infinite homogeneous distribution of electrons ($\rho={\rm cst}$, ${\bf u}={\bf 0}$, $\Phi=0$) is a steady state of the fluid equations (\ref{pep1})-(\ref{pep3}) provided that the condition of electroneutrality $\rho=\rho_i$ is satisfied.

Linearizing the Euler-Poisson system around an infinite homogeneous distribution of electrons and decomposing the perturbations in normal modes of the form $e^{i({\bf k}\cdot {\bf r}-\omega t)}$, we obtain the dispersion relation (Nicholson 1992):
\begin{eqnarray}
\label{pep7}
\omega^2=c_s^2k^2+\omega_p^2,
\end{eqnarray}
where $c_s^2=p'(\rho)$ is the velocity of sound and
\begin{eqnarray}
\label{pep5}
\omega_{p}^{2}\equiv \frac{4\pi \rho e^{2}}{m^{2}},
\end{eqnarray}
is the plasma pulsation. For an isothermal equation of state $p=\rho T$, we get
\begin{eqnarray}
\label{pep8b}
\omega^2=T k^2+\omega_p^2,
\end{eqnarray}
and for a polytropic equation of state $p=K\rho^{\gamma}$, we obtain
\begin{eqnarray}
\label{pep8c}
\omega^2=\gamma T k^2+\omega_p^2.
\end{eqnarray}
The dispersion relation (\ref{pep7}) shows that a plasma described by fluid equations is always stable. For $k=0$, the pulsation tends to a constant
\begin{eqnarray}
\label{pep9}
\omega=\pm\omega_p.
\end{eqnarray}
Therefore, in this limit of small wavenumbers, plasma oscillations do not propagate in space: $\rho({\bf r},t)=\rho({\bf r},0)e^{-i\omega_p t}$. This behavior is in marked contrast with all usual types of oscillating phenomena such as acoustic waves for which $\omega=ck$. For large wavenumbers, the dispersion relation approaches that of sound waves: $\omega=\pm c_s k$.

{\it Remark:} Since $\int\delta\rho\delta\Phi\, d{\bf r}=\frac{m^2}{4\pi e^2}\int (\Delta\delta\Phi)^2\, d{\bf r}>0$ for a plasma, the second variations (\ref{ep10b}) of the energy functional (\ref{ep9}) are always positive: $\delta^2{\cal W}>0$. Therefore, the energy functional has a unique (global) minimum and the plasma is nonlinearly dynamically stable with respect to the barotropic Euler-Poisson system.

\subsection{The Vlasov-Poisson system}
\label{sec_pvp}

The fluid equations are not very appropriate to describe a plasma because most plasmas are in a regime where the collisions between charges are negligible. Therefore, from now on, we shall describe Coulombian plasmas by the Vlasov-Poisson system
\begin{equation}
\label{pvp1}
{\partial f\over\partial t}+{\bf v}\cdot {\partial f\over\partial {\bf r}}-\nabla\Phi\cdot {\partial f\over\partial {\bf v}}=0,
\end{equation}
\begin{equation}
\Delta\Phi=-\frac{4\pi e^2}{m^2}(\rho-\rho_i),
\label{pvp2}
\end{equation}
which ignores collisions between charges. These equations are similar to the Vlasov-Poisson system (\ref{vh0})-(\ref{vh1}) describing  collisionless stellar systems  except that the sign of the interaction is reversed and that there is a neutralizing background. We shall investigate the linear dynamical stability of an infinite homogeneous medium. Therefore, we consider stationary solutions of the Vlasov equation of
the form $f=f({\bf v})$. The dispersion relation can be written (Nicholson 1992):
\begin{eqnarray}
\epsilon(k,\omega)\equiv 1-{4\pi e^2\over m^2k^2}\int_{C} {{f'(v)}\over v-\frac{\omega}{k}}dv=0.
\label{pvp3}
\end{eqnarray}
As in Sec. \ref{sec_lsv}, we have taken ${\bf k}$ along the $z$-axis and noted $v$ for $v_z$ and $f(v)$ for $\int f({\bf v})\, dv_xdv_y$.

In general, the dispersion relation (\ref{pvp3}) cannot be solved explicitly to obtain $\omega(k)$ except in some very simple cases. For example, for cold systems described by the distribution function $f=\rho \delta({v}-{v}_0)$, we obtain after an integration by parts
\begin{eqnarray}
\omega=v_0 k\pm \omega_p.
\label{lsv8s}
\end{eqnarray}
In particular, when $v_0=0$, we get
\begin{eqnarray}
\omega=\pm \omega_p.
\label{lsv9s}
\end{eqnarray}
The system is stable to all wavenumbers. We also note that the dispersion relation (\ref{lsv9s}) coincides with the dispersion relation  (\ref{pep7}) obtained with fluid equations with $c_s=0$.

Let us briefly recall important properties of the dispersion relation (\ref{pvp3}) obtained in plasma physics (Balescu 1963, Nicholson 1992). We consider a symmetric distribution and write the complex pulsation in the form $\omega=\omega_r+i\omega_i$. When $\omega_i\ll\omega_r$ (weakly damped perturbations), the complex pulsation is given by\footnote{A more accurate expression (Jackson 1960, Nicholson 1992) is obtained by replacing $f'(\omega_p/k)$ by $f'(\omega_r/k)$. This introduces a factor $e^{3/2}$ in the Landau formula (\ref{pvp5}).}
\begin{eqnarray}
\label{pvp4}
\omega_r^2=\omega_p^2+3Tk^2,\qquad \omega_{i}=\frac{\pi\omega_{p}^{3}}{2\rho k^2}f'\left (\frac{\omega_{p}}{k}\right ),
\end{eqnarray}
with $T=\langle v^2\rangle$ (where we recall that $v=v_z$ in the present case).  These relations are valid for $k/k_D\ll 1$. The real part of the pulsation satisfies a dispersion relation of the form (\ref{fe17}) with $c_s^2=3T$. Therefore, large wavelengths perturbations in a collisionless plasma correspond to one dimensional  isentropic perturbations with index $\gamma=3$ in a gas (see Appendix \ref{sec_fee}). This relation dispersion is called the Langmuir wave dispersion relation (Nicholson 1992). On the other hand, when $f'(\omega_p/k)<0$, the expression of the imaginary part of the pulsation corresponds to Landau damping. If $f'(v)\le 0$ for all $v\ge 0$, as for the Maxwellian, the plasma is stable. For the Maxwell distribution, we get the Landau formula
\begin{eqnarray}
\omega_{i}=-\sqrt{\frac{\pi}{8}}\omega_p \left (\frac{k_{D}}{k}\right )^{3}e^{-\frac{k_{D}^{2}}{2k^2}},
\label{pvp5}
\end{eqnarray}
for the damping rate, where we have introduced the Debye wavenumber
\begin{eqnarray}
\label{pep6}
k_{D}^{2}\equiv \frac{4\pi e^2\beta\rho}{m^2}=\beta  \omega_{p}^{2},
\end{eqnarray}
where $\beta=1/T$ is the inverse temperature. Alternatively, if there exists wavenumbers $k$ such that $f'(\omega_p/k)>0$, the perturbation grows and the plasma is unstable. Since the expression (\ref{pvp4}) is valid for small $k$, this implies that $f'(v)$ must be positive for large velocities. This corresponds for example to the ``bump-on-tail'' situation analyzed in plasma physics (Nicholson 1992).  The dispersion relation  (\ref{pvp4})-a was first obtained by Vlasov (1938,1945). However, the derivation given by Vlasov was not rigorous since he evaluated the integral (\ref{pvp3}) on the real axis on which there is a singularity. Landau (1946) performed a rigorous mathematical study and showed that, in addition to the oscillations, the waves  must be damped exponentially leading to equation (\ref{pvp4})-b for the damping rate. It is interesting to note that the Langmuir wave dispersion relation (\ref{pvp4})-a is rather independent on the precise form of the distribution function (it depends only on the variance $T=\langle v^2\rangle$) while the Landau damping (\ref{pvp4})-b is very sensitive to the distribution function. For example, the expression of Landau damping for polytropic distributions is given by (see Appendix E of Chavanis \& Delfini 2009):
\begin{eqnarray}
\omega_i=-\sqrt{\frac{\pi}{8}}B_n\omega_p\left (\frac{k_D}{k}\right )^3\left (n-\frac{1}{2}\right )\frac{1}{n+1}\nonumber\\
\times\left\lbrack 1-\frac{k_D^2}{2(n+1)k^2}\right\rbrack_+^{n-3/2}.
\label{lapol}
\end{eqnarray}

When $\omega_r\ll\omega_i$ (heavily damped perturbations), the pulsation is given for the Maxwellian distribution by
\begin{eqnarray}
\omega_r=\pm m\frac{\pi\omega_p}{2k_{D}}\frac{k}{\sqrt{\ln k}},  \qquad \omega_i=-\frac{2\omega_{p}}{k_{D}}k\sqrt{\ln k}.
\label{pvp6}
\end{eqnarray}
These expressions are valid for $k/k_D\rightarrow +\infty$. There
exists several branches of solutions parameterized by the odd integer
$m$ (but the branch $m=1$ is the most relevant). Expression
(\ref{pvp6}) of the complex pulsation was first obtained by Landau
(1946). The evolution of $\omega_r(k)$ and $\omega_i(k)$ with $k$ for
a Maxwellian distribution are represented in Fig. 3 of Jackson (1960)
and it can be compared with Fig. \ref{pulsation} for stellar
systems. For small $k$, the damping is weak but it increases rapidly
for increasing wavenumbers so that the waves are never observed at
sufficiently large wavenumbers.

{\it Remark:} Let us consider a plasma with a DF of the form $f=f(v^2)$ with $f'<0$. Since $\int\delta\rho\delta\Phi\, d{\bf r}>0$ for a plasma, the second variations (\ref{vh3b}) of the constrained pseudo entropy (\ref{vh2})  are always negative: $\delta^2{S}\le 0$. Therefore, the pseudo entropy has a unique (global) maximum at fixed mass and energy and the plasma is nonlinearly dynamically stable. These variational technics were introduced in the plasma literature by W. Newcomb in Bernstein (1958) and Gardner (1963); see also the review of Holm et al. (1985).

\subsection{The condition of marginal stability}

When $\omega_i=0$, the real and imaginary parts of the dielectric
function
$\epsilon(k,\omega_{r})=\epsilon_{r}(k,\omega_{r})+i\epsilon_{i}(k,\omega_{r})$
are
\begin{eqnarray}
\epsilon_{r}(k,\omega_{r})=1-{4\pi e^2\over m^2k^2}{P}\int_{-\infty}^{+\infty} {{f'(v)}\over v-\frac{\omega_{r}}{k}}dv,
\label{pny1}
\end{eqnarray}
\begin{eqnarray}
\epsilon_{i}(k,\omega_{r})=-{4\pi^2 e^2\over m^2k^2}f'(\omega_{r}/k).
\label{pny2}
\end{eqnarray}
The condition of marginal stability corresponds to $\epsilon(k,\omega)=0$ and $\omega_i=0$, i.e. $\epsilon_r({k},\omega_r)=\epsilon_i({k},\omega_r)=0$. We obtain therefore the equations
\begin{eqnarray}
 1-\frac{4\pi e^2}{m^2k^2}P\int_{-\infty}^{+\infty} {f'(v)\over {v}-{\omega_r}/{k}}d{v}=0,
\label{cms1bp}
\end{eqnarray}
\begin{eqnarray}
f'\left ({\omega_r}/{k}\right )=0.
\label{cms2bp}
\end{eqnarray}
The second condition (\ref{cms2bp}) imposes that the phase velocity $\omega_r/k=v_{ext}$ is equal to a velocity where $f(v)$ is extremum ($f'(v_{ext})=0$). The first condition (\ref{cms1bp})  then determines the wavenumber(s) corresponding to marginal stability. It can be written
\begin{eqnarray}
k_{c}=\left (\frac{4\pi e^2}{m^2}\int_{-\infty}^{+\infty} \frac{f'(v)}{v-v_{ext}}\, dv\right )^{1/2}.
\label{cms4p}
\end{eqnarray}
Finally, the pulsation(s) corresponding to marginal stability are $\omega_r=v_{ext}k_c$ and the perturbation behaves like $\delta f\sim e^{-i\omega_r t}$.

\subsection{Particular solutions of $\epsilon(k,\omega)=0$}
\label{sec_monbis}

We can look for a solution of the dispersion relation
$\epsilon(k,\omega)=0$ in the form $\omega=i\omega_{i}$ corresponding to
$\omega_{r}=0$. In that case, the perturbation grows ($\omega_{i}>0$)
or decays ($\omega_{i}<0$) without oscillating. Like in Sec. \ref{sec_mon}, we shall assume that
$f(v)$ is an even function. For $\omega_i>0$, the growth rate is given by
\begin{eqnarray}
1-{4\pi e^2\over m^2 k^{2}}\int_{-\infty}^{+\infty}\frac{v f'(v)}{v^2+\frac{\omega_{i}^2}{k^2}}\, dv=0.
\label{mon2plas}
\end{eqnarray}
For $\omega_{i}<0$, the decay rate is given by
\begin{eqnarray}
1-{4\pi e^2\over m^2 k^{2}}\int_{-\infty}^{+\infty}\frac{v f'(v)}{v^2+\frac{\omega_{i}^2}{k^2}}\, dv-i \frac{8\pi^2 e^2}{m^2k^2} f'\left (\frac{i\omega_{i}}{k}\right )=0.
\label{mon5bb}
\end{eqnarray}
Let us assume that the distribution satisfies
\begin{eqnarray}
\int_{-\infty}^{+\infty}\frac{f'(v)}{v}\, dv>0.
\label{mon2plasb}
\end{eqnarray}
This is not the generic case. For example, the Maxwell distribution does not satisfy this inequality. If inequality (\ref{mon2plasb}) is satisfied, then the marginal mode $\omega=0$ corresponds to the critical wavenumber
\begin{eqnarray}
k_c=\left ({4\pi e^2\over m^2}\int_{-\infty}^{+\infty}\frac{f'(v)}{v}\, dv\right )^{1/2}.
\label{mon2plasc}
\end{eqnarray}
Furthermore, repeating the steps of Sec. \ref{sec_mon}, we obtain for $k\rightarrow k_c$ that\footnote{In the derivation, we have assumed that $f''(0)\neq 0$. If $f''(0)=0$, we need to develop the Taylor expansion to the next order.}
\begin{eqnarray}
\omega_i=\frac{k_c^3m^2}{4\pi^2 e^2f''(0)}\left (1-\frac{k^2}{k_c^2}\right ), \qquad (k\rightarrow k_c).
\label{mon15b}
\end{eqnarray}
This formula leads to the following result. If the distribution is maximum at $v=0$, so that $f''(0)<0$, we find that the mode $\omega=i\omega_i$ is stable for $k<k_c$ and unstable for $k>k_c$. Alternatively, if the distribution is minimum at $v=0$, so that $f''(0)>0$, we find that  the mode $\omega=i\omega_i$ is stable for $k>k_c$ and unstable for $k<k_c$. This result  will be illustrated  in connection to Fig. \ref{PLASMAphasediag} for the symmetric double humped distribution. Note that this result implies that a symmetric distribution satisfying inequality (\ref{mon2plasb}) is always unstable (to some wavenumbers).

\subsection{The Nyquist method}
\label{sec_pny}

To apply the Nyquist method\footnote{Nyquist (1932) introduced  his graphical method in  relation to servomechanism theory. His method was first applied to plasma physics by Harris (1959), Penrose (1960) and Jackson (1960).}, we have to plot the curve
$(\epsilon_{r}(k,\omega_{r}),\epsilon_{i}(k,\omega_{r}))$ parameterized by
$\omega_{r}$ going from $-\infty$ to $+\infty$ (for a fixed wavenumber $k$). Let us consider the
asymptotic behavior for $\omega_{r}\rightarrow \pm\infty$. Since $f(v)$ is positive
and tends to zero for $v\rightarrow \pm\infty$, we conclude that
$\epsilon_{i}(k,\omega_{r})\rightarrow 0$ for $\omega_{r}\rightarrow
\pm\infty$ and that $\epsilon_{i}(k,\omega_{r})<0$ for
$\omega_{r}\rightarrow -\infty$ while $\epsilon_{i}(k,\omega_{r})>0$ for
$\omega_{r}\rightarrow +\infty$.  On the other hand, for
$\omega_{r}\rightarrow \pm\infty$, we obtain at leading order
\begin{eqnarray}
\epsilon_{r}(k,\omega_{r})\simeq 1-{4\pi^2 e^2\rho\over m^2\omega_{r}^{2}}, \qquad (\omega_{r}\rightarrow \pm\infty).
\label{pny3}
\end{eqnarray}
From these results, we conclude that the behavior of the curve close to the point $(1,0)$ is like the one represented in Fig. \ref{maxwellP}. For $\omega_r/k=v_{ext}$, where
$v_{ext}$ is a velocity at which the distribution is extremum
$(f'(v_{ext})=0)$, the imaginary part of the dielectric function
$\epsilon_i(k,k v_{ext})=0$ and the real part of the dielectric function
\begin{equation}
\epsilon_r(k,k v_{ext})= 1-{4\pi^2 e^2\over m^2k^2}\int_{-\infty}^{\infty}\frac{f'(v)}{v-v_{ext}}\, dv.
\end{equation}
Subtracting the value $f'(v_{ext})=0$ in the numerator of the integrand, and  integrating by parts, we obtain
\begin{eqnarray}
\epsilon_{r}(k,k v_{ext})=1+{4\pi^2 e^2\over m^2k^2}\int_{-\infty}^{+\infty} \frac{f(v_{ext})-f(v)}{(v-v_{ext})^{2}}\, dv.
\label{pny4}
\end{eqnarray}
If $v_{Max}$ denotes the velocity corresponding to the global maximum of the distribution, we clearly have
\begin{eqnarray}
\epsilon_{r}(k,k v_{Max})> 1.
\label{pny5}
\end{eqnarray}

\subsubsection{Single-humped distributions}
\label{sec_pshr}

Let us assume that the distribution $f(v)$ has a single maximum at
$v=v_{0}$ (so that $f'(v_0)=0$) and tends to zero for $v\rightarrow
\pm \infty$. Then, the Nyquist curve cuts the $x$-axis ($\epsilon_{i}(k,\omega_{r})$ vanishes) at the limit point
$(1,0)$ when $\omega_{r}\rightarrow \pm \infty$ and at the point
$(\epsilon_{r}(k,kv_0),0)$ when $\omega_{r}/k=v_{0}$. Due to its behavior
close to the limit point $(1,0)$, the fact that it rotates in the
counterclockwise sense, and the property that $\epsilon_r(k,k v_0)>1$ according to Eq. (\ref{pny5}),  the Nyquist curve must necessarily behave like in
Fig. \ref{maxwellP}.  Therefore, the Nyquist curve starts on the
real axis at $\epsilon_r(k,\omega_r) =1$ for $\omega_r \rightarrow
-\infty$, then going in counterclockwise sense it crosses the real
axis at the point $\epsilon_r(k,k v_0)>1$ and returns on the real axis at
$\epsilon_r(k, \omega_r) =1$ for $\omega_r \rightarrow +
\infty$. Therefore, it cannot encircle the origin. According to the Nyquist criterion exposed in
Sec. \ref{sec_n}, we conclude that a single-humped distribution
is always linearly dynamically stable\footnote{The fact that a single-humped distribution is always stable is known as Gardner (1963)'s theorem; see also Jackson (1960).}.

\begin{figure}[htbp]
\centerline{
\includegraphics[width=8cm,angle=0]{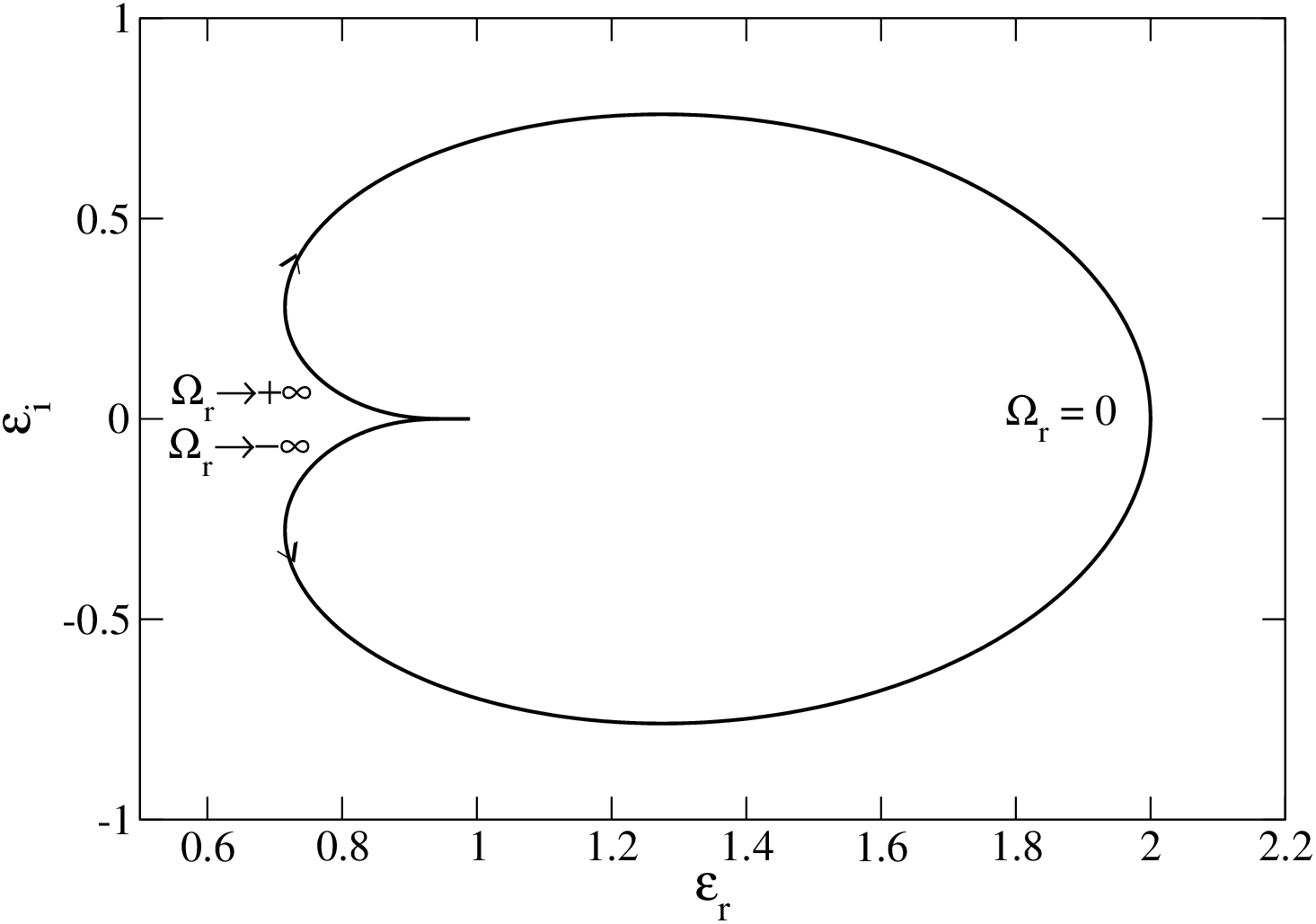}
} \caption[]{Nyquist curve for the Maxwell distribution. The DF is always stable.}
\label{maxwellP}
\end{figure}

Let us specifically consider the Maxwell distribution (\ref{md1}) for illustration. The dielectric function (\ref{pvp3}) associated to the Maxwell distribution is
\begin{eqnarray}
\epsilon(k,\omega)=1+\frac{4\pi e^2}{m^2k^2}\left (\frac{\beta}{2\pi}\right )^{1/2}\rho\int_{C}\frac{\beta v}{v-\frac{\omega}{k}}e^{-\beta {v^{2}\over 2}}\ dv.
\label{Pmd3}
\end{eqnarray}
Introducing the Debye wavenumber (\ref{pep6}), it can be rewritten
\begin{eqnarray}
\epsilon({k},\omega)=1+{k_{D}^{2}\over k^{2}} W\biggl ({\sqrt{\beta}\omega\over k}\biggr ).
 \label{Pmd4}
\end{eqnarray}
It will be convenient in the following to work with dimensionless quantities.
We introduce the dimensionless wavenumber and the dimensionless pulsation
\begin{eqnarray}
\eta=\frac{k_D^2}{k^2}, \qquad \Omega=\frac{\omega}{\omega_p}.
\label{Pmn1}
\end{eqnarray}
Noting that $\sqrt{\beta}\omega/k=\sqrt{\eta}\Omega$, the dielectric function (\ref{Pmd4}) can be rewritten
\begin{eqnarray}
\epsilon({\eta},\Omega)=1+\eta W(\sqrt{\eta}\Omega).
 \label{Pmn3}
\end{eqnarray}
When $\Omega_{i}=0$, the real and imaginary parts of the dielectric function  $\epsilon(\eta,\Omega_{r})=\epsilon_{r}(\eta,\Omega_{r})+i\epsilon_{i}(\eta,\Omega_{r})$ are given by
\begin{equation}
\label{Pmn4}
\epsilon_{r}(\eta,\Omega_{r})=1+\eta W_{r}\left (\sqrt{\eta}\Omega_r\right ),
\end{equation}
\begin{equation}
\label{Pmn5}
\epsilon_{i}(\eta,\Omega_{r})=\eta W_{i}\left (\sqrt{\eta}\Omega_r\right ).
\end{equation}
The Nyquist curve for the Maxwell distribution is represented in Fig. \ref{maxwellP}.

On the other hand, if we consider a polytrope of index $n=1/2$ (waterbag distribution) as in Sec. \ref{sec_fermi}, we find that the dielectric function is given by
\begin{eqnarray}
\epsilon(k,\omega)=1+\frac{k_D^2}{3k^2}W_{1/2}\left (\frac{\omega}{k\sqrt{T}}\right ),
 \label{dielanew}
\end{eqnarray}
where $W_{1/2}(x)$ is given by Eq. (\ref{dielb}).
The condition $\epsilon(k,\omega)=0$ determines the dispersion relation. We get
\begin{eqnarray}
\omega^2=\omega_p^2+3Tk^2.
 \label{dielcnew}
\end{eqnarray}
We note that, for the water-bag distribution, the general asymptotic
behavior (\ref{pvp4}) becomes exact. We also note that for the
specific index $n=1/2$ ($\gamma=3$), the dispersion relation in a
collisionless stellar system takes the same form as in a gas (see
Sec. \ref{sec_pep}).

\subsubsection{Double-humped distributions}
\label{sec_dhhr}

Let us consider a double-humped distribution with a global maximum at
$v_{Max}$, a minimum at $v_{min}$ and a local maximum at $v_{max}$. We
assume $v_{Max}<v_{min}<v_{max}$. The Nyquist curves starts at
$(1,0)$, progresses in the counterclockwise sense and crosses the
$x$-axis at $\epsilon_{r}(v_{Max})>1$, then at $\epsilon_{r}(v_{min})$
and $\epsilon_{r}(v_{max})$. We can convince ourselves by making drawings of the following results. If

$(+++)$: $\epsilon_{r}(v_{Max})>0$, $\epsilon_{r}(v_{min})>0$, $\epsilon_{r}(v_{max})>0$,

$(+--)$: $\epsilon_{r}(v_{Max})>0$, $\epsilon_{r}(v_{min})<0$, $\epsilon_{r}(v_{max})<0$,

$(++-)$: $\epsilon_{r}(v_{Max})>0$, $\epsilon_{r}(v_{min})>0$, $\epsilon_{r}(v_{max})<0$,

\noindent the Nyquist curve does not encircle the origin so the
system is stable. If

$(+-+)$: $\epsilon_{r}(v_{Max})>0$, $\epsilon_{r}(v_{min})<0$, $\epsilon_{r}(v_{max})>0$,

\noindent the Nyquist curve rotates one time
around the origin so that there is one mode of instability. Since
$\epsilon_{r}(v_{Max})>0$ there is no mode of marginal stability with
$\omega_{r}/k=v_{Max}$. Cases $(+++)$, $(+-+)$ and $(+--)$ are
observed in Sec. \ref{sec_rass} for an asymmetric double-humped
distribution made of two Maxwellians.

If the double-humped distribution is symmetric with respect to the
origin with two maxima at $\pm v_{*}$ and a minimum at $v=0$, we get
the same results as above with the additional properties
$\epsilon_{r}(v_{Max})=\epsilon_{r}(v_{max})=\epsilon_{r}(v_{*})>1$
and $\epsilon_{r}(v_{min})=\epsilon_{r}(0)$. We have only two cases
$(+++)$ and $(+-+)$. They are observed in Sec. \ref{sec_rsh} for a
symmetric double-humped distribution made of two Maxwellians. Since
$\epsilon_{r}(v_{*})>0$, there is no mode of marginal stability with
$\omega_{r}/k=\pm v_{*}$.

It can be shown that a plasma is unstable (to some wavelengths) iff $f(v)$ has a minimum at a value $v=v_{min}$ such that
\begin{eqnarray}
\int_{-\infty}^{+\infty} {{f'(v)}\over v-v_{min}}dv=\int_{-\infty}^{+\infty} \frac{f(v)-f(v_{min})}{(v-v_{min})^{2}}\, dv>0.
\label{penrose}
\end{eqnarray}
This result was proven by Penrose (1960) and it is sometimes called the Penrose criterion\footnote{In fact, we found that an equivalent criterion was obtained at the same period by Noerdlinger (1960).}.
This criterion can be deduced from the Nyquist method as follows. A double-humped distribution $f(v)$ is unstable if there exists a range of $k$ such that we are in the situation $(+-+)$, i.e. $\epsilon_r(v_{min})<0$ and $\epsilon_r(v_{max})>0$. The first condition can be satisfied (for sufficiently small $k$) iff  condition (\ref{penrose}) is realized, which is the Penrose criterion. Then, the range of unstable wavenumbers is
\begin{eqnarray}
\frac{4\pi e^2}{m^2}\int_{-\infty}^{+\infty} {{f'(v)}\over v-v_{max}}dv<k^2<\frac{4\pi e^2}{m^2}\int_{-\infty}^{+\infty} {{f'(v)}\over v-v_{min}}dv.\nonumber\\
\label{penroserange}
\end{eqnarray}
If the first integral is negative, the range of unstable wavenumbers is $k<k_c^{(min)}$. This corresponds to Fig. \ref{pdg1} and to the region $y>y_c$ in Fig. \ref{pdg2}. If the first integral is positive, the range of unstable wavenumbers is $k_c^{(max)}<k<k_c^{(min)}$. This corresponds to the region $y_*<y<y_c$ in Fig. \ref{pdg2} (indeed, $y_c$ corresponds precisely to the case where the first integral becomes equal to zero).

\subsection{The symmetric double-humped distribution}
\label{sec_rsh}

Let us now consider the symmetric double-humped distribution (\ref{e1}). In the plasma case, the dielectric function can be written
\begin{eqnarray}
\label{rsh1}
\epsilon(\eta,\Omega)= 1+\frac{\eta}{2} \left [ W(\sqrt{\eta}\Omega-\sqrt{y}))+W(\sqrt{\eta}\Omega+\sqrt{y}))  \right ],\nonumber\\
\end{eqnarray}
where $\eta$ (wavelength) and  $\Omega$ (pulsation) are defined by Eq. (\ref{Pmn1}) and $y$ (separation) is defined by Eq. (\ref{e5}).
The condition of marginal stability corresponds to $\epsilon(k,\omega)=0$ and $\omega_i=0$. The condition $\epsilon_i(k,\omega_r)=0$ is equivalent to $f'(\omega_r/k)=0$ so that the phase velocity   $\omega_r/k=v_0$ corresponds to the velocities where the distribution is extremum. Then, the wavenumbers at which the distribution  is marginally stable are obtained by solving $\epsilon_r(k,\omega_r=k v_0)=0$. Proceeding as in Sec. \ref{sec_vhq} and introducing the parameter $x=V_0/V_a$, we find that the equations determining the critical wavenumbers
$\eta_c(y)$ are given by
\begin{equation}
\label{rsh2}
y=\frac{1}{2x}\ln\left (\frac{1+x}{1-x}\right ),
\end{equation}
\begin{equation}
\label{rsh3}
\eta=\frac{-2}{\left [ W_{r}(\sqrt{y}(x - 1))+W_{r}(\sqrt{y}(x+ 1))   \right ]}.
\end{equation}
We note that only the sign in Eq. (\ref{rsh3}) changes with respect to the study of the gravitational case, so we can readily adapt the results of Sec. \ref{sec_vhq} to the present situation by simply reverting the sign. For  $\sqrt{\eta}\Omega_{r}=\pm V_*$, corresponding to $x\neq 0$, there is no physical solution to Eqs. (\ref{rsh2})-(\ref{rsh3}) with positive  $\eta>0$. Therefore, in the plasma case, there is no marginal mode with non zero pulsation for the symmetric double-humped distribution (in agreement with the general discussion of Sec. \ref{sec_dhhr}). We now consider the marginal mode with $\Omega_{r}=0$. This corresponds to the ``degenerate" solution  $x=0$ (for any $y$) for which Eqs. (\ref{rsh2})-(\ref{rsh3}) reduce to
\begin{equation}
\label{rsh4}
\eta_c^{(0)}(y)=\frac{-1}{W_{r}(\sqrt{y})}.
\end{equation}
According to Fig. \ref{w3}, physical solutions exist only for $y\ge
y_{max}=z_{c}^{2}=1.708$. We note that the range of parameters that was
forbidden in the gravitational case is now allowed in the plasma case
and vice versa. For $y<y_{max}$, the system is stable. This result is to be expected
since, for $y=0$, the distribution (\ref{e1}) reduces to the Maxwellian that is stable
for a repulsive interaction. We now consider the range $y\in \lbrack
y_{max},+\infty\lbrack$.
For $y\rightarrow y_{max}$, we have
\begin{equation}
\label{rsh5}
\eta_{c}^{(0)}(y)\sim \frac{2y_{max}}{y-y_{max}}, \qquad (y\rightarrow y_{max}).
\end{equation}
On the other hand, the curve $\eta_{c}^{(0)}(y)$ has a minimum at $(4.511,3.512)$ (see Appendix A of Chavanis \& Delfini 2009). Finally,
for $y\rightarrow +\infty$, we have
\begin{equation}
\label{rsh6}
\eta_{c}^{(0)}(y)\sim y\rightarrow +\infty, \qquad (y\rightarrow +\infty).
\end{equation}
Again this result is expected because, for $y\rightarrow
+\infty$, the two humps do not ``see" each other and behave as two
independent single-humps distributions that are stable in the repulsive
case.

The critical wavenumber  $\eta_c^{(0)}(y)$ corresponding to marginal stability determined previously is represented as a function of the separation $y$ in Fig. \ref{PLASMAphasediag}. We have also plotted  the line $y=1$. On the left of this line, the distribution has a single maximum at $V_0=0$ and on the right, the distribution has two maxima at $V_0=\pm V_*$ and a minimum at $V_0=0$. In order to investigate the stability of the solutions in the different regions, we have  used the Nyquist criterion.

\begin{figure}[htbp]
\centerline{
\includegraphics[width=8cm,angle=0]{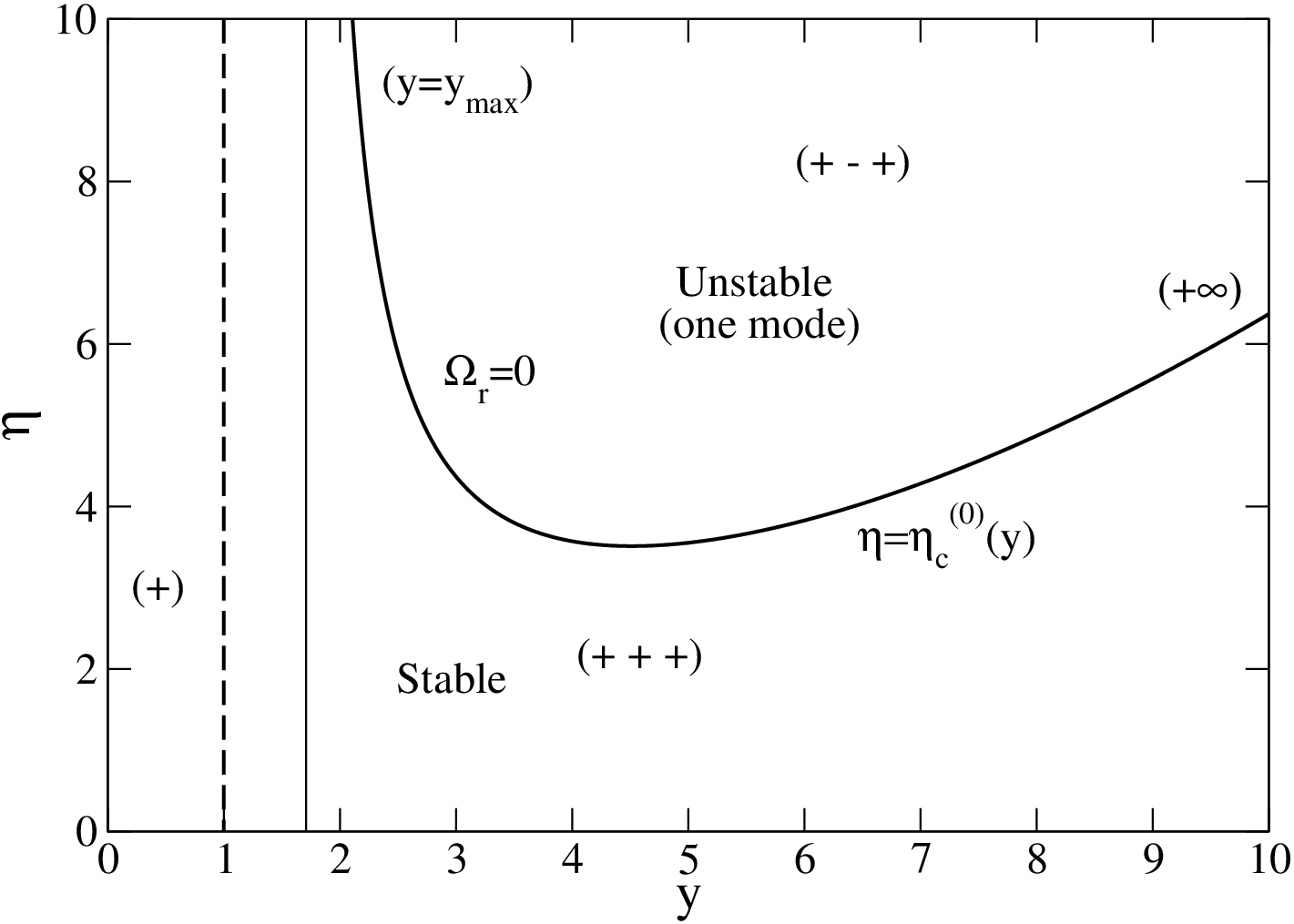}
} \caption[]{Stability diagram of the symmetric double-humped
distribution (\ref{e1}) in the case of plasmas.}
\label{PLASMAphasediag}
\end{figure}

For $y<1$, the DF has only one maximum at $V_0=0$. There is no marginal mode and the distribution is always
stable (see Fig. \ref{doublesym_y0.5_n6P}).

\begin{figure}[htbp]
\centerline{
\includegraphics[width=8cm,angle=0]{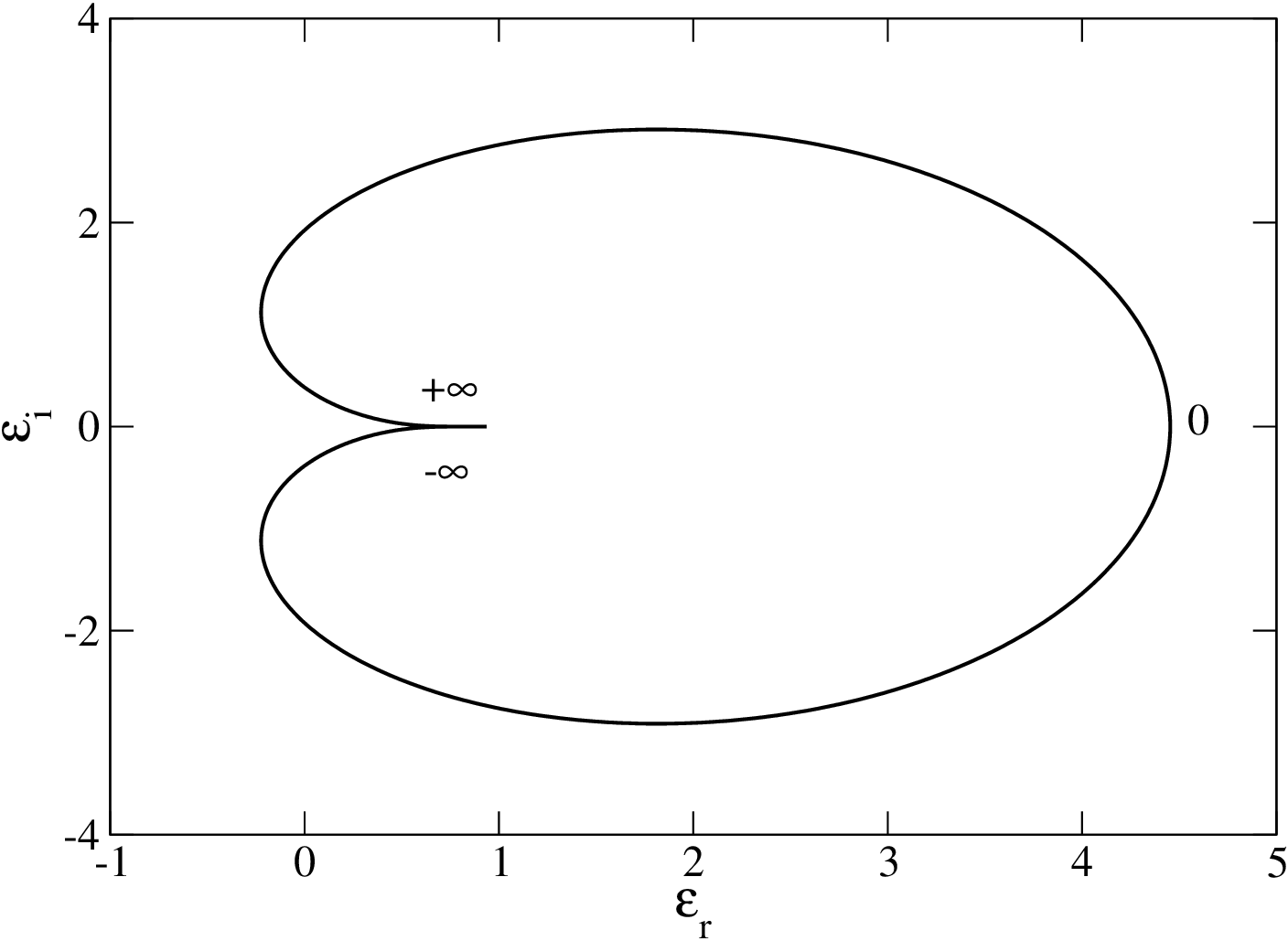}
} \caption[]{Nyquist curve for $y<1$ (specifically $y=0.5$ and $\eta=6$).  The DF is stable for any perturbation. Case (+).}
\label{doublesym_y0.5_n6P}
\end{figure}

For $1<y<y_{max}$, the DF has a minimum at $V_0=0$ and two maxima at $\pm V_*$. There is no marginal mode and the distribution is always
stable (see Figs. \ref{doublesym_y1.2_n6P}).

\begin{figure}[htbp]
\centerline{
\includegraphics[width=8cm,angle=0]{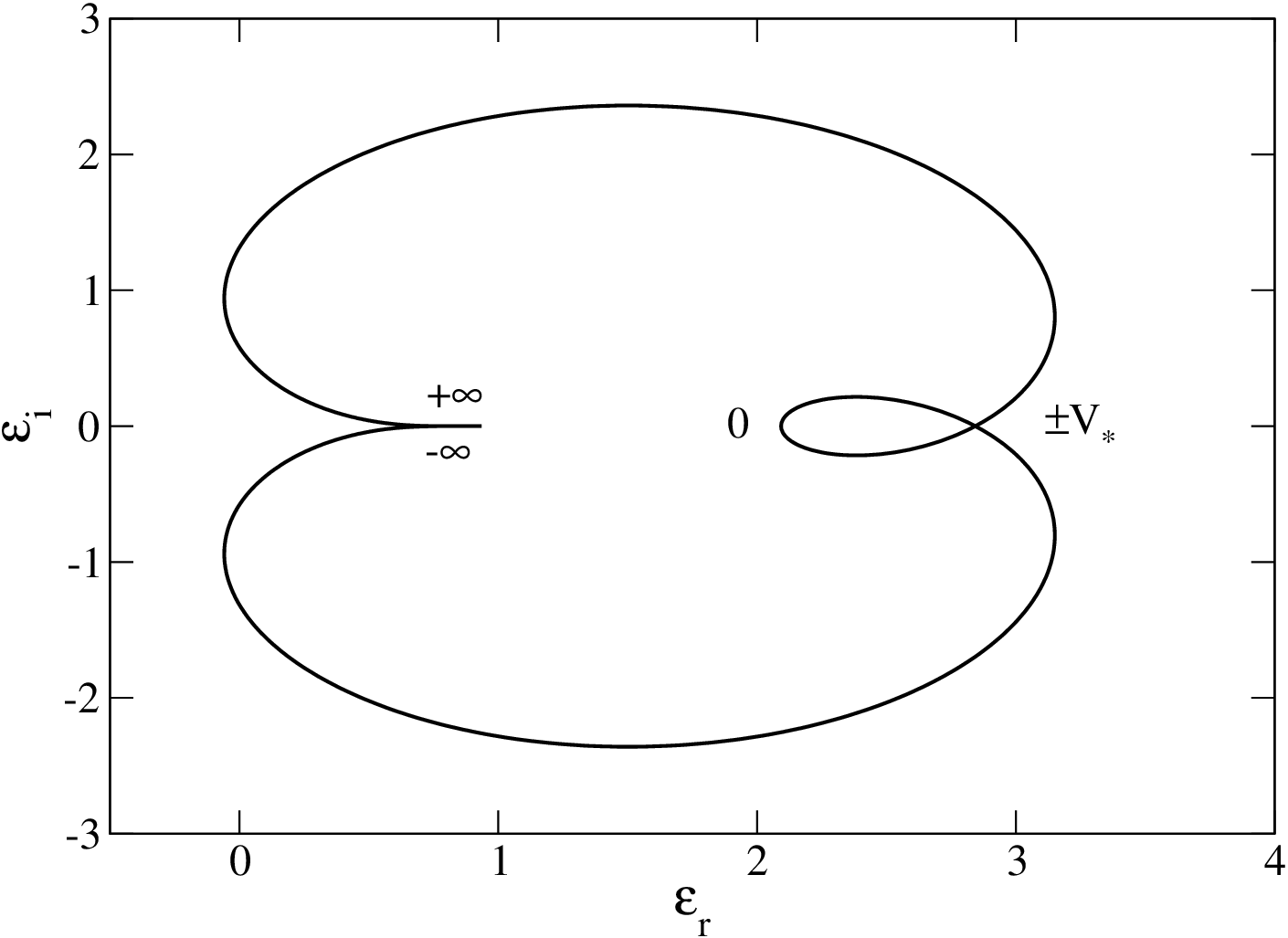}
} \caption[]{Nyquist curve for $1<y<y_{max}$ (specifically $y=1.2$ and $\eta=6$). The DF is stable for any perturbation. Case (+ + +).}
\label{doublesym_y1.2_n6P}
\end{figure}

For $y>y_{max}$, the distribution has
a minimum at $V_0=0$ and two maxima at $\pm V_*$. There exists one wavenumber $\eta_{c}^{(0)}$ at which
the distribution is marginally stable.  For $\eta=\eta_{c}^{(0)}$, the marginal
perturbation does not propagate ($\Omega_r=0$). By considering the
Nyquist curves in this region (see Figs. \ref{doublesym_y5_n2P}-\ref{doublesym_y5_n6P}), we
find that the DF is stable for $\eta<\eta_{c}^{(0)}$ and unstable
for $\eta>\eta_{c}^{(0)}$.

\begin{figure}[htbp]
\centerline{
\includegraphics[width=8cm,angle=0]{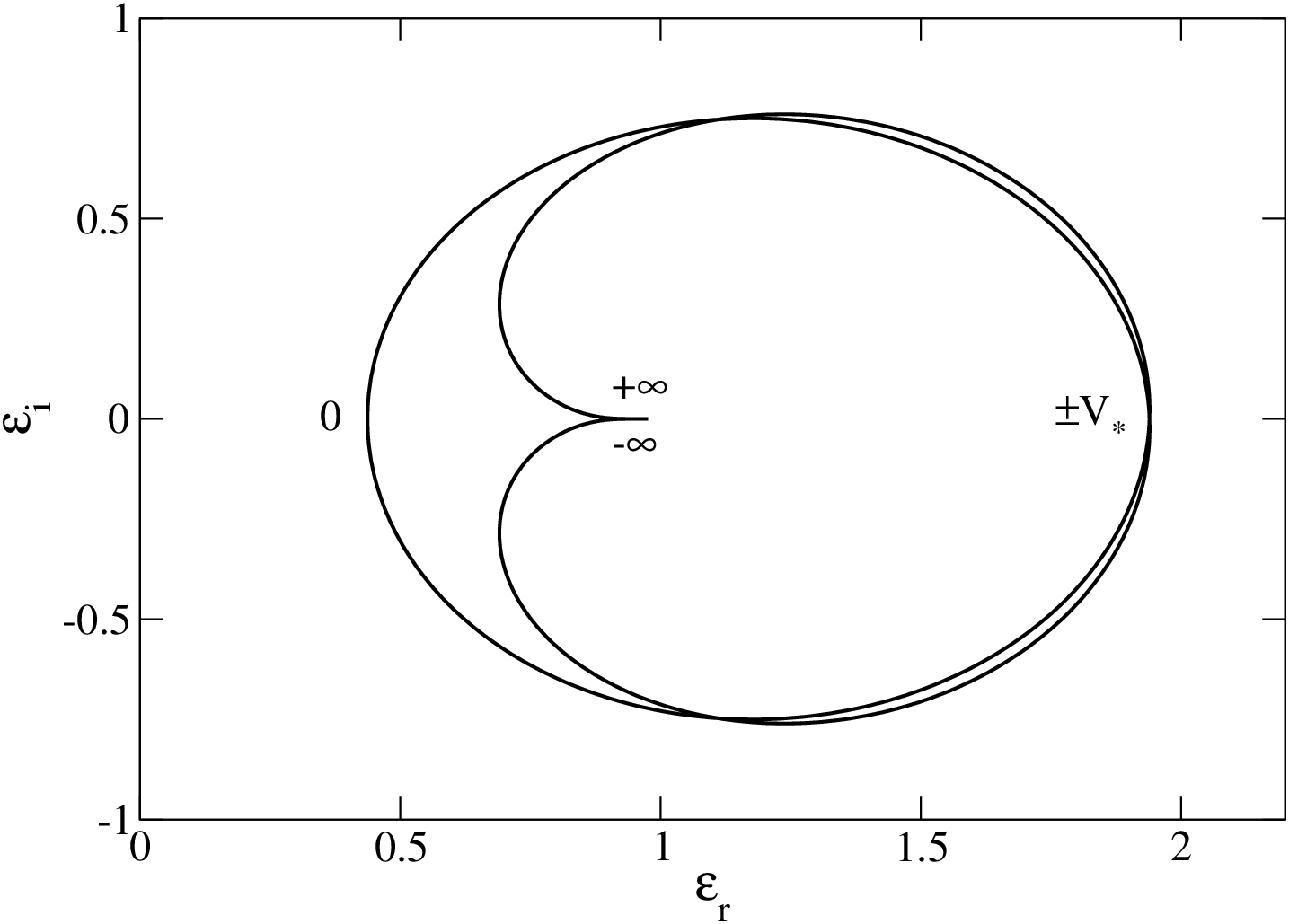}
} \caption[]{Nyquist curve for $y>y_{max}$ and $\eta<\eta_{c}^{(0)}$ (specifically $y=5$ and $\eta=2$). The DF is stable for this range of wavenumbers. Case (+ + +).}
\label{doublesym_y5_n2P}
\end{figure}

\begin{figure}[htbp]
\centerline{
\includegraphics[width=8cm,angle=0]{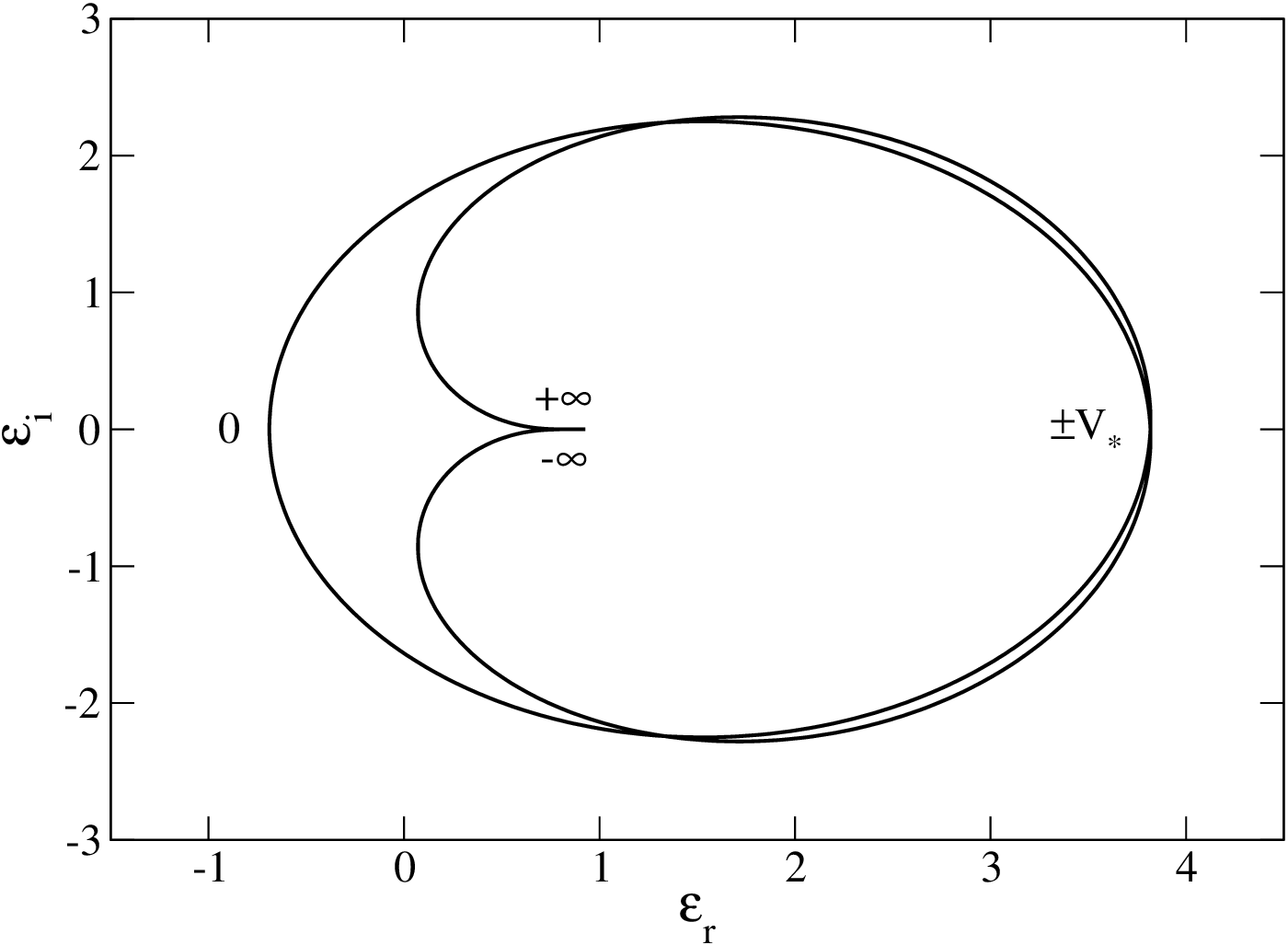}
} \caption[]{Nyquist curve for $y>y_{max}$ and $\eta>\eta_{c}^{(0)}$ (specifically $y=5$ and $\eta=6$). There is one mode of instability in this range of wavenumbers. Case (- + -).}
\label{doublesym_y5_n6P}
\end{figure}

Let us make some remarks:

1. If we consider all possible perturbations, Fig. \ref{PLASMAphasediag} shows that a symmetric double-humped distribution is stable if $y<y_{max}$ (i.e. $v_a<1.307 \, T^{1/2}$) and unstable if $y>y_{max}$ (i.e. $v_a>1.307 \, T^{1/2}$). Therefore, instability occurs when the drift velocity $v_a$ is sensibly larger than the thermal speed $\sqrt{T}$ so that the two humps are well separated. In particular, at $T=0$, a double-humped distribution is always unstable to some wavelengths (see Appendix \ref{sec_cold}). More precisely, a symmetric double-humped distribution with $y>y_{max}$ is stable for perturbations with  wavenumbers $k>(k_c)_*$ (corresponding to the solid line) and unstable for perturbations with wavenumbers $k<(k_c)_*$. The critical wavenumber $(k_c)_*$ corresponds to the marginal  perturbation for which the phase velocity $\omega/k$ coincides with the {\it minimum} of the velocity distribution: $v_0=0$.

2. For $y>y_{max}$, the mode $\omega=i\omega_i$ is stable for $\eta<\eta_c^{(0)}$ and it becomes unstable when we increase $\eta$ above the marginal line  $\eta=\eta_c^{(0)}$. This is consistent with the general result (\ref{mon15b}) since, for $y>1$, the DF is {\it minimum} at $v_0=0$.

3. For a double-humped distribution, the critical wavelength is infinite for  $v_a\le \sqrt{y_{max}T}$ (stable), then decreases, reaches a minimum and increases again. By contrast, if the plasma is modeled as a contrastreaming gas, the critical wavelength is infinite for  $v_a< c_s$ (stable), then jumps discontinuously to zero at $v_a=c_s$ and increases  for $v_a>c_s$ (Ikeuchi et al. 1974).

\subsection{The asymmetric double-humped distribution}
\label{sec_rass}

The dielectric function associated to the asymmetric double-humped
distribution (\ref{ae1}) in the plasma case is
\begin{eqnarray}
\label{rass1}
\epsilon(\eta,\Omega)= 1+\frac{\eta}{1+\Delta} \biggl [ W(\sqrt{\eta}\Omega - \sqrt{y}))\nonumber\\
+\Delta W(\sqrt{\eta}\Omega + \sqrt{y}))   \biggr ].
\end{eqnarray}
Proceeding as in Sec. \ref{sec_vha} and introducing the parameters
$x=V_0/V_a$ and $y=V_a^2$, we find that the equations determining the
critical wavenumbers $\eta_c(y)$ are given by
\begin{equation}
\label{rass2}
y=\frac{1}{2x}\ln\left (\frac{1+x}{1-x}\right
)+\frac{\ln(\Delta)}{2x},
\end{equation}
\begin{equation}
\label{rass3}
\eta=-\frac{1+\Delta}{\left [ W_{r}(\sqrt{y}(x - 1))+\Delta W_{r}(\sqrt{y}(x+ 1))   \right ]}.
\end{equation}
Equation (\ref{rass2}) determines the extrema of
the distribution $f(v)$ and Eq. (\ref{rass3}) determines the wavenumbers
corresponding to the modes of marginal stability. As in
Sec. \ref{sec_vha}, the curve $\eta_c(y)$ can be obtained by varying
$x$ between $-1$ and $+1$. In the plasma case, there exists
physical solutions with positive $\eta$ only for $0<x\le x_c$.
For $x\rightarrow 0^+$, using Eq. (\ref{rass2}), we find that
$y\rightarrow +\infty$. Then, we get
\begin{equation}
\label{rass4}
\eta_c^{(p)}(y)\sim y\rightarrow  +\infty, \qquad (y\rightarrow +\infty).
\end{equation}
For $x\rightarrow x_c$, we find that $y\rightarrow y_c$ and
\begin{equation}
\label{rass5}
\eta_c^{(s)}(y)\propto \frac{1}{y-y_c}\rightarrow +\infty, \qquad (y\rightarrow y_c),
\end{equation}
where $s=p$ if $\Delta<\Delta_{*}$ and $s=+$ if
$\Delta>\Delta_{*}$. For $\Delta<\Delta_{*}$, we get only one marginal branch
$\eta_{c}^{(p)}(y)$ corresponding to the mode $\Omega_r=V_{p}/\sqrt{\eta_c}>0$ (see
Fig. \ref{pdg1}). For $\Delta>\Delta_{*}$, we get two marginal
branches $\eta_{c}^{(p)}(y)$ and $\eta_{c}^{(+)}(y)$ corresponding to
the modes $\Omega_r=V_{p}/\sqrt{\eta_c}>0$ and $\Omega_r=V_{+}/\sqrt{\eta_c}>0$ (see
Fig. \ref{pdg2}).  They connect each other at $(y_{*},\eta_{*})$
corresponding to $x=x_{*}$. At that point they touch the line
$y=y_{*}$ separating distributions with one or two maxima.

\begin{figure}[htbp]
\centerline{
\includegraphics[width=8cm,angle=0]{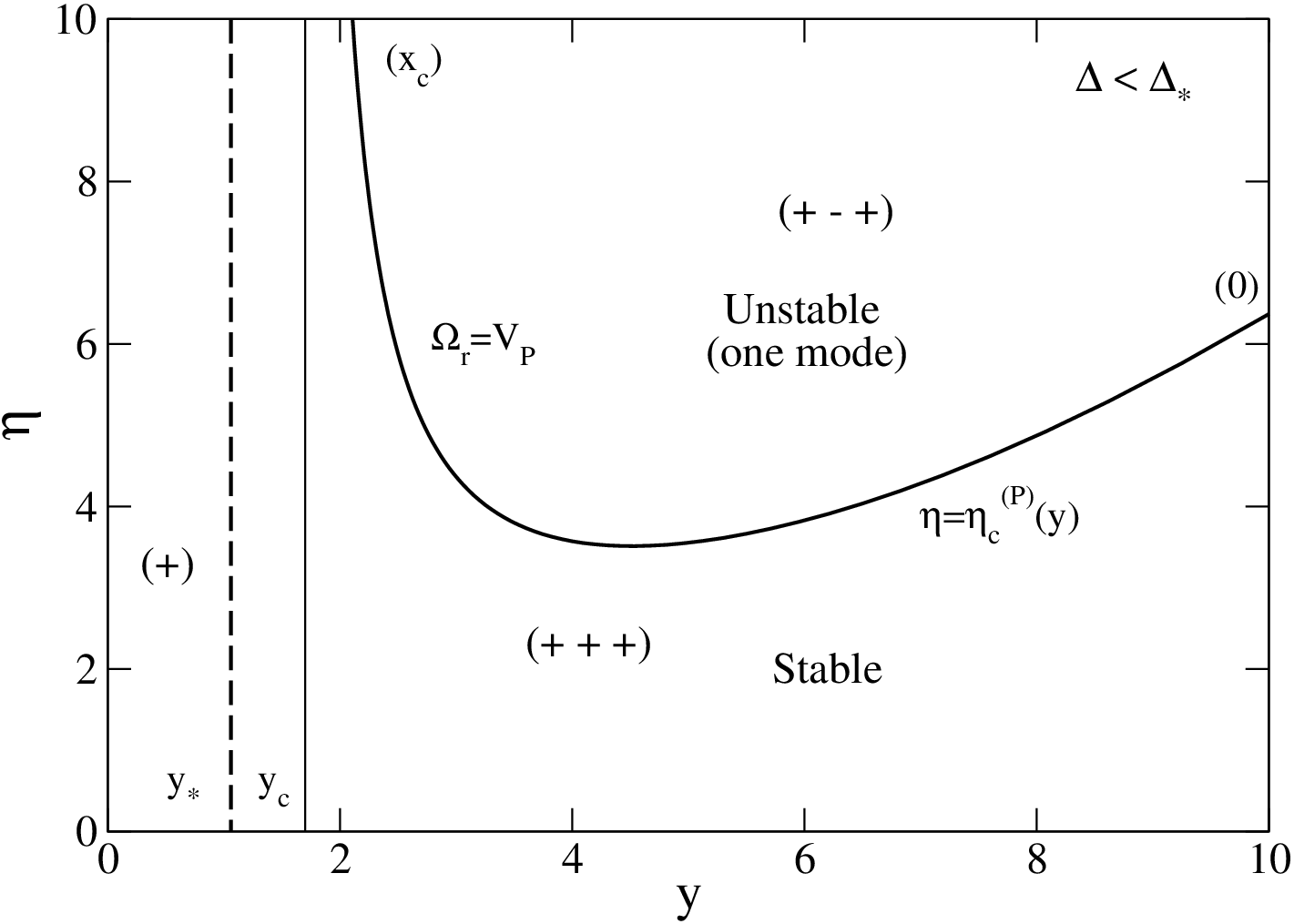}
} \caption[]{Stability diagram of the asymmetric double-humped distribution for $\Delta<\Delta_*$ (specifically $\Delta=1.02$). There exists one marginal branch $\eta_c^{(p)}(y)$.}
\label{pdg1}
\end{figure}

\begin{figure}[htbp]
\centerline{
\includegraphics[width=8cm,angle=0]{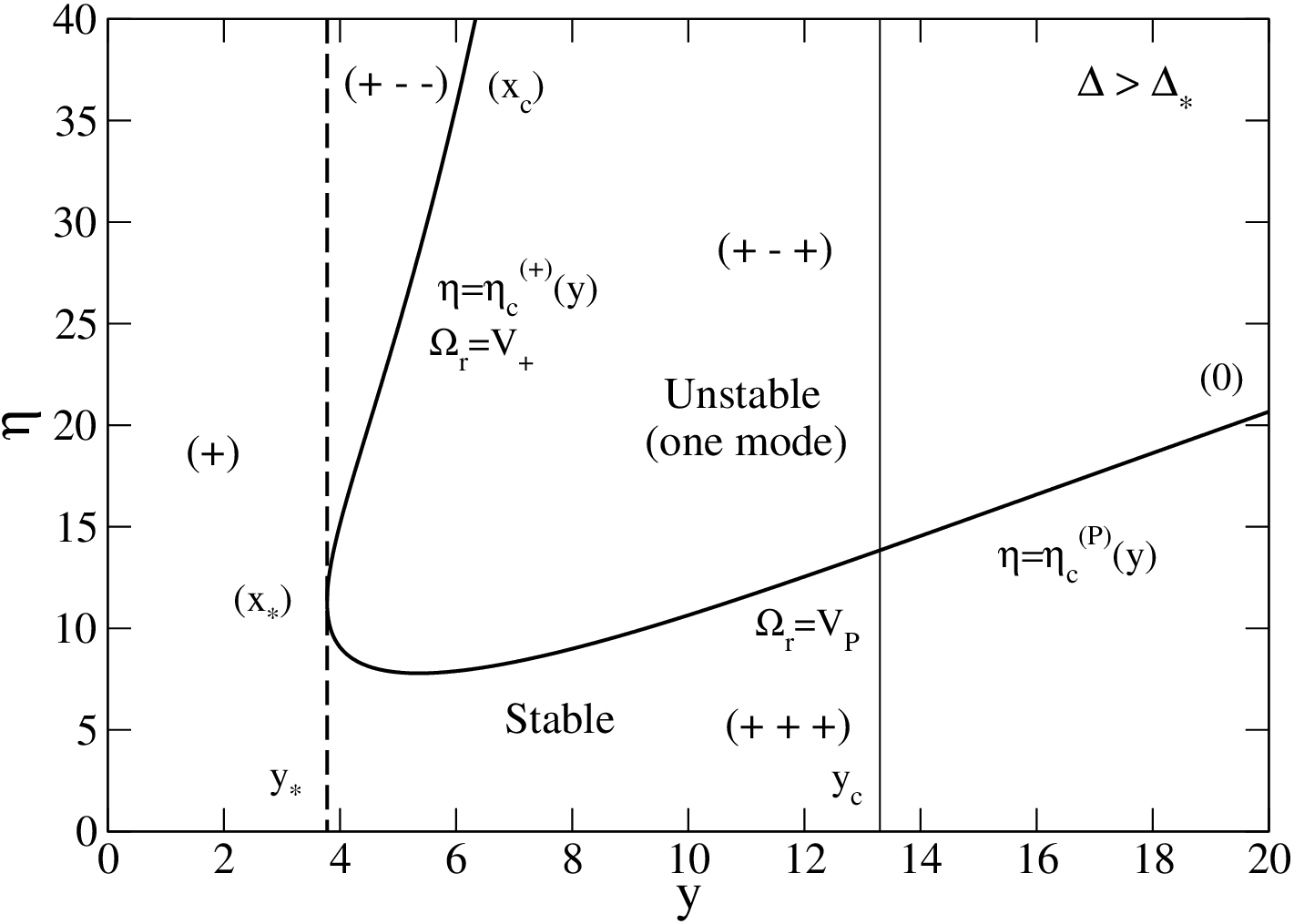}
} \caption[]{Stability diagram of the asymmetric double-humped distribution for $\Delta>\Delta_*$ (specifically $\Delta=50$). There exists two marginal branches $\eta_c^{(p)}(y)$ and $\eta_c^{(+)}(y)$.}
\label{pdg2}
\end{figure}

The stability diagrams corresponding to the asymmetric double-humped
distribution with $\Delta<\Delta_*$ and $\Delta>\Delta_*$ are represented in
Figs. \ref{pdg1} and \ref{pdg2}. To investigate the stability of the solutions in the different regions of the parameter space, we have used the Nyquist criterion. If $\Delta<\Delta_*$, the description of the stability diagram can be made in continuity with the case of a symmetric double-humped distribution: for $y<y_{c}$ (i.e. $v_a^2<y_c(\Delta)T$), the plasma is stable to all perturbations; for  $y>y_{c}$ (i.e. $v_a^2>y_c(\Delta)T$), the plasma is stable for  $k>k_c^{(p)}(v_a)$ and unstable for $k<k_c^{(p)}(v_a)$. If $\Delta>\Delta_*$,  the stability diagram changes: for $y<y_{*}$ (i.e. $v_a^2<y_*(\Delta)T$), the plasma is stable to all perturbations; for  $y_{*}<y<y_c$ (i.e. $y_*(\Delta)T<v_a^2<y_c(\Delta)T$), the plasma is stable for  $k>k_c^{(p)}(v_a)$, unstable for $k_c^{(+)}(v_a)<k<k_c^{(p)}(v_a)$ and stable for  $k<k_c^{(+)}(v_a)$: this is similar to a re-entrant phase; for  $y>y_c$ (i.e. $v_a^2>y_c(\Delta)T$), the plasma is stable for  $k>k_c^{(p)}(v_a)$ and  unstable for $k<k_c^{(p)}(v_a)$. In conclusion, instability occurs when the drift velocity is sensibly larger than the thermal speed so that the two humps are well separated. However, the precise threshold changes whether the asymmetry is smaller or larger than $\Delta_*$. For $\Delta<\Delta_*$, the plasma is stable iff $v_a^2<y_c(\Delta)T$ and for  $\Delta>\Delta_*$, the plasma is stable iff $v_a^2<y_*(\Delta)T$.

\begin{figure}[htbp]
\centerline{
\includegraphics[width=8cm,angle=0]{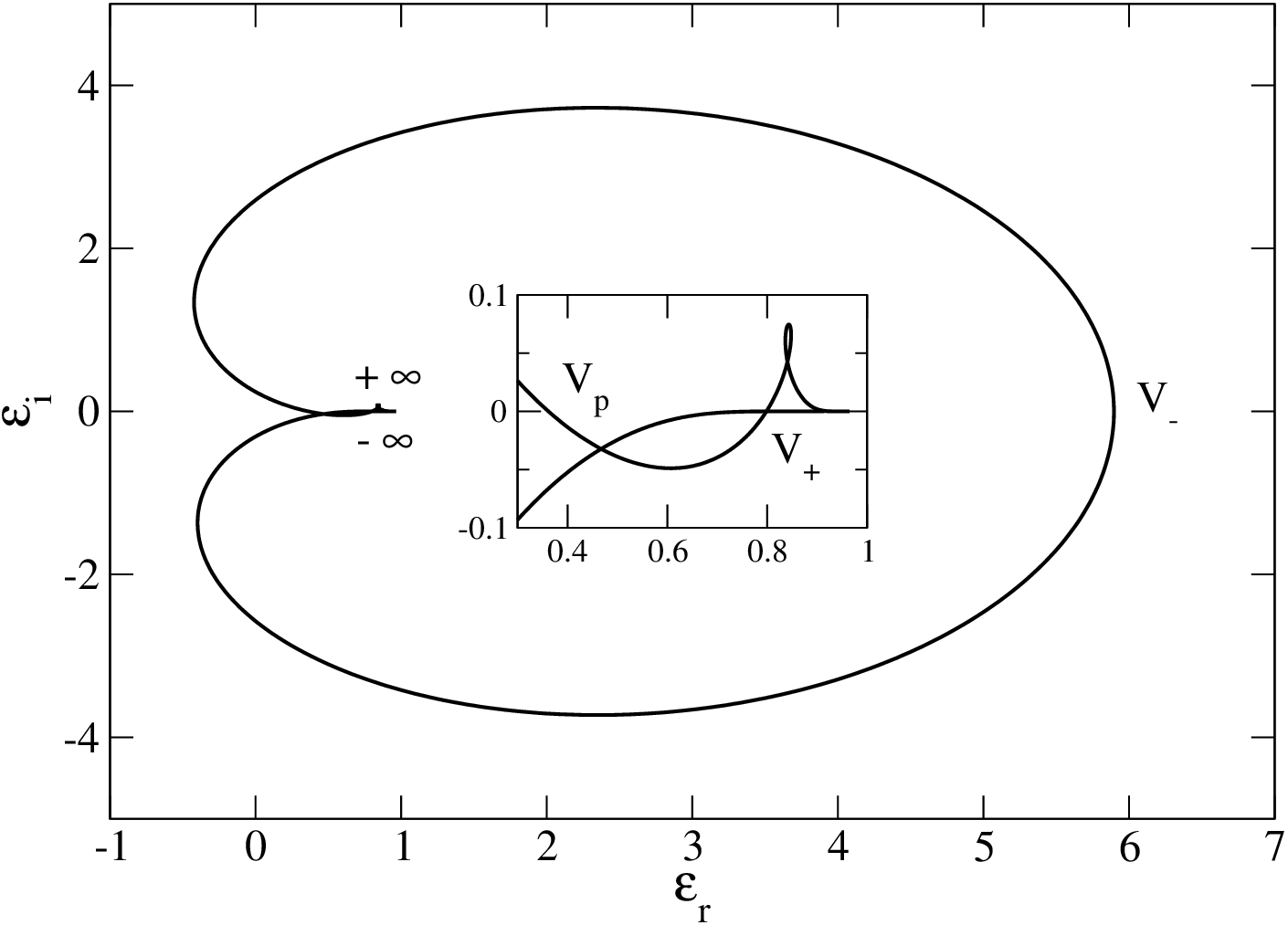}
} \caption[]{$\Delta>\Delta_*$: Nyquist curve for $y_*<y<y_c$ and $\eta<\eta_c^{(p)}$
(specifically $\Delta=50$, $y=5$ and $\eta=5$). The DF has a global maximum at $V_{-}$, a minimum at $V_p$ and a local maximum at $V_{+}$. The DF is stable (with respect to
this perturbation) because the Nyquist curve does not encircle the
origin.  Case (+ + +).}
\label{data_delta50_y5_n5}
\end{figure}

\begin{figure}[htbp]
\centerline{
\includegraphics[width=8cm,angle=0]{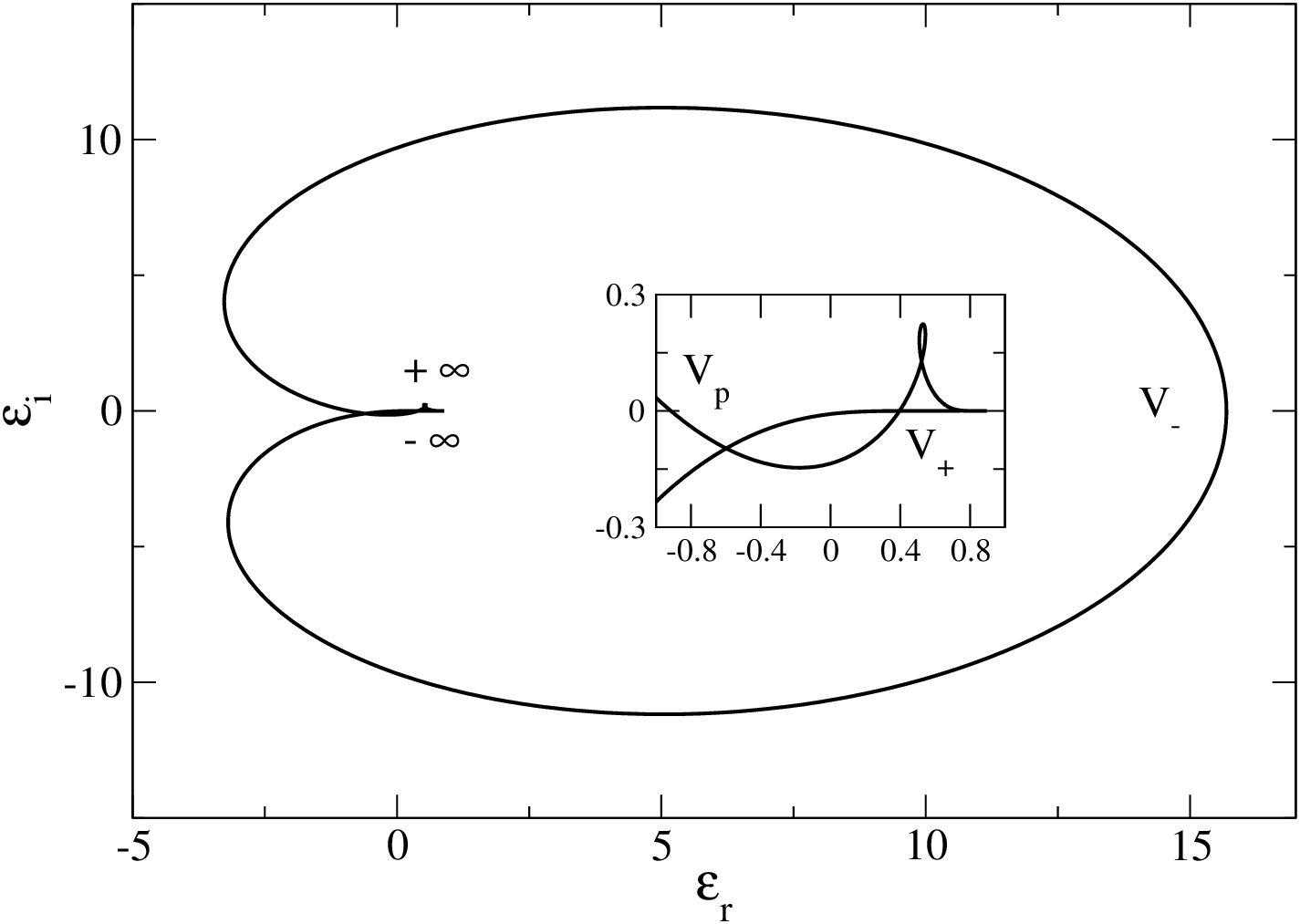}
} \caption[]{$\Delta>\Delta_*$: Nyquist curve for $y_*<y<y_c$ and $\eta_c^{(p)}<\eta<\eta_c^{(+)}$
(specifically $\Delta=50$, $y=5$ and $\eta=15$). The DF has a global maximum at $V_{-}$, a minimum at $V_p$ and a local maximum at $V_{+}$. The DF is unstable (with respect to
this perturbation) because the Nyquist curve encircles the
origin once. There is $N=1$ unstable mode. Case (+ - +).}
\label{data_delta50_y5_n15}
\end{figure}

\begin{figure}[htbp]
\centerline{
\includegraphics[width=8cm,angle=0]{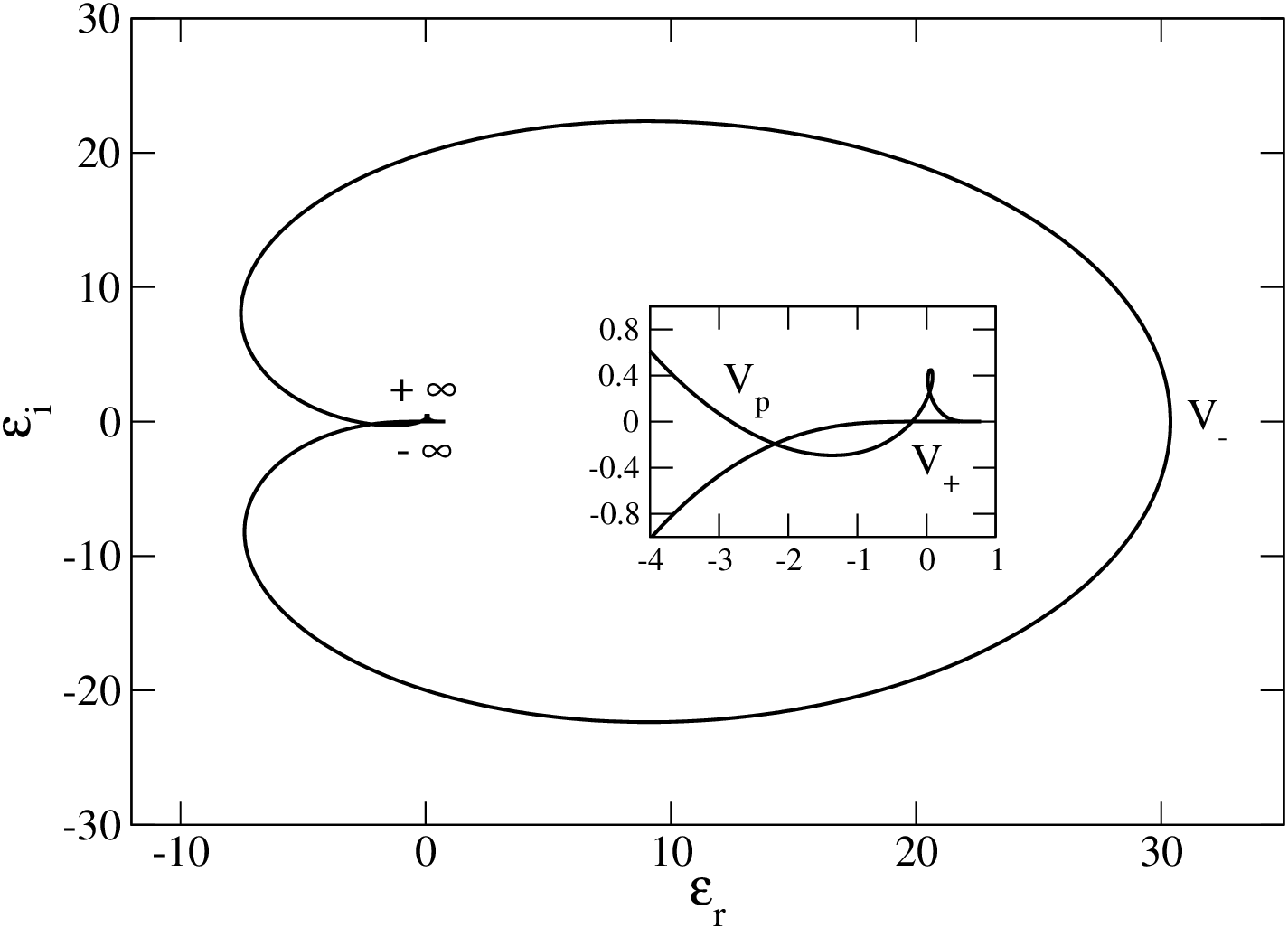}
} \caption[]{$\Delta>\Delta_*$: Nyquist curve for $y_*<y<y_c$ and $\eta>\eta_c^{(+)}$
(specifically $\Delta=50$, $y=5$ and $\eta=30$). The DF has a global maximum at $V_{-}$, a minimum at $V_p$ and a local maximum at $V_{+}$. The DF is stable (with respect to
this perturbation) because the Nyquist curve does not encircle the
origin.  Case (+ - -).}
\label{data_delta50_y5_n30}
\end{figure}

\section{Conclusion}

We have carried out an exhaustive study of the linear dynamical
stability/instability of an infinite and homogeneous self-gravitating
medium with respect to perturbations with wavenumber $k$.  This is the
classical Jeans problem that describes the growth of perturbations and
the emergence of structures (like stars and galaxies) in astrophysics
and cosmology. We have considered the case of a collision-dominated
gas described by the Euler equation or the case of a collisionless
stellar system described by the Vlasov equation. In the latter case,
we have studied single-humped distributions as well as symmetric and
asymmetric double-humped distributions. We have used the Nyquist
theorem to settle the stability of the system. Detailed stability
diagrams in the $(v_a,k)$ plane for different values of the asymmetry
parameter $\Delta$ have been obtained. We have also derived general
analytical results concerning the dispersion relation. Finally, we
have studied the case of Coulombian plasmas in parallel.  The general
methods developed here to study the stability of a homogeneous
self-gravitating medium can have applications in other domains, not
necessarily astrophysics, where the system is described by the Euler
or the Vlasov equations coupled to a mean field equation for the
potential. One example is the HMF model (Chavanis \& Delfini 2009).
Although the structure of the problem is the same, the results and the
stability diagrams differ because the potential of interaction is
different. Many generalizations of our work are possible, changing
the potential of interaction or the class of distribution functions,
but the present paper provides many methods and analytical results
that can be useful to investigate more general situations.

\appendix

\section{Equivalence between the variational problems  $(\ref{vh4})$ and
$(\ref{ep10})$}
\label{sec_eqa}

The equivalence between the variational problems $(\ref{vh4})$ and
$(\ref{ep10})$ has been discussed in previous works (Chavanis
2006a,2008b).  We shall here provide a new derivation of this
result. To that purpose, we shall show the equivalence between the
stability conditions (\ref{ep10b}) and (\ref{vh4b}).

First of all, using the identities $f'(\epsilon)=-\beta/C''(f)$ (see Sec. \ref{sec_vh}) and $p'(\rho)=-\rho/\rho'(\Phi)$ (see Sec. \ref{sec_ep}), we can rewrite the second variations (\ref{vh4b}) and (\ref{ep10b}) in the form
\begin{eqnarray}
\label{eqam1}
\delta^2 F=-\frac{1}{2}\int\frac{(\delta f)^{2}}{f'(\epsilon)}\, d{\bf r}d{\bf v}+\frac{1}{2}\int\delta\rho\delta\Phi\, d{\bf r},
\end{eqnarray}
\begin{eqnarray}
\label{eqam2}
\delta^2 {\cal W}=-\frac{1}{2}\int\frac{(\delta \rho)^{2}}{\rho'(\Phi)}\, d{\bf r}+\frac{1}{2}\int\delta\rho\delta\Phi\, d{\bf r}.
\end{eqnarray}
We shall determine the perturbation $\delta f_{*}({\bf r},{\bf v})$ that minimizes $\delta^{2}F[\delta f]$ given by Eq. (\ref{eqam1}) with the constraint $\delta \rho=\int \delta f\, d{\bf v}$. Since the specification of  $\delta\rho$ determines $\delta\Phi$, hence the second integral in Eq. (\ref{eqam1}), we can write the variational problem in the form
\begin{eqnarray}
\label{eqa1}
\delta\left (-\frac{1}{2}\int\frac{(\delta f)^2}{f'(\epsilon)}\, d{\bf r}d{\bf v}\right )-\int\lambda({\bf r})\delta\left (\int \delta f\, d{\bf v}\right )\, d{\bf r}=0,\nonumber\\
\end{eqnarray}
where $\lambda({\bf r})$ is a Lagrange multiplier. This gives
\begin{eqnarray}
\label{eqa2}
\delta f_*=-\lambda({\bf r}) f'(\epsilon),
\end{eqnarray}
and it is a global minimum of $\delta^{2}F[\delta f]$ with the previous constraint since $\delta^2(\delta^2 F)=-\frac{1}{2}\int \frac{\delta(\delta f)^2}{f'(\epsilon)}\, d{\bf r}d{\bf v}\ge 0$ (the constraint is linear in $\delta f$ so its second variations vanish). The Lagrange multiplier is determined from the constraint $\delta \rho=\int \delta f\, d{\bf v}$  yielding $\delta\rho=-\lambda\int f'(\epsilon)\, d{\bf v}$. Therefore, the optimal perturbation (\ref{eqa2}) can be finally written
\begin{eqnarray}
\label{eqa3}
\delta f_*=\frac{\delta\rho}{\int f'(\epsilon)\, d{\bf v}}f'(\epsilon).
\end{eqnarray}
Therefore, $\delta^{2}F[\delta f]\ge \delta^{2}F[\delta f_*]$. Explicating $\delta^{2}F[\delta f_*]$ using Eqs. (\ref{eqam1}) and (\ref{eqa3}), we obtain
\begin{eqnarray}
\label{eqa3b}
\delta^{2}F[\delta f]\ge -\frac{1}{2}\int \frac{(\delta\rho)^2}{\int f'(\epsilon)\, d{\bf v}}\, d{\bf r}+\frac{1}{2}\int\delta\rho\delta\Phi\, d{\bf r}.
\end{eqnarray}
Finally, using $\rho'(\Phi)=\int f'(\epsilon)\, d{\bf v}$, the foregoing inequality can be rewritten
\begin{eqnarray}
\label{eqa4}
\delta^{2}F[\delta f]\ge -\frac{1}{2}\int \frac{(\delta\rho)^2}{\rho'(\Phi)}\, d{\bf r}+\frac{1}{2}\int\delta\rho\delta\Phi\, d{\bf r}
\equiv \delta^{2}{\cal W}[\delta\rho],\nonumber\\
\end{eqnarray}
where the r.h.s. is precisely the functional (\ref{eqam2}). Furthermore, there is equality in Eq. (\ref{eqa4}) iff $\delta f=\delta f_*$. This proves that the stability criteria (\ref{ep10b}) and (\ref{vh4b}) are equivalent. Indeed: (i) if inequality (\ref{ep10b}) is fulfilled for all perturbations $\delta\rho$ that conserve mass, then according to Eq. (\ref{eqa4}), we know that inequality (\ref{vh4b}) is fulfilled for all perturbations $\delta f$ that conserve mass. (ii) if there exists a perturbation $\delta\rho_*$ that makes $\delta^{2}{\cal W}[\delta\rho]<0$, then the perturbation $\delta f_*$ given by Eq. (\ref{eqa3}) with $\delta\rho=\delta\rho_*$ makes $\delta^{2}F[\delta f]=\delta^{2}{\cal W}[\delta\rho]<0$. In conclusion, the stability criteria (\ref{ep10b}) and (\ref{vh4b}) are equivalent.

\section{Fluid equations}
\label{sec_fee}

In a collisional gas, the evolution of the distribution function $f({\bf r},{\bf v},t)$ is governed by Boltzmann's transport equation
\begin{equation}
\label{fe1}
{\partial f\over\partial t}+{\bf v}\cdot {\partial f\over\partial {\bf r}}-\nabla\Phi\cdot {\partial f\over\partial {\bf v}}=\left ({\partial f\over\partial t}\right )_{coll},
\end{equation}
where $(\partial f/\partial t)_{coll}$ is the collision operator. This operator locally conserves the mass, the impulse and the kinetic energy. The Boltzmann equation relaxes towards the Maxwellian distribution  for $t\rightarrow +\infty$ due to collisions. From Eq. (\ref{fe1}), we can easily derive a hierarchy of hydrodynamic equations satisfied by the local moments of the velocity distribution. Defining the density $\rho=\int f\, d{\bf v}$, the local velocity ${\bf u}=\int f{\bf v}\, d{\bf v}$ and the kinetic temperature $\frac{d}{2}\rho T=\frac{1}{2}\int f w^2\, d{\bf v}$ (where ${\bf w}={\bf v}-{\bf u}$ is the relative velocity and $d$ the dimension of space), the first three equations of this hierarchy are
\begin{equation}
\label{fe2}
{\partial\rho\over\partial t}+\nabla \cdot (\rho {\bf u})=0,
\end{equation}
\begin{equation}
\label{fe3}
{\partial {u_i}\over\partial t}+u_j\frac{\partial {u}_i}{\partial x_j}=-{1\over\rho}\frac{\partial P_{ij}}{\partial x_j}-\frac{\partial\Phi}{\partial x_i},
\end{equation}
\begin{equation}
\label{fe4}
\frac{d}{2}\rho\left (\frac{\partial T}{\partial t}+{\bf u}\cdot \nabla T\right )=-\nabla {\bf q}-P_{ij}\frac{\partial u_j}{\partial x_i},
\end{equation}
where $P_{ij}=\int fw_iw_j\, d{\bf v}$ is the pressure tensor and ${\bf q}=\frac{1}{2}\int fw^2{\bf w}\, d{\bf v}$ is the current of heat. These equations are known as the Maxwell equations of transfer (van Kampen 1957) or as the Jeans equations (Binney \& Tremaine 1987). For a collision-dominated gas, we can close this hierarchy of hydrodynamic equations by using a Chapman-Enskog expansion which is valid when the mean free path in the gas is short. The zeroth-order approximation amounts to making the local thermodynamic equilibrium (L.T.E.) assumption
\begin{eqnarray}
\label{fe5}
f_{L.T.E.}({\bf r},{\bf v},t)=\left (\frac{1}{2\pi T({\bf r},t)}\right )^{d/2}\rho({\bf r},t) e^{-\frac{  ({\bf v}-{\bf u}({\bf r},t))^2}{2T({\bf r},t)}}.\nonumber\\
\end{eqnarray}
This yields ${\bf q}={\bf 0}$ and $P_{ij}=p\delta_{ij}$ where $p=\frac{1}{d}\int f w^2\, d{\bf v}=\rho T$ is the pressure. In that case, we obtain the Euler equations (Huang 1966)
\begin{equation}
\label{fe6}
{\partial\rho\over\partial t}+\nabla \cdot (\rho {\bf u})=0,
\end{equation}
\begin{equation}
\label{fe7}
{\partial {\bf u}\over\partial t}+{\bf u}\cdot \nabla {\bf u}=-{1\over\rho}\nabla p-\nabla\Phi,
\end{equation}
\begin{equation}
\frac{d}{2}\left (\frac{\partial T}{\partial t}+{\bf u}\cdot \nabla T\right )+T\nabla\cdot {\bf u}=0.
\label{fe8}
\end{equation}
\begin{equation}
\label{fe9}
p=\rho T.
\end{equation}
Equation (\ref{fe8}) can be rewritten in the form
\begin{equation}
\frac{d}{dt}\left (\frac{\rho}{T^{d/2}}\right )=0,
\label{fe10}
\end{equation}
where $d/dt=\partial/\partial t+{\bf u}\cdot \nabla$ is the material derivative. This equation expresses the conservation of the entropy by the Euler equation (dissipative effects appear at the next order in the Chapman-Enskog expansion leading to the Navier-Stokes equations). We have therefore a closed system of hydrodynamic equations. The steady states correspond to ${\bf u}={\bf 0}$ and to the condition of hydrostatic equilibrium (\ref{ep7}). However, at the level of the Euler equations, nothing determines $T({\bf r})$ at equilibrium, i.e. there exists steady states of Eqs. (\ref{fe6})-(\ref{fe9}) with any temperature profile $T({\bf r})$. Two cases are particularly relevant. If we consider that the collisions had time to establish a statistical equilibrium state, then $T({\bf r})=T$ is uniform at equilibrium yielding an isothermal equation of state $p({\bf r})=\rho({\bf r})T$. By contrast, if collisions had not time to establish a statistical  equilibrium state, we can consider a steady state where the specific entropy is uniform in the whole system: $s({\bf r})=s$. This is equivalent to having  $\rho({\bf r})/T({\bf r})^{d/2}$, $p({\bf r})/T({\bf r})^{(d+2)/2}$ or $p({\bf r})/\rho({\bf r})^{(d+2)/d}$ spatially uniform at equilibrium. We then obtain a polytropic equation of state  $p({\bf r})=K\rho({\bf r})^{\gamma}$ with
\begin{equation}
\label{fe11}
\gamma=\frac{c_p}{c_v}=\frac{d+2}{d}.
\end{equation}
In that case, the temperature behaves like $T({\bf r})=K\rho({\bf r})^{\gamma-1}$. The isothermal distribution is justified by taking into account dissipative effects (e.g. using the Navier-Stokes equations) and the polytropic distribution is justified by completely ignoring dissipative effects. However, even if an isothermal distribution has been established under the effect of collisions, when we investigate the dynamical stability of such a distribution, we can usually ignore dissipative effects and use the Euler equations (\ref{fe6})-(\ref{fe9}). This is because the dynamical time is in general much shorter than the collisional relaxation time.

If we consider the stability of an infinite homogeneous self-gravitating system or plasma described by the hydrodynamic equations (\ref{fe6})-(\ref{fe9}), we find that the dispersion relation is of the form (\ref{jeans1}) or (\ref{pep7}) with
\begin{equation}
\label{fe12}
c_s^2=\frac{\delta p}{\delta\rho}.
\end{equation}
In the present case, $\delta p/\delta\rho\neq p'(\rho)$ since the gas is not barotropic. Linearizing Eq. (\ref{fe10}) around the steady state, it reduces to
\begin{equation}
\label{fe13}
\frac{\partial}{\partial t}\delta \left (\frac{\rho}{T^{d/2}}\right )=0,
\end{equation}
so that $\rho/T^{d/2}$, $p/T^{(d+2)/2}$ or $p/\rho^{(d+2)/d}$ are independent on time in the linear regime. Therefore, the perturbations are adiabatic and we have
\begin{equation}
\label{fe14}
\frac{\delta p}{p}=\frac{d+2}{d}\frac{\delta \rho}{\rho},
\end{equation}
yielding $\delta p=\gamma T\delta \rho$ hence
\begin{equation}
\label{fe15}
c_s^2=\gamma T.
\end{equation}
Regrouping all these results, the dispersion relation for a self-gravitating system or a plasma described by the (non barotropic) Euler equations (\ref{fe6})-(\ref{fe9}) are
\begin{eqnarray}
\omega^{2}= \gamma T k^{2}-4\pi G\rho,
\label{fe16}
\end{eqnarray}
\begin{eqnarray}
\label{fe17}
\omega^2= \gamma T  k^2+\omega_p^2.
\end{eqnarray}
The are valid for any equilibrium distribution, e.g. isothermal or polytropic.

Let us now consider the case of a collisionless system described by the Vlasov equation obtained from Eq. (\ref{fe1}) by neglecting the collision term. The hydrodynamic equations (\ref{fe2})-(\ref{fe4}) remain valid but, in the present case, the L.T.E assumption is not justified anymore so that the hierarchy of equations cannot be closed. However, for $k\rightarrow 0$, the dispersion relation based on the Vlasov equation is given for a self-gravitating system by (see Sec. \ref{sec_mon}):
\begin{eqnarray}
\omega_{i}^2=4\pi G\rho-3Tk^2-...
\label{mon7}
\end{eqnarray}
This dispersion relation is consistent with Eq. (\ref{fe16}) with $\gamma=3$ (corresponding to $d=1$).
Therefore, for large wavelengths, a collisionless stellar system can be treated by fluid equations experiencing a one dimensional adiabatic perturbation. For $k\rightarrow 0$, the dispersion relation based on the Vlasov equation are given for a plasma by (see Sec. \ref{sec_pvp}):
\begin{eqnarray}
\label{pvp4b}
\omega_r^2=\omega_p^2+3Tk^2+...,\qquad \omega_{i}=\frac{\pi\omega_{p}^{3}}{2\rho k^2}f'\left (\frac{\omega_{r}}{k}\right ).
\end{eqnarray}
The Landau damping is due to collective effects and cannot be obtained in a hydrodynamic approximation. However, the dispersion relation for the pulsation $\omega_r$ is consistent with Eq. (\ref{fe17}) with $\gamma=3$ (corresponding to $d=1$). Therefore, for large wavelengths, the oscillations of a collisionless plasma  can be treated by fluid equations experiencing a one dimensional adiabatic compression\footnote{As noted by van Kampen (1957), this differs from the results of Thomson \& Thomson (1933) whose considered isothermal perturbations and from the results of Gross (1951) who considered three dimensional adiabatic perturbations.}.

\section{Calculation of $I$ for isothermal and polytropic distributions}
\label{sec_cali}

Let us define
\begin{eqnarray}
I=\int_{-\infty}^{+\infty}\frac{f'(v)}{v}\, dv.
\label{cali1}
\end{eqnarray}
For the isothermal distribution (\ref{md1}), the integration is straightforward and yields
\begin{eqnarray}
I_{iso}=-\beta \rho.
\label{cali2}
\end{eqnarray}
For the polytropic distribution (\ref{pdr1}), the integrals can be expressed in
terms of Gamma functions. After simplification, using
$\Gamma(x+1)=x\Gamma(x)$, we obtain
\begin{eqnarray}
I_{poly}=-\frac{\beta \rho}{\gamma},
\label{cali3}
\end{eqnarray}
which includes the isothermal distribution (\ref{cali2}) as a special case corresponding to $\gamma=1$.

\section{Direct calculation of $W(ix)$}
\label{sec_wi}

Let us compute $W(z)$ defined by Eq. (\ref{md5}) when $z=ix$. When $x>0$, using Eq. (\ref{lsv5}), we have
\begin{eqnarray}
W(ix)=\frac{1}{\sqrt{2\pi}}\int_{-\infty}^{+\infty}\frac{t}{t-ix}e^{-t^2/2}\, dt.
\label{wi1}
\end{eqnarray}
Multiplying the numerator by $t+ix$ and noting that the imaginary part vanishes by symmetry, we obtain
\begin{eqnarray}
W(ix)=\frac{1}{\sqrt{2\pi}}\int_{-\infty}^{+\infty}\frac{t^2}{t^2+x^2}e^{-t^2/2}\, dt.
\label{wi2}
\end{eqnarray}
For any real $x$, we have the identity
\begin{eqnarray}
\frac{1}{\sqrt{2\pi}}\int_{-\infty}^{+\infty}\frac{t^2}{t^2+x^2}e^{-t^2/2}\, dt=\nonumber\\
1-\sqrt{\frac{\pi}{2}}|x|e^{x^2/2}\left \lbrack 1-{\rm erf}\left (\frac{|x|}{\sqrt{2}}\right )\right \rbrack.
\label{wi3}
\end{eqnarray}
Therefore, for $x>0$, we obtain
\begin{eqnarray}
W(ix)=1-\sqrt{\frac{\pi}{2}}x e^{x^2/2}{\rm erfc}\left (\frac{x}{\sqrt{2}}\right ).
\label{wi4}
\end{eqnarray}
When  $x<0$, using Eq. (\ref{lsv7}), we have
\begin{eqnarray}
W(ix)=\frac{1}{\sqrt{2\pi}}\int_{-\infty}^{+\infty}\frac{t}{t-ix}e^{-t^2/2}\, dt-\sqrt{2\pi}x e^{x^2/2},
\label{wi5}
\end{eqnarray}
which is the same as
\begin{eqnarray}
W(ix)=\frac{1}{\sqrt{2\pi}}\int_{-\infty}^{+\infty}\frac{t^2}{t^2+x^2}e^{-t^2/2}\, dt-\sqrt{2\pi}x e^{x^2/2}.
\label{wi6}
\end{eqnarray}
Using identity (\ref{wi3}) and the fact that ${\rm erf}(-x)=-{\rm erf}(x)$, we finally obtain for $x<0$:
\begin{eqnarray}
W(ix)=1-\sqrt{\frac{\pi}{2}}x e^{x^2/2}{\rm erfc}\left (\frac{x}{\sqrt{2}}\right ).
\label{wi7}
\end{eqnarray}
Comparing Eqs. (\ref{wi4}) and (\ref{wi7}), and noting that $W(0)=1$, we see that such expressions are valid for any real $x$.

\section{The case of two cold streams}
\label{sec_cold}

We consider a distribution function of the form
\begin{equation}
f(v)=\frac{\rho}{2}\left (\delta(v+v_a)+\delta(v-v_a)\right),
\label{app1}
\end{equation}
corresponding to two symmetric cold streams ($T=0$) moving in opposite direction with velocity $\pm v_a$.
We shall consider the stability of this distribution with respect to the Vlasov equation for self-gravitating systems, plasmas and for the HMF model.

{\it Self-gravitating systems:} integrating by parts, the dispersion relation (\ref{lsv4}) can be rewritten
\begin{equation}
1+\frac{4\pi G}{k^2}\int \frac{f(v)}{\left (v-\frac{\omega}{k}\right )^2}\, dv=0.\label{app2}
\end{equation}
For the distribution (\ref{app1}), we obtain\footnote{The same dispersion relation is obtained if we consider two gaseous cold streams described by the Euler equations (cf Talwar \& Kalra 1966).}
\begin{equation}
k^2=-2\pi G\rho\left \lbrack \frac{1}{\left (v_a+\frac{\omega}{k}\right )^2}+\frac{1}{\left (v_a-\frac{\omega}{k}\right )^2}\right\rbrack.\label{app3}
\end{equation}
This is a $4$th degree equation for $\omega$ with real coefficients. Therefore, it admits $4$ complex solutions that occur in pairs which are complex conjugate of one another, i.e. $\omega_{1,2}=\omega_r\pm i\omega_i$ and  $\omega_{3,4}=\omega'_r\pm i\omega'_i$. Accordingly, the system is stable for a perturbation of wavenumber $k$ iff the four roots are real ($\omega_i=\omega'_i=0$). However, it is clear that Eq. (\ref{app3}) has no solution with real $\omega$ so that the distribution (\ref{app1}) is always unstable (to any wavenumber). In fact, the roots of this equation can be obtained analytically. They are given by
\begin{equation}
\omega^2=v_a^2k^2-2\pi G\rho\left (1\pm \sqrt{1-\frac{2k^2v_a^2}{\pi G\rho}}\right ).\label{app4}
\end{equation}
Let us introduce the wavenumber
\begin{equation}
k_0=\left (\frac{\pi G\rho}{2}\right )^{1/2}\frac{1}{v_a}.\label{app5}
\end{equation}
If $k<k_0$, then $\omega^2$ is real negative so that $\omega$ is purely imaginary. In that case, the perturbation grows without oscillating ($\omega_r=0$, $\omega_i>0$). On the other hand, if $k>k_0$, then $\omega^2$ and $\omega$ are complex. In that case, the perturbation grows and  oscillates ($\omega_r\neq 0$, $\omega_i>0$). This is called overstability. Therefore, two cold streams are unstable for $k<k_0$ and overstable for $k>k_0$. If we consider an asymmetric distribution, we find that the system is always overstable.

{\it Plasmas:} in the case of plasmas, performing the transformation $G\rightarrow -e^2/m^2$, the dispersion relation becomes\footnote{The same dispersion relation is obtained if we consider two gaseous cold streams described by the Euler equations (cf Kahn 1958).}
\begin{equation}
k^2=\frac{2\pi e^2\rho}{m^2}\left \lbrack \frac{1}{\left (v_a+\frac{\omega}{k}\right )^2}+\frac{1}{\left (v_a-\frac{\omega}{k}\right )^2}\right\rbrack.\label{app6}
\end{equation}
From the graph of the function $k^2=F(\omega/k)$, we find that Eq. (\ref{app6}) has $4$ real roots if $k>k_c$ where
\begin{equation}
k_c=\left (\frac{4\pi e^2\rho}{m^2}\right )^{1/2}\frac{1}{v_a}.\label{app7}
\end{equation}
Therefore, the distribution (\ref{app1}) is stable for $k>k_c$ and unstable for $k<k_c$. The roots of Eq. (\ref{app6}) are given by
\begin{equation}
\omega^2=v_a^2k^2+\frac{2\pi e^2\rho}{m^2}\left (1\pm \sqrt{1+\frac{2k^2v_a^2m^2}{\pi e^2\rho}}\right ).\label{app8}
\end{equation}
The solutions with the sign $+$ correspond to $\omega^2$ real positive (pure oscillations). The solutions with the sign $-$ correspond to $\omega^2$ real positive if $k>k_c$ (pure oscillations) and to $\omega^2$ real negative if $k<k_c$. In that case, $\omega$ is purely imaginary so that the perturbation grows without oscillating ($\omega_r=0$, $\omega_i>0$). There is no overstable mode. If we consider an asymmetric distribution with asymmetry $\Delta$, we find that $k_c^2=4\pi e^2\rho/m^2v_a^2$ as for the symmetric case. For $k<k_c$, the system is overstable.

{\it HMF model:} for the attractive HMF model, performing the transformation $k\rightarrow 1$  and $4\pi G\rightarrow k/2$, the dispersion relation becomes
\begin{equation}
1=-\frac{k\rho}{4}\left \lbrack \frac{1}{\left (v_a+\omega\right )^2}+\frac{1}{\left (v_a-\omega\right )^2}\right\rbrack,\label{app9}
\end{equation}
where $k$ is the coupling constant. The distribution (\ref{app1}) is always unstable. The roots of Eq. (\ref{app9}) are given by
\begin{equation}
\omega^2=v_a^2-\frac{k\rho}{4}\left (1\pm \sqrt{1-\frac{16 v_a^2}{k\rho}}\right ).\label{app10}
\end{equation}
Let us introduce the velocity
\begin{equation}
v_0=\left (\frac{k\rho}{16}\right )^{1/2}.\label{app11}
\end{equation}
If $v_a<v_0$, then $\omega^2$ is real negative so that $\omega$ is purely imaginary. In that case, the perturbation increases without oscillating ($\omega_r=0$, $\omega_i>0$). On the other hand, if $v_a>v_0$, then $\omega^2$ and $\omega$ are complex. In that case, the perturbation increases and  oscillates ($\omega_r\neq 0$, $\omega_i>0$). Therefore, two cold streams are unstable for $v_a<v_0$ and overstable for $v_a>v_0$. If we consider an asymmetric distribution, we find that the system is always overstable.

For the repulsive HMF model, performing the transformation $k\rightarrow -k$, the dispersion relation becomes
\begin{equation}
1=\frac{k\rho}{4}\left \lbrack \frac{1}{\left (v_a+\omega\right )^2}+\frac{1}{\left (v_a-\omega\right )^2}\right\rbrack.\label{app12}
\end{equation}
From the graph of the curve $F(\omega)=1$, we find that Eq. (\ref{app12}) has $4$ real roots if $v_a>v_c$ where
\begin{equation}
v_c=\left (\frac{k\rho}{2}\right )^{1/2}.\label{app13}
\end{equation}
Therefore, the distribution (\ref{app1}) is stable for $v_a>v_c$ and unstable for $v_a<v_c$. The roots of Eq. (\ref{app12}) are given by
\begin{equation}
\omega^2=v_a^2+\frac{k\rho}{4}\left (1\pm \sqrt{1+\frac{16 v_a^2}{k\rho}}\right ).\label{app14}
\end{equation}
The solutions with the sign $+$ correspond to $\omega^2$ real positive (pure oscillations). The solutions with the sign $-$ correspond to $\omega^2$ real positive if $v_a>v_c$ (pure oscillations) and to $\omega^2$ real negative if $v_a<v_c$. In that case, $\omega$ is purely imaginary so that the perturbation increases without oscillating ($\omega_r=0$, $\omega_i>0$). There is no overstable mode. If we consider an asymmetric distribution with asymmetry $\Delta$, we find that $v_c^2=(\Delta+1)k\rho/4$. For $v<v_c$, the system is overstable.

\end{document}